\documentclass[twocolumn,tighten]{aastex63}

\usepackage{natbib}
\usepackage{epsfig}
\usepackage{amsmath}
\usepackage{xcolor}
\usepackage{hyperref}
\usepackage[encapsulated]{CJK}
\usepackage{ucs}
\usepackage[utf8x]{inputenc}
\usepackage{url}
\usepackage{bm}

\graphicspath{{./Figures/}}

\newcommand{\shortminus}{\scalebox{0.75}[1.0]{\ensuremath{-}}}
\newcommand{\coone}{\textup{CO}\,\ensuremath{(1{\shortminus}0)}}
\newcommand{\cotwo}{\textup{CO}\,\ensuremath{(2{\shortminus}1)}}
\newcommand{\cothree}{\textup{CO}\,\ensuremath{(3{\shortminus}2)}}
\newcommand{\hcnone}{\textup{HCN}\,\ensuremath{(1{\shortminus}0)}}
\newcommand{\hcopone}{\textup{HCO}\ensuremath{^{+}\,(1{\shortminus}0)}}

\makeatletter
\DeclareRobustCommand{\ion}[2]{%
  \text{#1\,\check@mathfonts\fontsize\sf@size\z@\selectfont #2}%
}
\makeatother

\newcommand{\acounits}{\mbox{M$_\odot$ pc$^{-2}$ (K km s$^{-1}$)$^{-1}$}}

\newcommand{\SFR}{\mbox{\textit{SFR}}}

\newcommand{\pairs}{152}

\newcommand{\OSU}{\affil{Department of Astronomy, The Ohio State University, 140 West 18th Avenue, Columbus, Ohio 43210, USA}}

\newcommand{\Alberta}{\affil{Department of Physics, University of Alberta, Edmonton, AB T6G 2E1, Canada}}

\newcommand{\CCAPP}{\affil{Center for Cosmology and Astroparticle Physics, 191 West Woodruff Avenue, Columbus, OH 43210, USA}}

\newcommand{\CfA}{\affil{Harvard-Smithsonian Center for Astrophysics, 60 Garden Street, Cambridge, MA 02138, USA}}

\newcommand{\Heidelberg}{\affil{Astronomisches Rechen-Institut, Zentrum f\"{u}r Astronomie der Universit\"{a}t Heidelberg, M\"{o}nchhofstra\ss e 12-14, D-69120 Heidelberg, Germany}}

\newcommand{\IRAM}{\affil{Institut de Radioastronomie Millim\'{e}trique (IRAM), 300 Rue de la Piscine, F-38406 Saint Martin d'H\`{e}res, France}}

\newcommand{\ITA}{\affil{Universit\"{a}t Heidelberg, Zentrum f\"{u}r Astronomie, Institut f\"{u}r Theoretische Astrophysik, Albert-Ueberle-Str 2, D-69120 Heidelberg, Germany}}

\newcommand{\IWR}{\affil{Universit\"{a}t Heidelberg, Interdisziplin\"{a}res Zentrum f\"{u}r Wissenschaftliches Rechnen, Im Neuenheimer Feld 205, D-69120 Heidelberg, Germany}}

\newcommand{\Maryland}{\affil{Department of Astronomy, University of Maryland, College Park, MD 20742, USA}}

\newcommand{\MPE}{\affil{Max-Planck-Institut f\"{u}r extraterrestrische Physik, Giessenbachstra{\ss}e 1, D-85748 Garching, Germany}}

\newcommand{\MPIA}{\affil{Max-Planck-Institut f\"{u}r Astronomie, K\"{o}nigstuhl 17, D-69117, Heidelberg, Germany}}

\newcommand{\OAN}{\affil{Observatorio Astron\'{o}mico Nacional (IGN), C/Alfonso XII, 3, E-28014 Madrid, Spain}}

\newcommand{\ObsParis}{\affil{Sorbonne Universit\'{e}, Observatoire de Paris, Universit\'{e} PSL, CNRS, LERMA, F-75014, Paris, France}}

\newcommand{\Toulouse}{\affil{Universit\'{e} de Toulouse, UPS-OMP, IRAP, F-31028 Toulouse cedex 4, France}}

\newcommand{\UBonn}{\affil{Argelander-Institut f\"ur Astronomie, Universit\"at Bonn, Auf dem H\"ugel 71, 53121 Bonn, Germany}}

\newcommand{\UCSD}{\affil{Center for Astrophysics and Space Sciences, Department of Physics,  University of California,\\ San Diego, 9500 Gilman Drive, La Jolla, CA 92093, USA}}

\newcommand{\UGent}{\affil{Sterrenkundig Observatorium, Universiteit Gent, Krijgslaan 281 S9, B-9000 Gent, Belgium}}

\newcommand{\UMass}{\affil{University of Massachusetts—Amherst, 710 N. Pleasant Street, Amherst, MA 01003, USA}}

\newcommand{\UWyoming}{\affil{Department of Physics and Astronomy, University of Wyoming, Laramie, WY 82071, USA}}

\newcommand{\LAM}{\affil{
Aix Marseille Universit\'{e}, CNRS, CNES, LAM (Laboratoire d’Astrophysique de Marseille), F-13388 Marseille,
France}}

\newcommand{\INAF}{\affil{INAF -- Osservatorio Astrofisico di Arcetri, Largo E. Fermi 5, I-50157, Firenze, Italy}}

\newcommand{\Tamkang}{\affil{Department of Physics, Tamkang University, No.151, Yingzhuan Rd., Tamsui Dist., New Taipei City 251301, Taiwan}}

\newcommand{\MPIFR}{\affil{Max-Planck-Institut f\"ur Radioastronomie, Auf dem H\"ugel 69, 53121 Bonn, Germany}}

\newcommand{\IGN}{\affil{Centro de Desarrollos Technol\'{o}gicos, Observatorio de Yebes (IGN), 19141 Yebes, Guadalajara, Spain}}

\shorttitle{Low-\textit{J} CO Line Ratios for Local Galaxies}
\shortauthors{Leroy, Rosolowsky, Usero et al.}

\begin{document}

\title{Low-\textit{J} CO Line Ratios From Single Dish CO Mapping Surveys and PHANGS--ALMA}

\author[0000-0002-2545-1700]{Adam~K.~Leroy}
\OSU \CCAPP

\author[0000-0002-5204-2259]{Erik~Rosolowsky}
\Alberta

\author[0000-0003-1242-505X]{Antonio~Usero}
\OAN

\author[0000-0002-4378-8534]{Karin~Sandstrom}
\UCSD

\author[0000-0002-3933-7677]{Eva~Schinnerer}
\MPIA

\author{Andreas~Schruba}
\MPE

\author[0000-0002-5480-5686]{Alberto D. Bolatto}
\Maryland

\author[0000-0003-0378-4667]{Jiayi~Sun \begin{CJK*}{UTF8}{gbsn}(孙嘉懿)\end{CJK*}}
\OSU

\author[0000-0003-0410-4504]{Ashley.~T.~Barnes}
\UBonn

\author[0000-0002-2545-5752]{Francesco Belfiore}
\INAF

\author[0000-0003-0166-9745]{Frank Bigiel}
\UBonn

\author[0000-0002-8760-6157]{Jakob~S. den Brok}
\UBonn

\author[0000-0001-5301-1326]{Yixian Cao}
\LAM

\author[0000-0003-2551-7148]{I-Da Chiang \begin{CJK*}{UTF8}{bkai}(江宜達)\end{CJK*}}
\UCSD

\author[0000-0002-5635-5180]{M\'elanie Chevance}
\Heidelberg

\author[0000-0002-5782-9093]{Daniel A. Dale}
\UWyoming

\author[0000-0002-1185-2810]{Cosima Eibensteiner}
\UBonn

\author[0000-0001-5310-467X]{Christopher M. Faesi}
\UMass

\author[0000-0001-6708-1317]{Simon C. O. Glover}
\ITA

\author[0000-0002-9181-1161]{Annie Hughes}
\Toulouse

\author[0000-0002-9165-8080]{Mar\'ia J. Jim\'enez Donaire}
\OAN
\IGN

\author[0000-0002-0560-3172]{Ralf S.\ Klessen}
\ITA
\IWR

\author[0000-0001-9605-780X]{Eric~W.~Koch}
\CfA

\author[0000-0002-8804-0212]{J.~M.~Diederik Kruijssen}
\Heidelberg

\author[0000-0001-9773-7479]{Daizhong~Liu}
\MPE

\author[0000-0002-6118-4048]{Sharon~E.~Meidt}
\UGent

\author[0000-0002-1370-6964]{Hsi-An Pan}
\MPIA
\Tamkang

\author[0000-0003-3061-6546]{Jérôme Pety}
\IRAM
\ObsParis

\author[0000-0003-1111-3951]{Johannes~Puschnig}
\UBonn

\author[0000-0002-0472-1011]{Miguel~Querejeta}
\OAN

\author[0000-0002-2501-9328]{Toshiki~Saito}
\MPIA

\author[0000-0002-5783-145X]{Amy Sardone}
\OSU \CCAPP

\author[0000-0002-7365-5791]{Elizabeth J. Watkins}
\Heidelberg

\author[0000-0003-4678-3939]{Axel Weiss}
\MPIFR

\author[0000-0002-0012-2142]{Thomas~G.~Williams}
\MPIA

\begin{abstract}
We measure the low-$J$ CO line ratio $R_{21} \equiv \cotwo/\coone$, $R_{32} \equiv \cothree/\cotwo$, and $R_{31} \equiv \cothree/\coone$ using whole-disk CO maps of nearby galaxies. We draw \cotwo\ from PHANGS--ALMA, HERACLES, and follow-up IRAM surveys; \coone\ from COMING and the Nobeyama CO Atlas of Nearby Spiral Galaxies; and \cothree\ from the JCMT NGLS and APEX LASMA mapping. Altogether this yields $76$, $47$, and $29$ maps of $R_{21}$, $R_{32}$, and $R_{31}$ at $20'' \sim 1.3$~kpc resolution, covering $43$, $34$, and $20$ galaxies. Disk galaxies with high stellar mass, $\log (M_\star/\textrm{M}_\odot) = 10.25{-}11$ and star formation rate, $\SFR = 1{-}5$~M$_\odot$~yr$^{-1}$, dominate the sample. We find galaxy-integrated mean values and $16\%{-}84\%$ range of $R_{21} = 0.65\ (0.50{-}0.83)$, $R_{32}=0.50\ (0.23{-}0.59)$, and $R_{31}=0.31\ (0.20{-}0.42)$. We identify weak trends relating galaxy-integrated line ratios to properties expected to correlate with excitation, including $\SFR/M_\star$ and $\SFR/L_{\rm CO}$. Within galaxies, we measure central enhancements with respect to the galaxy-averaged value of ${\sim} 0.18^{+0.09}_{-0.14}$~dex for $R_{21}$, $0.27^{+0.13}_{-0.15}$~dex for $R_{31}$, and $0.08^{+0.11}_{-0.09}$~dex for $R_{32}$. All three line ratios anti-correlate with galactocentric radius and positively correlate with the local star formation rate surface density and specific star formation rate, and we provide approximate fits to these relations. The observed ratios can be reasonably reproduced by models with low temperature, moderate opacity, and moderate densities, in good agreement with expectations for the cold ISM. Because the line ratios are expected to anti-correlate with the \coone-to-H$_2$ conversion factor, $\alpha_{\rm CO}^{1-0}$, these results have general implications for the interpretation of CO emission from galaxies.
\end{abstract}


\section{Introduction}
\label{sec:intro}

Rotational line emission from carbon monoxide (CO) represents the main way to trace the distribution, kinematics, and physical conditions in the molecular interstellar medium in external galaxies \citep[ISM, e.g., see reviews by][]{BOLATTO13B,KLESSEN16}. After several decades focused primarily on the fundamental $^{12}\coone$ transition at $\nu \approx 115$~GHz \citep[e.g.,][]{YOUNG91,YOUNG95,HELFER03}, improvements in (sub)millimeter facilities over the last fifteen years have enabled extensive mapping of nearby galaxies in $^{12}$\cotwo\ and $^{12}$\cothree\ at $\nu \approx 231$~GHz and $\nu \approx 345$~GHz \citep[e.g.,][]{LEROY09,WILSON12}. In the last decade, the Atacama Large Millimeter/\linebreak[0]{}submillimeter Array (ALMA) has come online and revolutionized our view of molecular line emission from galaxies, while also accelerating the trend towards observing multiple CO lines. Thanks to its excellent site and submillimeter sensitivity, ALMA can often map \cotwo\ and \cothree\ several times faster than \coone\ at matched resolution and sensitivity. As a result, ALMA surveys of nearby galaxies have targeted all three low-$J$ CO lines, \cothree, \cotwo, and \coone \citep[e.g.,][]{GARCIABURILLO14,HIROTA18,LEROY21b}.

Meanwhile, studies of redshifted CO emission have also become common, tracing the molecular gas at earlier cosmic epochs. Driven by similar technical considerations, these studies currently focus on \cothree, \cotwo\ or even higher~$J$ transitions \citep[e.g., see reviews by][]{CARILLI13,HODGE20,TACCONI20}. In the near future, observations at high redshift may become even more diverse as the proposed next generation Very Large Array \citep[ngVLA;][]{MURPHY18} will vastly improve our ability to observe \coone\ emission at intermediate and high redshift.

This increased diversity of CO line observations at low and high~$z$ makes the ability to translate between results obtained using these different CO lines crucial. Despite the proliferation of \cotwo\ and \cothree\ studies, many surveys still target \coone, including xCOLD GASS \citep{SAINTONGE17} and CARMA EDGE \citep{BOLATTO17}, the largest low-$z$ single dish and interferometric CO surveys to date. Critical work informing our interpretation of CO emission has also built on observations of a single transition, e.g., \citet{DONOVANMEYER13} focused on \coone\ emission, \citet{SANDSTROM13} studied \cotwo , and \citet[][]{WILSON08} employed \cothree. Well-understood, observationally-tested translations between the various low-$J$ CO lines are required to link these efforts.

Indeed, translations between the different transitions are not straightforward because the ratios among \cothree, \cotwo, and \coone\ also reflect physical conditions in the molecular gas. The observed ratios emerge from an interplay among the distributions of collider density, kinetic temperature, $T_\textrm{kin}$, and column density per line width (see \S\ref{sec:expect}). These in turn depend on the structure, kinematics, and heating mechanisms at play in the cold ISM. The ratios of low-$J$ CO lines thus represent a potentially powerful observational probe of the local physical conditions in the molecular ISM. This potential is complicated by degeneracies in their interpretation and the modest dynamic range in their observed values. This limited dynamic range places relatively strict requirements on observations aiming to measure these line ratios.

In contrast to commonly used ``dense gas tracers'' like \hcnone\ and \hcopone , the CO lines are bright and can be studied across a range of environments \citep[see][regarding relative line strengths]{USERO15}. Numerical simulations can now resolve CO chemistry and predict CO line emission over whole molecular clouds, large parts of a spiral galaxy, or even entire dwarf galaxies \citep[e.g.,][]{GLOVER12,PENAZOLA17,PENALOZA18,GONG20,HU21}, but such calculations remain extremely challenging for tracers of higher density gas \citep[e.g.,][]{ONUS18}. A combined observational, numerical, and analytic approach that leverages ratios among the low-$J$ CO lines and their isotopologues represents a promising path forward to diagnose physical conditions in the molecular gas of galaxies. This approach can become even more powerful when paired with high resolution imaging of the CO emission \citep[e.g., see][]{GALLAGHER18B}, which places constraints on the mean density and kinematics of the cold gas \citep[e.g., see][]{SUN18,SUN20B,ROSOLOWSKY21}. Of course, spectroscopy targeting multiple CO transitions or CO isotopologues has a long history \cite[e.g.,][]{PAPADOPOULOS99,ISRAEL01,ISRAEL03,ISRAEL15,BAYET04,BAYET06,KAMENETZKY14,KAMENETZKY17}. However, most previous work has focused on single-pointing or galaxy-integrated measurements, with a heavy emphasis on galaxy centers and starburst galaxies, including many ultraluminous and luminous infrared galaxies (U/LIRGs). Resolved studies that measure the ratios among multiple CO lines over the full area of ``normal'' star-forming main sequence galaxies remain relatively scarce.

This paper presents new measurements of the $\cotwo/\coone$, $\cothree/\cotwo$, and $\cothree/\coone$ line ratios for nearby galaxies ($D < 40$~Mpc, median $\sim 14$~Mpc) based on maps of CO emission from \citet[][hereafter the ``NRO Atlas'']{KUNO07}, HERACLES \citep{LEROY09}, the JCMT NGLS \citep{WILSON12}, COMING \citep{SORAI19}, PHANGS--ALMA \citep{LEROY21b}, new IRAM 30{-}m \cotwo\ observations (P.I.\ A.~Schruba), and new APEX LASMA \cothree\ observations (P.I.\ A.~Weiss). We measure both resolved and integrated CO line ratios using mapping surveys. All of these surveys except PHANGS--ALMA use receiver arrays on single dish telescopes to cover large areas quickly \citep[e.g., see][]{SCHUSTER07}. Restricting our focus to mapping data allows us to construct identical matched apertures when measuring integrated ratios. This avoids the common issue of mismatched beams, which plagued some earlier studies that relied on pointed observations. These surveys also target many of the largest, closest, best studied galaxies, so the ratios for individual targets are of particular interest. Finally, because we analyze maps, we can measure line ratios associated with distinct regions to, e.g., test for a dependence of excitation on galactocentric radius or star formation rate surface density \citep[e.g., following][]{DENBROK21,YAJIMA21}.

\citet{LEROY09}, \citet{WILSON12}, and \citet{LEROY13} calculated these ratios based on first versions of HERACLES and the JCMT NGLS, and \citet{YAJIMA21} have recently combined HERACLES and COMING. But the number and quality of CO maps of galaxies have grown significantly compared to any study currently in the literature, particularly with the release of PHANGS--ALMA. Quite a few studies have examined these ratios in individual galaxies and noted local variations in individual ratios \citep[e.g.,][]{CROSTHWAITE07,KODA12,VLAHAKIS13,UEDA12,DRUARD14,LAW18,KODA20}, but so far there has been relatively little attempt to synthesize these mapping measurements \citep[though see the beam-matched, single-pointing measurements by][]{SAINTONGE17,LAMPERTI20}.

As a practical matter, we present our study in the context of the PHANGS--ALMA \cotwo\ survey. PHANGS--ALMA mapped \cotwo\ across $90$ nearby galaxies at ${\lesssim} 150$~pc resolution. To aid in the interpretation of these data, we also aim to improve our empirical understanding of the $\cotwo/\coone$ and $\cothree/\cotwo$ ratios. Ultimately, we expect this to improve our ability to estimate the molecular mass and infer an appropriate CO-to-H$_2$ conversion factor for these data. This work complements three other recent or forthcoming studies. \citet{DENBROK21} use $9$ new, high quality $22''$ resolution \coone\ maps from the IRAM \mbox{30-m} telescope to investigate the resolved $\cotwo/\coone$ ratio.  This work also complements the study by T.~Saito et al.\ (in preparation), which investigates the behavior of the \cotwo/\coone\ ratio at higher $4{-}8''$ resolution in four PHANGS--ALMA targets. In scope, our study resembles the recent thorough investigation by \citet{YAJIMA21}, but we take advantage of a larger database of \cotwo\ maps and include \cothree\ in our analysis.

After framing some theoretical and observational expectations (\S\ref{sec:expect}), we describe the data that we use and our measurements (\S\ref{sec:meas}). Then we measure galaxy-integrated line ratios (\S\ref{sec:global}) and compare them to galaxy's integrated properties (\S\ref{sec:intcorr}). Then we examine the resolved behavior of the ratio as a function of galactocentric radius, local \SFR, and stellar mass surface density (\S\ref{sec:local}). Finally, we discuss the implications of our measurements and next steps (\S\ref{sec:discussion}) and then summarize our results (\S\ref{sec:summary}).

\section{Expectations}
\label{sec:expect}

\begin{figure*}[ht!]
\begin{center}
\includegraphics[width=0.49\textwidth]{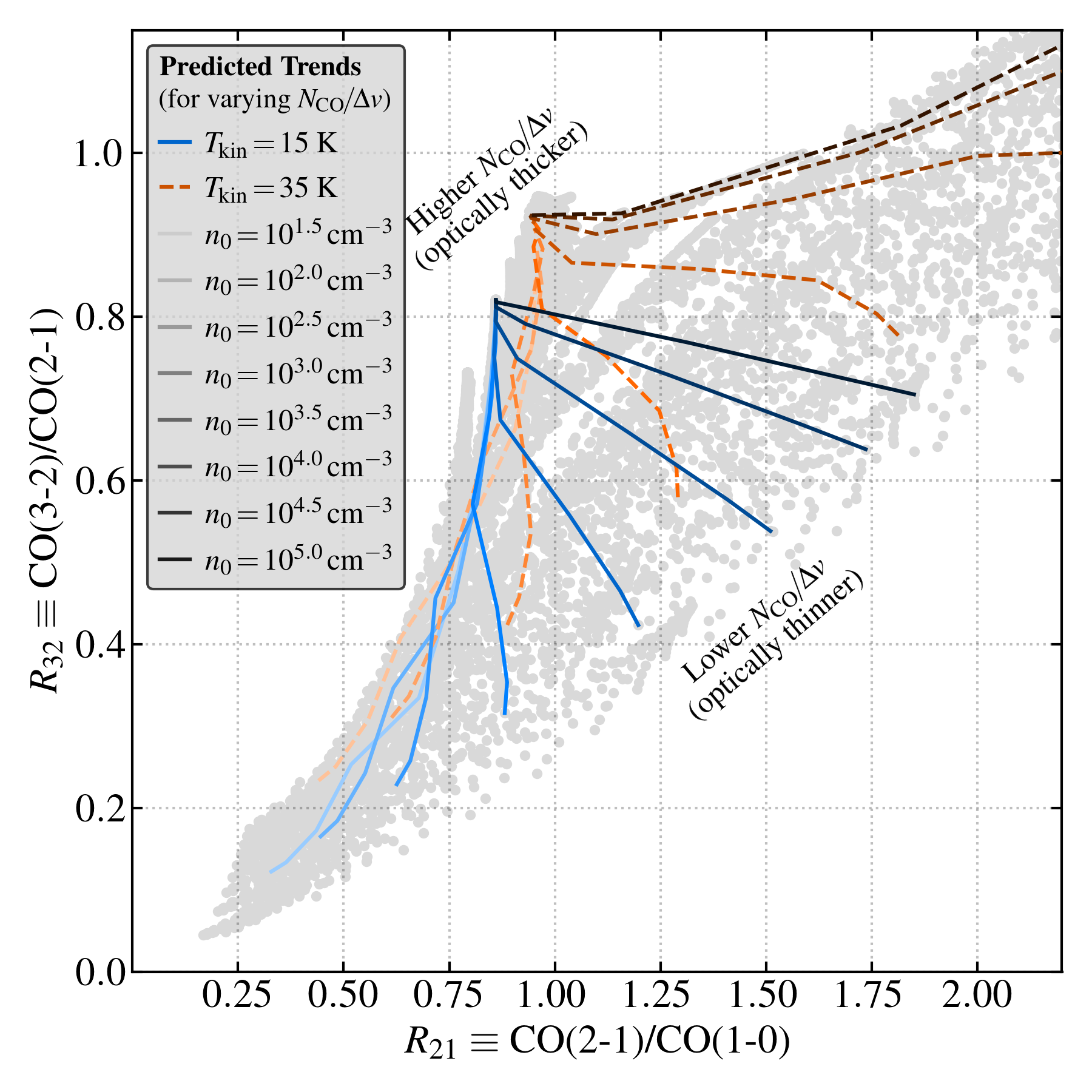}
\includegraphics[width=0.49\textwidth]{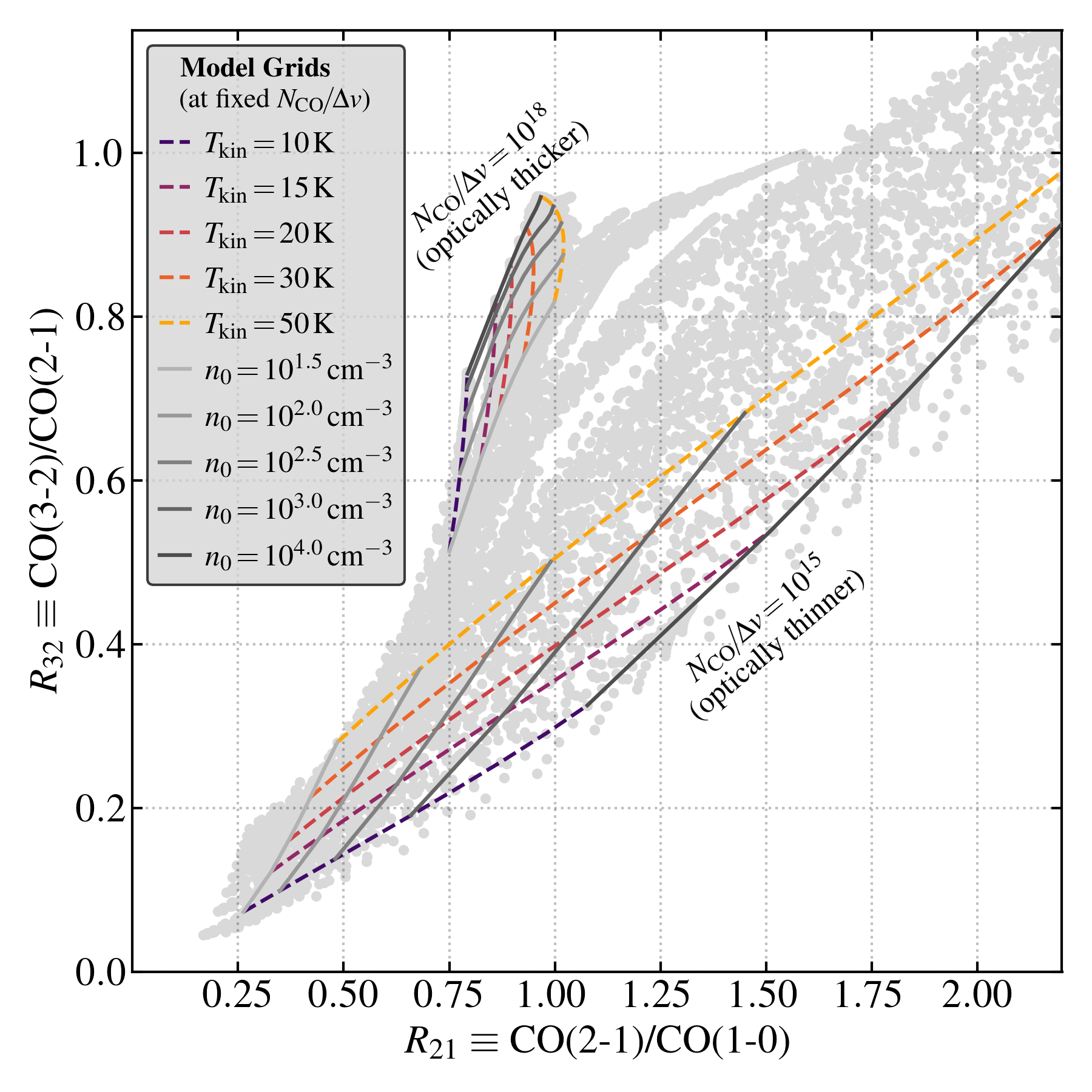}
\end{center}
\caption{\textit{CO line ratios produced by model calculations (gray points) with key trends illustrated.} Predicted line ratios $R_{32}$ vs.\ $R_{21}$ for model calculations using RADEX \citep{VANDERTAK07} and lognormal density distributions following \citet[][]{LEROY17B}. Each gray point shows an individual model. As described in Appendix~\ref{sec:comodels}, each model is characterized by a kinetic temperature ($T_{\rm kin}$), a total CO column density per line width ($N_{\rm CO}/\Delta v$), a lognormal density distribution width ($\sigma$), and a mean H$_2$ collider density ($n_0$). To illustrate how the line ratios change when varying density ($n_0$), temperature ($T_{\rm kin}$), and opacity (set by $N_{\rm CO}/\Delta v$) while holding other parameters fixed, a subset of the data is shown by colored lines for reference. Both panels show the same model grid using gray points while changing the reference. In the left panel, the colored lines show the effect of varying $N_{\rm CO}/\Delta v$, which sets opacity and escape probability, while holding temperature and density constant at a few representative values. In the right panel, dashed lines show the result of varying temperature ($T_{\rm kin}$), and solid lines show the results of varying the mean density ($n_0$) while holding $N_{\rm CO}/\Delta v$ fixed. For all reference lines, we fix $\sigma=0.6$~dex. Note that these lines are intended to be illustrative, not spanning the full phase space of the modeling. The full model grid spanning a plausible range of temperatures and densities is available as a machine readable table in Appendix~\ref{sec:comodels}.
\label{fig:model}}
\end{figure*}

Throughout this paper, we refer to the line ratios as
\begin{equation}\label{eq:rdef}
\begin{split}
R_{21} &\equiv I_{2-1}/I_{1-0} \\
R_{32} &\equiv I_{3-2}/I_{2-1} \\
R_{31} &\equiv I_{3-2}/I_{1-0}
\end{split}
\end{equation}
\noindent Here $I_{2-1}$, for example, refers to the velocity-integrated specific intensity of the \cotwo\ line, with analogous definitions for the other lines. All intensities, luminosities, and line ratios in this paper are calculated in Rayleigh--Jeans brightness temperature units. Line-integrated intensities are presented in K~km~s$^{-1}$ and luminosities given in K~km~s$^{-1}$~pc$^{2}$.

In these ``Kelvin'' units, we expect a line ratio of~$1$ for all ratios for an optically thick source in local thermodynamic equilibrium (LTE) when both transitions sit securely on the Rayleigh--Jeans tail given the source temperature, $T_{\rm kin}$. Note however, that under real conditions the Rayleigh--Jeans criterion, $h \nu \ll k T_{\rm kin}$, may not be satisfied. This will happen, for example considering emission in high frequency transitions from low temperature sources. When the Rayleigh--Jeans criterion is not met, because $T_{\rm kin}$ has low enough values relative to the frequencies of the observed transitions, this will drive the ``thermal'' value of the line ratio observed from an optically thick source to values below $1$. In Appendix~\ref{sec:corrections}, we illustrate the expected opaque LTE value for the relevant ratios, and also show the effects of the cosmic microwave background. For purposes of reading the observational results in this paper, the key point is that at relevant temperatures, $T_{\rm kin} \sim 10{-}20$~K, the expected ratio for opaque, thermalized gas can be as low as ${\sim}0.7$.

\medskip

\noindent \textit{Expectations from models:} Theoretically, the observed line ratios depend on the distributions of temperature, $T_{\rm kin}$, collider density, $n_{\mathrm{H_2}}$, and column density of CO per line width in the gas, $N/\Delta v$. A full discussion of the interplay of these quantities with $R_{32}$, $R_{21}$, and $R_{31}$ lies beyond the scope of this work, and we refer the reader to \citet{BOLATTO13A}, \citet{SHIRLEY15}, \citet{LEROY17B}, and \citet{PENALOZA17}, each of which touches on some aspects of the topic. 

As a brief summary, we illustrate the behavior of $R_{32}$ and $R_{21}$ in Figure~\ref{fig:model}, which plots results from a set of model calculations following \citet{LEROY17B}. In the figure, each point shows the line ratios predicted from a model that has a lognormal distribution of collider densities described by a mean density, $n_0$, and a width, $\sigma$. Each model also has a single fixed $T_{\rm kin}$ and a single value of $N_{\rm CO}/\Delta v$, which we adopt for each individual density layer. We use RADEX \citep{VANDERTAK07} with data from the Leiden Atomic and Molecular Database \citep[LAMDA;][]{SCHOIER05} to calculate predicted emission. The calculations generally follow \citet{LEROY17B} with the distinction that here we fix $N_{\rm CO}/\Delta v$ rather than fixing the optical depth, $\tau$, of a particular line as in that paper. Because these exact calculations may be of general use and are not fully reported in \citet{LEROY17B}, we report the model grid as a machine readable table in Appendix~\ref{sec:comodels}.

Figure~\ref{fig:model} illustrates the combined effects of temperature, density, and optical depth on the line ratios. In the left panel, each line shows fixed $T_{\rm kin}$ and a fixed density distribution, while we vary $N_{\rm CO}/\Delta v$, the total column density of CO molecules normalized to the line width. $N_{\rm CO}/\Delta v$ affects the optical depth and escape probability, and through these also affects the critical density and level populations. The figure shows how the low opacities yielded by low $N_{\rm CO}/\Delta v$ can lead to high, ``super-thermal'' line ratios with values $>1$ in the case of low opacity gas in LTE. Alternatively, for low density gas, low $N_{\rm CO}/\Delta v$ can yield very low line ratios, indicating sub-critically excited gas. Meanwhile higher $N_{\rm CO}/\Delta v$ tends to drive gas closer to optically thick LTE and towards line ratios of ${\sim}1$.

The right panel shows how at fixed $N_{\rm CO}/\Delta v$, the density distribution and temperature also play important roles. Their exact impact depends on the $N_{\rm CO}/\Delta v$. In general, higher density and higher temperature at fixed $N_{\rm CO}/\Delta v$ generally drive both ratios towards higher values. The variations for optically thin gas are more extreme, even allowing line ratios above~$1$, while optically thicker gas shows more dynamic range in $R_{32}$ at the densities illustrated because of the higher excitation requirements of those transitions.

\medskip 

\noindent \textit{Expectations from previous observations:} Previous observations establish some basic expectations for low-$J$ CO line ratios in nearby galaxies:

\begin{enumerate}
\item Normal star-forming galaxies show $R_{21}$ in the range ${\sim} 0.4{-}0.9$ \citep[e.g.,][]{LEROY13,DENBROK21,YAJIMA21}. $R_{31}$ likely shows lower values, sometimes as low as ${\sim} 0.2$, in normal galaxies \citep[e.g.,][]{MAO10,WILSON12}, but also a larger range of reported values \citep[e.g.,][]{MAUERSBERGER99,MAO10,LAMPERTI20}.
\item Starburst galaxies and active galaxies show higher, closer to thermal (i.e., ${\sim} 1$) ratios \citep[e.g.,][]{MAUERSBERGER99,WEISS05,MAO10,LAMPERTI20,YAJIMA21}, consistent with having both higher densities and hotter gas.
\item The central parts of normal star-forming galaxies show systematically higher $R_{21}$ \citep[e.g.,][]{BRAINE92,BRAINE93,LEROY09,LEROY13,ISRAEL20,DENBROK21,YAJIMA21}, consistent with higher densities and hotter gas in the central parts of these galaxies \citep[e.g.,][among many others]{MANGUM13,SUN20B} and with observations showing high temperatures and densities in the center of our own Milky Way \citep[e.g.,][]{AO13,GINSBURG16,KRIEGER17}.
\item In addition to the contrast between normal galaxies and starbursts, and between disks and galaxy centers, there is statistical evidence that regions with hotter dust, higher star formation rate surface density, or shorter depletion times show higher line ratios within normal star-forming galaxies \citep[e.g.,][]{LAMPERTI20,DENBROK21,YAJIMA21}.
\item Given the modest dynamic range of the observed ratios and the need to combine multiple instruments, calibration uncertainties can imply significant scatter in line ratio measurements \citep[see excellent discussions in][]{DENBROK21,YAJIMA21}. For single-pointing observations with single dish telescopes, uncertain aperture corrections also represent a significant source of uncertainty. These systematics, in addition to the limited sensitivity of mm-wave telescopes before ALMA, may help explain why many results in the literature show substantial scatter.
\end{enumerate}

These general trends are largely born out by detailed studies of individual galaxies \citep[e.g.,][]{KODA12,KODA20}, though there remains disagreement in the literature about the behavior of the ratios, e.g., relative to spiral arms or within individual targets. Some of this may reflect that at high resolution, line ratios can show detailed variations that track the location of individual heating sources or vary across spiral arms and bars \citep[e.g.,][and T.~Saito et al.\ in preparation]{UEDA12,LAW18}.

\section{Measurements}
\label{sec:meas}

\begin{figure*}[t!]
\begin{center}
\includegraphics[width=0.45\textwidth]{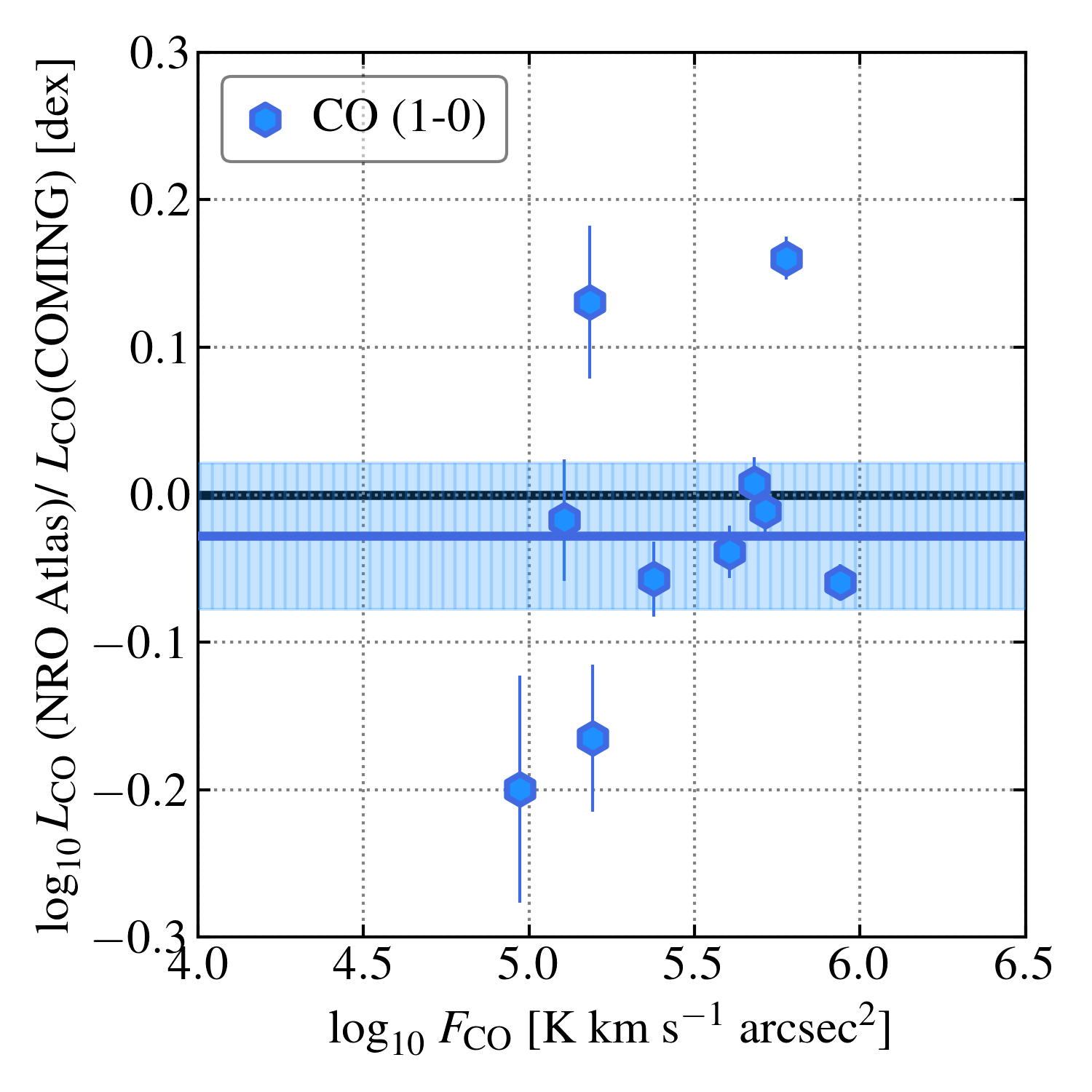}
\includegraphics[width=0.45\textwidth]{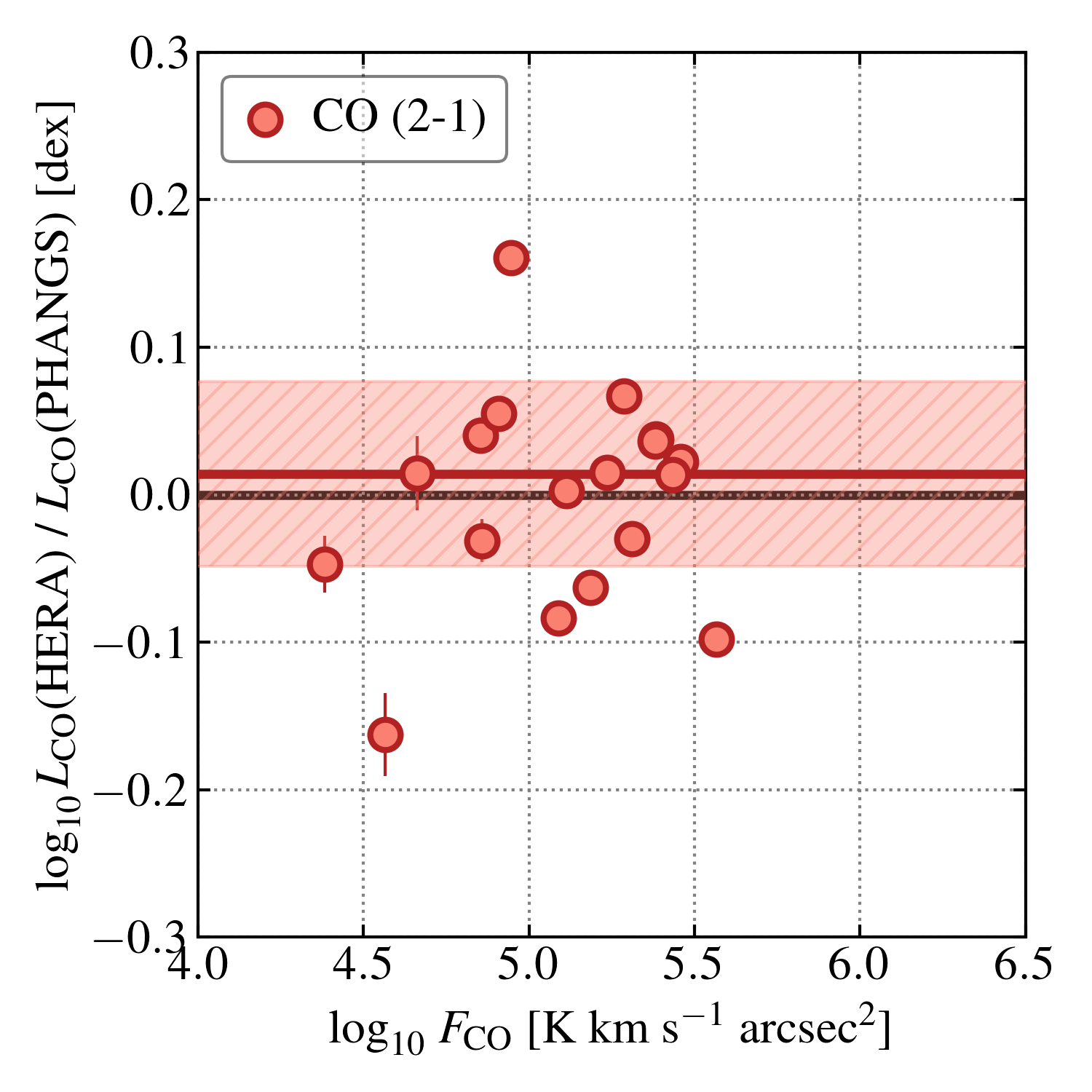}
\end{center}
\caption{\textit{Consistency among galaxy-integrated CO fluxes in repeat observations.} Each panel shows the ratio among the galaxy-integrated CO luminosity, $L_{\rm CO}$, estimated using different surveys targeting the same galaxy. Each point indicates one galaxy, with the error bar showing only statistical uncertainties. The $x$-axis shows the CO flux associated with the denominator (i.e., COMING or PHANGS-ALMA). A solid gray line in both panels indicates the expectation for perfect agreement and the shaded region and colored line show $\pm 1$ standard deviation about the median ratio.  The NRO atlas shows $0.91$ times the CO luminosity of COMING, with $\pm 12\%$ robustly estimated scatter. Where the two samples overlap in this study, IRAM \mbox{30-m} HERA mapping shows median $1.03$ times higher luminosity than PHANGS with $\pm 12\%$ robustly estimated scatter.
\label{fig:consistency}}
\end{figure*}

\begin{figure}[t!]
\begin{center}
\includegraphics[width=0.49\textwidth]{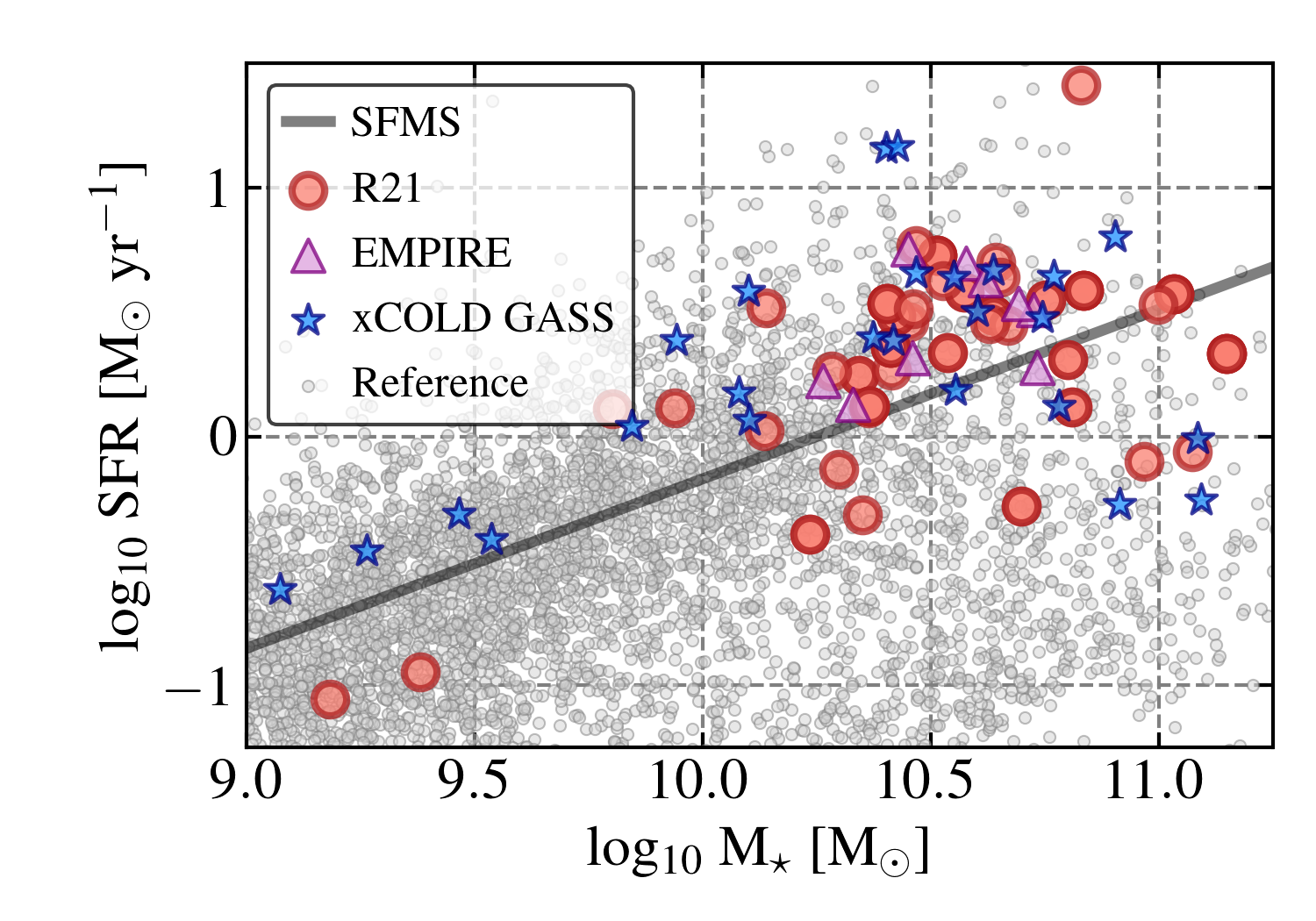}
\includegraphics[width=0.49\textwidth]{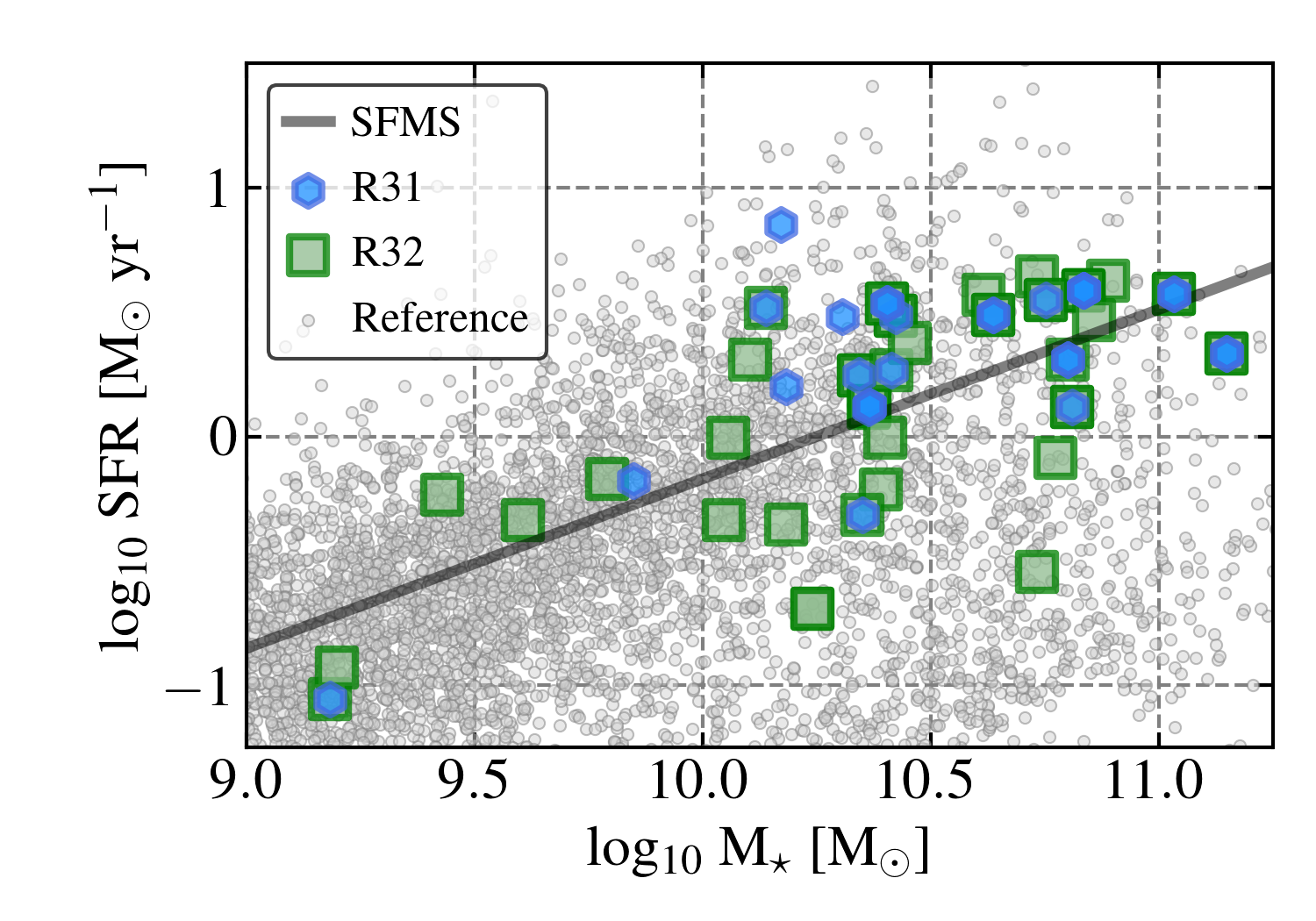}
\end{center}
\caption{\textit{Galaxies with beam-matched line ratio measurements in \SFR--$M_\star$ space.} Location of our targets in \SFR--$M_\star$ space, plotted over the larger sample of galaxies within $d < 50$~Mpc with \SFR\ and $M_\star$ calculated in a consistent way \citep{LEROY19}. The black line shows an estimate of the \SFR--$M_\star$ relationship for local star-forming galaxies \citep[][in excellent agreement with \citealt{CATINELLA18}]{LEROY19}. Red circles, blue hexagons, and green squares show galaxies with measurements (not limits) in this paper, with the top panel showing galaxies with $R_{21}$ measured and the bottom panel showing $R_{32}$ and $R_{31}$. For comparison, we also plot the properties of EMPIRE targets from \citet{DENBROK21} and the $R_{21}$ properties of the subset of xCOLD~GASS with beam-matched $R_{21}$ measurements from \citet{SAINTONGE17}. Despite the relatively large number of measurements, our $R_{21}$ measurements remain highly clustered in this space, biased towards high-mass and high-\SFR\ galaxies. The measurements involving the \cothree\ line show slightly more homogeneous sampling of parameter space despite their overall lower number and higher typical uncertainty.
\label{fig:sfms}}
\end{figure}

\begin{deluxetable}{lccc}[ht!]
\tabletypesize{\small}
\tablecaption{Summary of map pairs in this paper. \label{tab:sample}}
\tablewidth{0pt}
\tablehead{
\colhead{Survey Pair} & 
\colhead{Sample Size} \\
\colhead{} &
\colhead{(Meas./LL/UL)\tablenotemark{a}}
}
\startdata
\hline
\multicolumn{2}{c}{$R_{21} \equiv \cotwo/\coone$} \\
\hline
PHANGS--ALMA~+~COMING & 10/1/0 \\
PHANGS--ALMA~+~NRO Atlas & 18/0/0 \\
HERA~+~COMING & 23/1/0 \\
HERA~+~NRO Atlas & 23/0/0 \\
Total map pairs & 76 \\
Unique galaxies & 43 \\
\hline
\multicolumn{2}{c}{$R_{32} \equiv \cothree/\cotwo$} \\
\hline
NGLS~+~PHANGS--ALMA & 11/0/2 \\
NGLS~+~HERA & 22/0/5 \\
APEX~+~PHANGS--ALMA & 5/0/0 \\
APEX~+~HERA & 2/0/0 \\
Total map pairs & 47 \\
Unique galaxies & 34 \\
\hline
\multicolumn{2}{c}{$R_{31} \equiv \cothree/\coone$} \\
\hline
NGLS~+~COMING & 13/0/1 \\
NGLS~+~NRO Atlas & 11/0/0 \\
APEX~+~COMING & 2/0/0 \\
APEX~+~NRO Atlas & 2/0/0 \\
Total map pairs & 29 \\
Unique galaxies & 20 \\
\hline
\enddata
\tablenotetext{a}{Entries report number of map pairs yielding a measured line ratio (Meas.), a lower limit (LL), or an upper limit (UL).}
\tablecomments{Surveys: COMING is described by \citet{SORAI19}. ``NRO Atlas'' refers to the survey presented by \citet{KUNO07}. PHANGS--ALMA is described by \citet{LEROY21b}. ``HERA'' refers to HERACLES \citep{LEROY09,LEROY13} and a follow up Virgo Cluster survey (P.I.\ A.~Schruba). ``NGLS'' refers to the JCMT survey by \citet{WILSON12} supplemented by a few follow-up or archival JCMT observations. ``APEX'' refers to APEX LASMA mapping by J.~Puschnig et al.\ (in preparation).}
\end{deluxetable}

Table~\ref{tab:sample} summarizes the survey combinations and targets for each line ratio. We use new \cotwo\ maps from the PHANGS--ALMA survey \citep{LEROY21b}, \cotwo\ maps from HERACLES on the IRAM \mbox{30-m} telescope \citep{LEROY09}, and another set of IRAM \mbox{30-m} \cotwo\ maps that cover mostly Virgo Cluster targets (A.~Schruba et al.\ in preparation). We draw \coone\ maps from two Nobeyama \mbox{45-m} surveys, the CO Multiline Imaging of Nearby Galaxies (COMING) Survey \citep{SORAI19} and the Nobeyama CO Atlas of Nearby Spiral Galaxies \citep[hereafter the ``NRO Atlas'';][]{KUNO07}. We compare these to \cothree\ maps from the JCMT Nearby Galaxy Legacy Survey \citep[hereafter the NGLS;][]{WILSON12} and from a new APEX LASMA mapping project (J.~Puschnig et al.\ in preparation). In total, as summarized in Table~\ref{tab:sample} this leads to $\pairs$ map pairs with $43$ unique galaxies mapped in both \cotwo\ and \coone , $34$ mapped in \cothree\ and \cotwo , and $20$ mapped in \cothree\ and \coone . A total of $16$ unique galaxies have a measurement, not a limit, for all three lines.

To consider a line ratio measurement, we require that CO emission be securely detected in at least one transition, so that we can at least obtain a limit on the line ratio. For each target that meets this criterion, we estimate the integrated line ratios and compare these to the integrated properties of the galaxy. For targets with high enough signal-to-noise, we also measure the line ratio in individual $20''$ regions, with this $20''$ scale picked because it represents the common angular resolution achievable by all of the mapping surveys used in our analysis, with APEX LASMA being the limiting data set. The median distance to the target across all of our measurements is ${\sim}14$~Mpc, where this scale corresponds to ${\sim}1.3$~kpc, and the $16{-}84^{\rm th}$ percentile range of distance is 9 to 18~Mpc, implying physical beam sizes of ${\sim}0.9{-}1.8$~kpc. This resolution is typically sufficient to resolve the disk of the galaxy but not isolate individual molecular clouds or resolve features like spiral arms or bars. Using these measurements, we correlate the line ratios with local conditions in the galaxy disk, including galactocentric radius, $r_{\rm gal}$, local star formation rate surface density, $\Sigma_{\rm SFR}$, and stellar mass surface density, $\Sigma_\star$.

We make separate line ratio measurements for each galaxy and specific survey pair. This can lead to the case where we measure the same line ratio multiple times for a single galaxy, e.g., NGC~0628 appears in both HERACLES and PHANGS--ALMA. We use these duplicated observations to help assess the systematic uncertainty, confirming that key uncertainties in the field still often relate to calibration differences among telescopes \citep[see below, Figure~\ref{fig:consistency}, and more discussion in][]{DENBROK21}. We report all line ratio pairs, but when searching for possible correlations and fitting scaling relations we adopt only a single value of a line ratio per galaxy, using the following priority: PHANGS--ALMA over HERACLES; COMING over the NRO Atlas; APEX over the JCMT NGLS.

\medskip 

\textit{Conventions:} We correct all quoted surface densities for the effects of inclination. Our stellar mass and star formation rate maps assume a \citet{CHABRIER03} initial mass function (IMF) and are calibrated to be on the same scale as the \textit{GALEX}--\textit{WISE}--SDSS Legacy Survey \citep{SALIM16,SALIM18}.

\subsection{CO data}
\label{sec:codata}

Table~\ref{tab:sample} summarizes the sources of our line ratios, which come from combining data from ALMA, the IRAM \mbox{30-m} telescope, the Nobeyama Radio Observatory (NRO) \mbox{45-m} telescope, and the James Clerk Maxwell Telescope (JCMT). Specifically, we use the following individual surveys.

\medskip

\textit{PHANGS--ALMA CO\,(2\shortminus1) Data:} \citet{LEROY21b} describe the selection and observations of PHANGS--ALMA and \citet{LEROY21a} describe the data processing, imaging, and data product creation. Here we use the combined interferometric and total power \cotwo\ cubes convolved to our common resolution of $20''$. These PHANGS--ALMA \cotwo\ data have median native resolution $1.3''$, much higher than our working resolution. Because they include total power data, as well as short spacing 7{-}m array data, we expect them to have the correct global flux scale and to be sensitive to extended emission, and so to be well-suited to this analysis after convolution. \citet{LEROY21b} confirm an overall good agreement between the PHANGS--ALMA \cotwo\ and lower resolution single dish measurements, which we also show below.

ALMA provides a total power calibration based on regular monitoring of quasars, and the gain uncertainty associated with ALMA at these frequencies is nominally $5{-}10$\%. We take $10\%$ as a conservative estimate, though we note that this likely overestimates the true uncertainty. In \citet{LEROY21a}, we verified that the internal stability of the PHANGS--ALMA total power data appears very good, with fluxes repeatable at the ${\sim} 3\%$ level from day-to-day. A few cubes do suffer from $2{-}7$\% gain uncertainties due to issues described in \citet{LEROY21a}. The PHANGS--ALMA data have extremely good sensitivity compared to the other data in this paper, but their field of view tends to be more limited than the other maps, with PHANGS--ALMA typically covering 70\% of the total mid-IR emission from its target galaxy. As described in Section~\ref{sec:methods}, we account for this issue in our analysis.

\medskip

\textit{IRAM \mbox{30-m} HERA CO\,(2\shortminus1) Data:} We also analyze \cotwo\ maps from HERACLES \citep{LEROY09} and another $21$ galaxies observed by the IRAM \mbox{30-m} telescope as part of a survey focused on the Virgo Cluster (P.I.\ Schruba; A.~Schruba et al.\ in preparation). These new data were observed in a manner identical to HERACLES and reduced following the same procedures. Both data sets have native FWHM beam size of $13\farcs3$ and large extent, usually covering out beyond the optical radius of the galaxy. The calibration uncertainty associated with the HERA data is ${<}20\%$ based on a detailed gain analysis presented in \citet{DENBROK21}, a bootstrapping analysis in \citet{LEROY09}, and comparing to PHANGS--ALMA \citep{LEROY21b}. This calibration uncertainty can also apply within maps, reflecting uneven gains among the receiver array. \citet{DENBROK21} note the two most extreme cases, NGC~3627 and NGC~5194, were both observed early in the life of the HERA instrument \citep{SCHUSTER07} when observing procedures had not yet been optimized. To be conservative, we adopt a nominal uncertainty $20\%$, near the upper limit of the plausible calibration uncertainty. This adopted calibration uncertainty agrees with the consistency check shown in Figure~\ref{fig:consistency} and the results of the checks in \citet{LEROY21b}.

\medskip

\textit{NRO CO\,(1\shortminus0) Data:} We utilize \coone\ maps obtained by the Nobeyama Radio Observatory from the CO Multi-line Imaging of Nearby Galaxies (COMING) survey \citep{SORAI19}. These data have native resolution $17''$ and cover large areas in each target. We also compare to \coone\ maps from the Nobeyama CO Atlas of Nearby Spiral Galaxies (``NRO Atlas'') presented by \citet{KUNO07}. These have $15''$ resolution and higher sensitivity than COMING, but the \citet{KUNO07} data suffer from more visible mapping artifacts and poor baselines compared to \citet{SORAI19}. Following \citet{YAJIMA21}, we take the gain uncertainty for both data sets to be $25\%$, reflecting a combination of true calibration uncertainties and pointing errors. Similar to the HERA case noted above, these systematic uncertainties can apply within a galaxy, and do not only reflect an overall scaling from galaxy to galaxy. 

\medskip

\textit{JCMT CO\,(3\shortminus2) Data:} We also compare to maps of \cothree\ emission from the JCMT NGLS \citep{WILSON12} and individual galaxy follow-up programs (P.I.\ E.~Rosolowsky) observed under projects M09BC15, M10BC06, and M12AC03. The follow-up program data were calibrated using the {\sc starlink} software package \citep{CURRIE14} using the observatory-recommended pipelines. The JCMT data initially had $14\farcs5$ resolution. We convolved them to a resolution of $20''$ for further processing and translated them into the main beam temperature scale using an efficiency of $0.6$ at $345$~GHz \citep[following][]{WILSON12}. 

After this convolution, we inspected the data and found that they could be improved by fitting and subtracting low order baselines. For each line of sight, we defined a velocity range of interest based on the velocity range of CO emission seen in the other CO data for the galaxy. Specifically, we defined a reference cube used to define the baseline region, with the priority given to PHANGS--ALMA, then IRAM HERA data, then COMING data, then NRO Atlas data. Because the \cothree\ data tend to have lower signal to noise than the lower-$J$ cubes, using the other cubes as a template to fit the baseline should be well-defined and impose little or no bias. Then, we considered each line of sight in the JCMT cube. Along that line of sight, we excluded velocities detected above $\mathrm{S/N}=3$ in either the reference CO cube or the JCMT data themselves from the fit, and focused the baseline fit on regions near the detected line emission. We defaulted to a $\pm 50$~km~s$^{-1}$ fitting window but adjusted this slightly from galaxy-to-galaxy. Then we fit and subtracted a baseline from each line of sight, using iterative outlier rejection of the JCMT data to refine the fit. We used an order~$0$ fit, i.e., we subtracted the mean, for all galaxies except NGC~2403, where we used a linear fit. In a few galaxies, we also blanked regions of the JCMT cube that were clearly dominated by artifacts. Finally, after inspecting the data, we dropped a few potential targets where the data remained clearly dominated by artifacts despite our baseline fitting. By virtue of using the other CO data as a prior for baseline fits, we also effectively required that all JCMT targets were detected in another CO transition.

We lack a detailed characterization of the calibration and pointing uncertainty for the JCMT data, but according to the JCMT web pages\footnote{\url{https://www.eaobservatory.org/jcmt/instrumentation/heterodyne/calibration/}} the nominal calibration accuracy of the JCMT is 10\% before accounting for uncertainties in pointing or sub-optimal conditions. Empirically, the integrated fluxes that we measure frequently vary by ${\gtrsim} 10{-}20\%$ when we compare results before and after our rebaselining. We adopt an overall calibration uncertainty 20\% as a conservative estimate, consistent with, e.g., \citet{YAJIMA21,SORAI19} for NRO and \citet{DENBROK21,LEROY09} for IRAM HERA maps. As with the IRAM HERA maps, we consider this to represent the upper envelope of plausible uncertainties.

\medskip

\textit{APEX CO\,(3\shortminus2) Data:} We also compare to five maps of \cothree\ emission obtained using the Large APEX Sub-Millimetre Array (LASMA) receiver array on the Atacama Pathfinder Experiment (APEX) telescope \citep{GUESTEN08}\footnote{And see \url{http://www.mpifr-bonn.mpg.de/5278286/lasma}}. LASMA is a seven pixel, single polarization array receiver that can observe from $\nu = 268{-}375$~GHz. At the $\nu \approx 345$~GHz of \cothree, the array has a beam size of ${\sim}18.5''$. After reduction and convolution during gridding, these maps end up having a beam size of ${\sim}20''$, which sets the common resolution of our data. These maps were obtained as part of APEX projects \texttt{m-0103.f-9520a-2019} and \texttt{m-0104.f-9516a-2019} (PI: A.~Weiss) and will be presented in detail in J.~Puschnig et al.\ (in preparation). They target galaxies that have PHANGS--ALMA imaging, and the areal extent is designed to match that of the PHANGS--ALMA maps almost exactly. The final data cubes have been gridded to our common velocity resolution of 10~km~s$^{-1}$. At $20''$ resolution and with 10~km~s$^{-1}$ channels, the cubes have rms noise of ${\sim}9{-}11$~mK. The baselines and data quality appear excellent, with few visible artifacts and emission visible in most individual channels. Based on advice from the APEX team, we scale the maps assuming that LASMA has $0.9\times$ the nominal efficiency of APEX at these frequencies in order to account for a partial shadowing of two of the LASMA receivers. In most respects, the data resemble the other array receiver data. We expect the telescope to recover all the flux from the source, and the data should be well suited to stacking. The overall calibration of the data represents the main source of uncertainty for many of our calculations, and we assume LASMA+APEX to have rms gain uncertainty of $\pm 20\%$, in line with the other facilities. We defer more details of these data to J.~Puschnig et al.\ (in preparation).

\subsection{CO processing} 
\label{sec:processing}

We downsample all of the CO data to have velocity resolution ${\sim}10$~km~s$^{-1}$. Then we convolve all data cubes to a common angular resolution of $20''$ and reproject them onto the astrometric grid of the stellar mass maps described below.

For each \coone\ and \cotwo\ cube we produce a three dimensional high completeness ``signal mask'' that includes the volume of the cube where CO emission is detected in either \cotwo\ or \coone , as well as some surrounding volume. As discussed above, the JCMT cubes tend to have lower signal-to-noise and more artifacts than the other data, so at this stage we rebaseline them as described above. Given this situation, we apply the mask constructed based on the \cotwo\ and \coone\ to \cothree . Because we have relatively few APEX maps, we treat them the same as the JCMT data. On visual inspection, we do not see any evidence that this approach causes us to miss \cothree\ emission in our targets. Moreover, as described below our checks show the masks to have very high completeness for \cotwo\ and \coone ; given that the \cotwo\ data are more sensitive than the \cothree\ data we do not expect this choice to bias our results in any important way.

We construct these signal masks following a variation of the masking scheme in \citet{ROSOLOWSKY06} and the ``broad masking'' approach in \citet{LEROY21b,LEROY21a}. First, we estimate the noise from signal-free regions of each cube and then calculate the signal-to-noise ($\mathrm{S/N}$) for each pixel in each cube. We then construct individual ``signal'' masks for each \coone\ and \cotwo\ data cube. Each mask began with a high significance core identified based on a threshold $\mathrm{S/N}$ value. We expanded this initial mask to include all adjacent regions of the cube with lower, but still significant signal. Then, the masks are further dilated by $20$~km~s$^{-1}$ in velocity and several beam sizes in each spatial direction. Most of the masking was done at $30''$ resolution, to improve the $\mathrm{S/N}$, but we also included any bright emission seen only at $20''$ resolution in the mask. We adjusted the exact $\mathrm{S/N}$ thresholds used in the masking for each data set based on visual inspection until they yielded a mask that encompassed all visible CO emission in all cubes with a comfortable margin in both velocity and spatial extent. Based on this visual inspection, we also slightly lowered the core $\mathrm{S/N}$ threshold in the case of a few compact galaxies with faint CO emission.

For each galaxy we created a final mask for each galaxy by combining all signal masks from all individual \cotwo\ and \coone\ surveys. Any pixel included in any mask is included in the final mask. We adopt this approach aiming at high completeness and minimal bias, i.e., we try to include all likely CO emission in the mask, even if this increases the noise \citep[i.e., this is a ``broad'' mask following][]{LEROY21a}. In addition to visual inspection, we verified the completeness of the maps by comparing the integrated CO flux to a direct integral of the cube over the velocity width of the galaxy. For COMING, the masks include a median 97\% of the CO emission with ${<}0.03$~dex scatter. For PHANGS--ALMA, the masks include 100\% of the CO emission on average with ${<}0.01$~dex scatter.

Finally, we applied this combined mask to all cubes for that galaxy.  We collapse this masked cube to construct an integrated intensity (``moment~0'') map for each data cube. We calculate the associated statistical uncertainty from error propagation using the rms noise estimated from the signal-free parts of the cube.

\medskip

\textit{CO luminosities:} For comparing with the integrated properties of galaxies, we also calculate the CO luminosity, $L_\mathrm{CO}$, implied by each map. To do this, we adopt the distances compiled in \citet{ANAND21} for PHANGS--ALMA and follow \citet{LEROY19} for other targets. 

In some cases, this calculation is complicated by the fact that the CO line maps do not cover the entire area of the galaxy. In particular, this often affects PHANGS--ALMA \citep[see above and][]{LEROY21a}. In these cases, we apply an aperture correction that uses WISE3 emission as the template for CO emission. This approach is discussed in more detail in \citet{LEROY21b}, who show that WISE3 offers the best available template to construct such aperture corrections \citep[consistent with the findings by][that WISE3 correlates very strongly with CO emission]{CHOWN21}.

Note that when we report CO luminosities in Table~\ref{tab:meas}, we give only a single value of $L_{\coone}$, $L_{\cotwo}$, and $L_{\cothree}$ for each galaxy. In choosing which CO luminosity to report, we prefer COMING values over NRO Atlas values, PHANGS--ALMA values over IRAM \mbox{30-m} HERA values, and APEX values over JCMT values. Because we provide only a single luminosity, and because the luminosities include aperture corrections while the reported ratios use exactly matched apertures, we note that dividing our quoted CO luminosities will not yield exactly the same value as the line ratio reported in the table. That is, our reported $R_{21}$ uses a matched field of view and is calculated for each survey pair (see Section~\ref{sec:methods}), while the CO luminosities are aperture corrected and we report only one value for each transition.

\subsection{Stellar masses and star formation rates}

We estimate star formation rates and stellar masses based on \textit{GALEX} \citep{MARTIN05} far-ultraviolet (FUV) and near-ultraviolet (NUV) images, \textit{Spitzer} IRAC near-infrared (NIR) maps \citep{FAZIO04}, and \textit{WISE} \citep{WRIGHT10} NIR and mid-infrared (MIR) imaging. The \textit{GALEX} and \textit{WISE} maps were created as part of the $z=0$ Multiwavelength Galaxy Synthesis \citep{LEROY19}. The IRAC maps were obtained mostly by the S$^4$G survey \citep{SHETH10}. \citet{LEROY19} and \citet{LEROY21b} give details of the conversion from these bands to \SFR\ and $M_\star$. We use the same calculations described in \citet{LEROY21b}, which we carried out for PHANGS--ALMA, the targets of the HERA surveys, and the targets of the Nobeyama surveys in a self-consistent way. Figure~\ref{fig:sfms} shows the resulting estimated \SFR\ and $M_\star$ for our targets plotted over a large set of local galaxies \citep[from][]{LEROY19}.

Briefly, to estimate the \SFR, we use the best available combination of ultraviolet and mid-infrared data, preferring more stable combinations of tracers whenever available. In order of most preferred to least preferred, we use: FUV+WISE4, NUV+WISE4, FUV+WISE3, NUV+WISE3, WISE4-only, WISE3-only. We adopt the conversions reported in Table~7 of \citet{LEROY19} and apply them as detailed in \S 3 of that paper. These conversions are calibrated to reproduce galaxy-integrated \SFR\ values calculated for the SDSS main galaxy sample based on UV-to-IR CIGALE SED modeling by \citet{SALIM16} and \citet{SALIM18}. As discussed in that paper, these agree well with previous calibrations using similar bands \citep[e.g.,][]{SALIM07,KENNICUTT12,LEROY12,JANOWIECKI17}, usually within ${\sim} 0.1$~dex. In \citet{LEROY21b}, we show that the resolved $\Sigma_{\rm SFR}$ estimates agree with high quality Balmer decrement-based $\Sigma_{\rm SFR}$ measurements from PHANGS--MUSE (E.~Emsellem et al.\ A\&A submitted) within ${\sim} 20\%$ on average but that the UV+IR maps likely overestimate $\Sigma_{\rm SFR}$ in regions of low \SFR, with the most likely explanation being contamination by IR cirrus \citep[see][]{GROVES12,LEROY12}, but other effects like stochastic sampling of the initial mass function or issues with extinction correction also remain possible. The magnitude of the effect may reach up to a factor of $2$ for $\Sigma_{\rm SFR} \lesssim 10^{-3}$~M$_{\odot}$~yr$^{-1}$~kpc$^{-2}$.

We base our stellar mass estimates on near-infrared (near-IR) emission at $3.6~\mu$m (IRAC1) or $3.4~\mu$m (WISE1). After subtracting a background, we flag stars and replace them with interpolated values from similar galactocentric radii. Then, we convert from near-IR intensity to stellar mass surface density using a mass-to-light ratio that depends on the ratio of \SFR-to-WISE1. This quantity serves as a proxy for the specific star formation rate, $\SFR/M_\star$, which is a strong predictor of the WISE1 mass-to-light ratio in the \citet{SALIM16,SALIM18} work. \citet{LEROY21b} describe the detailed calculations and present comparisons to results from resolved stellar mass estimates from optical spectral mapping by PHANGS--MUSE (E.~Emsellem et al.\ A\&A submitted). \citet{LEROY19} present the motivation for the approach based on matching the \citet{SALIM16,SALIM18} estimates.

We measure integrated $M_\star$ and integrated \SFR\ by directly integrating all pixels within $2r_{25}$. Based on comparisons among different methods and bands, we adopt uncertainties of $0.1$~dex for both $M_\star$ and \SFR\ estimates. When relevant, we calculate offsets from the star-forming main sequence exactly as described by \citet{LEROY21b}.

\subsection{Line ratio measurements} 
\label{sec:methods}

Before proceeding, we reproject all data, which have already been convolved to $20''$, onto a grid with pixel size equal to the $20''$, i.e., we work with pixels equal to the FWHM beam size. This leads to a moderate undersampling of the maps in exchange for rendering the individual measurements mostly independent. We consider that the convolution to $20''$ has removed most sampling effects present in the on-the-fly single dish maps \citep[e.g., see][]{MANGUM07}.

\subsubsection{Integrated line ratios} 
\label{sec:measint}

We calculate each integrated line ratio over the area where both surveys involved have coverage and where the combined mask described in Section~\ref{sec:processing} indicates the presence of CO emission. For a given galaxy, we denote this matched area $\mathcal{A}$, and we calculate the line ratio, $R_{ul}$, as the ratio of the sum of emission from each line: 
\begin{equation}
    R_{ul} = \frac{\sum_{x,y \in \mathcal{A}} I_u(x,y)}{\sum_{x,y \in \mathcal{A}} I_l(x,y)}~.
\end{equation}
\noindent This ratio-of-sums approach weights the calculated $R_{ul}$ by intensity and when the maps and mask cover the whole galaxy, $R_{ul}$ will match the result expected from an unresolved, single pointing measurement.

We follow standard error propagation to estimate the statistical uncertainty on the measurement. The uncertainty in each measurement is the sum in quadrature of this statistical uncertainty with the calibration uncertainties for both telescopes: $\sigma^2 = \sigma^2_\mathrm{stat} + \sigma^2_{\mathrm{cal},u} + \sigma^2_{\mathrm{cal}, l}$.  The calibration uncertainty frequently dominates the total error budget.

We require that both lines be detected at a statistical $\mathrm{S/N}>4$ to report a ratio, i.e., $\mathrm{S/N}>4$ before accounting for the calibration uncertainties. For cases where only one line is detected at the required significance, we estimate and report an upper or lower limit using the $4\sigma$ statistical uncertainty in the undetected line to define the limit.

\medskip

\noindent \textit{Literature data:}  We compare our galaxy-integrated measurements to recent measurements combining IRAM \mbox{30-m} \coone\ maps from the EMPIRE survey with PHANGS--ALMA, HERACLES, and a new M51 \cotwo\ map \citep{DENBROK21}. In that case, we use the same procedure to calculate $M_\star$ and \SFR\ described above.

We also compare to the $28''$ single dish APEX+IRAM \mbox{30-m} line ratio measurements presented by \citet{SAINTONGE17}. These have closely matched beams and their stellar masses and \SFR\ values are calculated on a system similar to our own. They do not report their exact aperture correction for the IRAM \mbox{30-m} data, but note it to be between~2 and~10\%. We apply a 5\% upward correction to all IRAM \mbox{30-m} luminosities and include a 15\% overall calibration uncertainty in addition to their reported statistical error.

\medskip

\noindent \textit{Effect of the Cosmic Microwave Background:} The observed brightness temperature reflects only the contrast against the cosmic microwave background (CMB), such that the measured brightness temperature, $T_\mathrm{b}$, will be $T_\mathrm{b} = (1 - e^{-\tau}) (T_\mathrm{ex} - T_\mathrm{CMB})$ for each transition \citep[e.g., see][among many other discussions]{ECKART90,BOLATTO13B,ZSCHAECHNER18} with $T_\mathrm{ex}$ the relevant excitation temperature. This can imply modest corrections to the line ratios, especially for cold clouds. However, this radiative transfer proceeds only at the scale of molecular clouds themselves. The brightness temperatures in our current work are heavily affected by beam dilution. Without measuring clumping of CO emission at sub-resolution, we cannot calculate an appropriate correction for the CMB. These values can, in principle, be measured for PHANGS--ALMA \citep[e.g., following][]{LEROY13B} but we lack similar high resolution templates for the other data and the measurements for PHANGS--ALMA represent future work. We note the effect, do not apply any CMB correction, and leave an improved treatment for future works. See Appendix~\ref{sec:corrections} for more details.

\medskip

\noindent \textit{Consistency among integrated measurements:} Because surveys targeting the same line sometimes share targets, we make repeated measurements for several galaxies ratio pairs. Figure~\ref{fig:consistency} checks for internal consistency within our measurements. The left panel shows the ratio of \coone\ luminosity estimated from the NRO Atlas to COMING and the right panel shows the ratio of \cotwo\ luminosity estimated using HERA to that from PHANGS. In both panels we use the integrated galaxy luminosity, and so trust the aperture corrections described above to account for any differences in area covered. We do not show a panel for \cothree . Only one galaxy is detected in both JCMT and APEX, NGC~3627, and there the luminosity inferred from the APEX data is ${\sim}1.4$ times that calculated from the JCMT.

Overall, Figure~\ref{fig:consistency} illustrates that the \cotwo\ measurements are mostly consistent across the two surveys, with a median ratio only a few percent different from $1.0$. We do observe significant scatter, with rms variation of about $\pm 12\%$, much larger than the statistical noise. This agrees with \citet{LEROY21b} and mostly validates the calibration uncertainties adopted above. The situation for \coone\ is similar, with the NRO Atlas ${\sim}8\%$ lower than COMING on average and a scatter of about $\pm 12\%$ from a relatively low sample size. Based on \citet{SORAI19}, we expect that COMING has better overall calibration compared to the NRO Atlas.

Overall, Figure~\ref{fig:consistency} shows that systematic uncertainties related to calibration, pointing, etc. impose an uncertainty that has rms of order $10{-}20\%$ on individual CO line measurements from galaxies. We will see in the rest of the paper that this uncertainty is comparable to the range of variation in the line ratios across the galaxy population.

\subsubsection{Resolved, binned, normalized line ratios} 
\label{sec:measres}

In \S\ref{sec:local}, we compare line ratios to location within a galaxy. The challenges here are the limited signal-to-noise of individual measurements and the need to account for the substantial galaxy-to-galaxy calibration uncertainties.

We focus on three quantities: galactocentric radius, $r_{\rm gal}$; $\Sigma_{\rm SFR}$, the star formation rate per unit area; and $\Sigma_{\rm SFR}/\Sigma_\star$, the local specific star formation rate. We consider the area covered by both surveys and extract measurements of both relevant CO lines, $r_{\rm gal}$, $\Sigma_{\rm SFR}$, and $\Sigma_{\rm SFR}/\Sigma_\star$ for all $20''$ pixels in this overlap region. We calculate $r_{\rm gal}$ using the orientations and distances in \citet{LEROY21b}, drawing on \citet{LANG20} and \citet{ANAND21}. For cases outside PHANGS--ALMA, we prefer orientation parameters from S$^4$G \citep[][]{SHETH10,MUNOZMATEOS15} where available and follow the compilation in \citet{LEROY19} otherwise. We calculate bins for both physical $r_{\rm gal}$, expressed in units of kpc, and $r_{\rm gal}$ normalized to the effective half-mass radius, $r_{\rm eff}$, calculated in \citet{LEROY21b}. We also calculate $\Sigma_{\rm SFR}$ and $\Sigma_{\rm SFR}/\Sigma_\star$ as described above and in \citet{LEROY21b}. 

To account for the limited signal-to-noise of individual measurements, we define a set of bins in each quantity of interest. Then, within each galaxy we identify the pixels belonging to each bin and then sum all data for each line. As with the global line ratios, we divide the summed, binned values by one another to estimate the line ratio in that bin. As above, we propagate statistical uncertainties following standard error propagation, and we use a signal-to-noise threshold of~4 to determine whether a bin is a detection (both numerator and denominator have $\mathrm{S/N}>4$), an upper limit (only denominator has $\mathrm{S/N}>4$), or a lower limit (only numerator has $\mathrm{S/N}>4$). After calculating the line ratio, we account for uncertainty associated with the calibration, we normalize each measured line ratio by the galaxy-average value. That is, in \S\ref{sec:local} we measure the enhancement or depression of the ratio relative to its mean value in any given galaxy. This should remove any global gain calibration uncertainty term, though not local calibration uncertainties, e.g., due to pointing uncertainties or pixel-to-pixel gain variations. This also removes any real galaxy-to-galaxy scatter in the mean line ratio, so that this analysis focuses on how these variables drive relative changes in each line ratio within a galaxy.

We note the following details regarding bin construction:

\begin{enumerate}
\item When considering physical galactocentric radius, in units of kpc, we use linearly spaced bins $1$~kpc in width with the first bin centered at $r_{\rm gal} = 0$~kpc and the last one centered at $10$~kpc. Note that as discussed above, this implies some slight over- or under-sampling of the data because the range of distances to the galaxies means that our $20''$ resolution corresponds to different physical resolution across the sample.
\item When considering normalized galactocentric radius, we normalize by the half-mass radius, $r_{\rm eff}$, calculated following \citet{LEROY21b}. Our bins have width $0.5$ times $r_{\rm eff}$ with the first bin centered at $r_{\rm gal} = 0$ and the outermost bin centered at $r_{\rm gal} = 3~r_{\rm eff}$.
\item When considering $\Sigma_{\rm SFR}$, we bin the data by $\log_{10} \Sigma_{\rm SFR} / \langle \Sigma_{\rm SFR} \rangle$. Here $\langle \Sigma_{\rm SFR} \rangle$ is the galaxy averaged star formation rate surface density. We calculate via $\langle \Sigma_{\rm SFR} \rangle = 0.5 \, \SFR / ( \pi r_{\rm eff}^2 )$, i.e., the surface density implied by placing half of the galaxy-integrated star formation within the effective radius, $r_{\rm eff}$ measured for the mass by \citet{LEROY21b}. Normalizing in this way allows us to focus on how the internal structure of the line ratio tracks the local \SFR\ with fewer concerns about how the overall amplitude of \SFR\ or the calibration of our \SFR\ tracer varies from galaxy to galaxy. This makes sense given our similar galaxy-by-galaxy normalization of the CO line ratio for this analysis.
\item For specific star formation rate, we calculate $\Sigma_{\rm SFR}/\Sigma_\star$, normalize by the integrated galaxy-averaged $\textit{sSFR}=\SFR /M_\star$, and then bin $\log_{10} ((\Sigma_{\rm SFR}/\Sigma_\star) / \textit{sSFR})$ in bins of $0.25$~dex from $-1.25$ to $+1.25$~dex about the galaxy average.
\end{enumerate}

This binning procedure is functionally equivalent to a stacking procedure within each bin similar to that used by, e.g., \citet{CORMIER18}, \citet{JIMENEZDONAIRE19}, or \citet{DENBROK21}. It has the advantage of retaining information from individual pixels with modest signal-to-noise and so avoids some biases present in direct pixel-by-pixel analysis. We record bins in which both lines are detected at $\mathrm{S/N} > 4$ as measurements and record upper and lower limits using the $4\sigma$ value for the limiting line. Typically for $R_{21}$ any limits are lower limits because the \cotwo\ maps are more sensitive than the \coone\ maps. For $R_{32}$ and $R_{31}$, our limits are mostly upper limits because the \cothree\ maps lack sensitivity compared to the \cotwo\ and \coone\ maps.

\section{Results}
\label{sec:results}

 \begin{deluxetable*}{lccccccccc} 
 \tablecaption{Integrated CO Line Ratios and Galaxy Properties \label{tab:meas}} 
 \tablewidth{0pt} 
 \tabletypesize{\footnotesize} 
 \tablehead{ 
 \colhead{Galaxy} & 
 \colhead{Line Ratio} & 
 \colhead{Survey pair} & 
 \colhead{$\log_{10} R$} & 
 \colhead{Dist.} & 
 \colhead{$\log_{10} M_\star$} & 
 \colhead{$\log_{10} \textit{SFR}$} & 
 \colhead{$\log_{10} L_{\rm CO}^{1-0}$} & 
 \colhead{$\log_{10} L_{\rm CO}^{2-1}$} & 
 \colhead{$\log_{10} L_{\rm CO}^{3-2}$}  
 \\ 
 \colhead{} & 
 \colhead{} & 
 \colhead{} & 
 \colhead{} & 
 \colhead{(Mpc)} & 
 \colhead{(M$_\odot$)} & 
 \colhead{(M$_\odot$ yr$^{-1}$)} & 
 \multicolumn{3}{c}{(K km s$^{-1}$ pc$^2$)} 
 }  
\startdata 
ic0750 & R31 & JCMTCOMING & $-0.44\pm  0.04$ & 17.10 & 10.18 &  0.20 &  8.72 & \nodata &  8.27\\ 
ngc0253 & R21 & PHANGSNROATLAS & $-0.28\pm  0.00$ &  3.70 & 10.64 &  0.70 &  9.26 &  8.96 & \nodata\\ 
ngc0337 & R21 & HERACOMING & $-0.18\pm  0.08$ & 19.50 &  9.80 &  0.11 &  8.19 &  7.98 & \nodata\\ 
ngc0628 & R21 & HERACOMING & $-0.23\pm  0.02$ &  9.84 & 10.34 &  0.24 &  8.93 &  8.66 &  8.14\\ 
ngc0628 & R21 & PHANGSCOMING & $-0.30\pm  0.01$ &  9.84 & 10.34 &  0.24 &  8.93 &  8.66 &  8.14\\ 
ngc0628 & R31 & JCMTCOMING & $-0.78\pm  0.03$ &  9.84 & 10.34 &  0.24 &  8.93 &  8.66 &  8.14\\ 
ngc0628 & R32 & JCMTHERA & $-0.56\pm  0.03$ &  9.84 & 10.34 &  0.24 &  8.93 &  8.66 &  8.14\\ 
ngc0628 & R32 & JCMTPHANGS & $-0.43\pm  0.02$ &  9.84 & 10.34 &  0.24 &  8.93 &  8.66 &  8.14\\ 
ngc0925 & R32 & JCMTHERA & $-0.06\pm  0.07$ &  9.16 &  9.79 & -0.17 & \nodata &  7.52 &  7.52\\ 
ngc1068 & R21 & PHANGSNROATLAS & $-0.15\pm  0.01$ & 13.97 & 10.91 &  1.64 &  9.47 &  9.34 & \nodata\\ 
\enddata 
\tablecomments{This table is a stub. The full version of the table appears as a machine readable table 
in the online version of the paper. Columns give: 
Galaxy --- the name of the galaxy; 
Line Ratio --- the reported line ratio; 
Survey Pair --- shorthand for the pair of surveys used to make the measurement; 
$\log_{10} R$  --- the log of the measured ratio, with uncertainty. In the case of limits, we report the $4\sigma$ upper or lower limit as the value; 
$D$  --- the adopted distance in Mpc, following \citet{ANAND21}; 
$\log_{10} M_\star$  --- log of the stellar mass; 
$\log_{10} \textit{SFR}$  --- log of the star formation rate; 
$\log_{10} L_{\rm CO1-0}$, $L_{\rm CO2-1}$, and $L_{\rm CO3-2}$ --- log$_{10}$ of the best-estimate CO luminosity in the noted transition with aperture corrections applied. For the luminosity we report only one best-estimate value per galaxy. That is, this the single best estimate of $L_{\rm CO}$. We give preference to COMING over the NRO Atlas and PHANGS--ALMA over IRAM HERA data. Note that because the ratios $R$ are measured over matched apertures inside the galaxies they do \textit{not} match the ratios of luminosities by construction.}
\end{deluxetable*}

\begin{deluxetable}{lcccc}
\tabletypesize{\small}
\tablecaption{Galaxy-Integrated CO Line Ratios from Mapping Surveys \label{tab:intrats}}
\tablewidth{0pt}
\tablehead{
\colhead{Ratio} &  
\colhead{Mean} &
\colhead{Median} &
\colhead{$16^\mathrm{th}$ \%ile} &
\colhead{$84^\mathrm{th}$ \%ile}
}
\startdata
$R_{21}$\tablenotemark{a} & 0.65 & 0.61 & 0.50 & 0.83 \\
$R_{32}$ & 0.50 & 0.46 & 0.23 & 0.59 \\
$R_{31}$ & 0.31  & 0.29 & 0.20 & 0.42 \\
\enddata
\tablenotetext{a}{Includes EMPIRE  measurements from \citet{DENBROK21}.}
\tablecomments{See Figure~\ref{fig:hist}.}
\end{deluxetable}

\subsection{Global line ratios}
\label{sec:global}

\begin{figure*}[ht!]
\begin{center}
\includegraphics[width=0.875\textwidth]{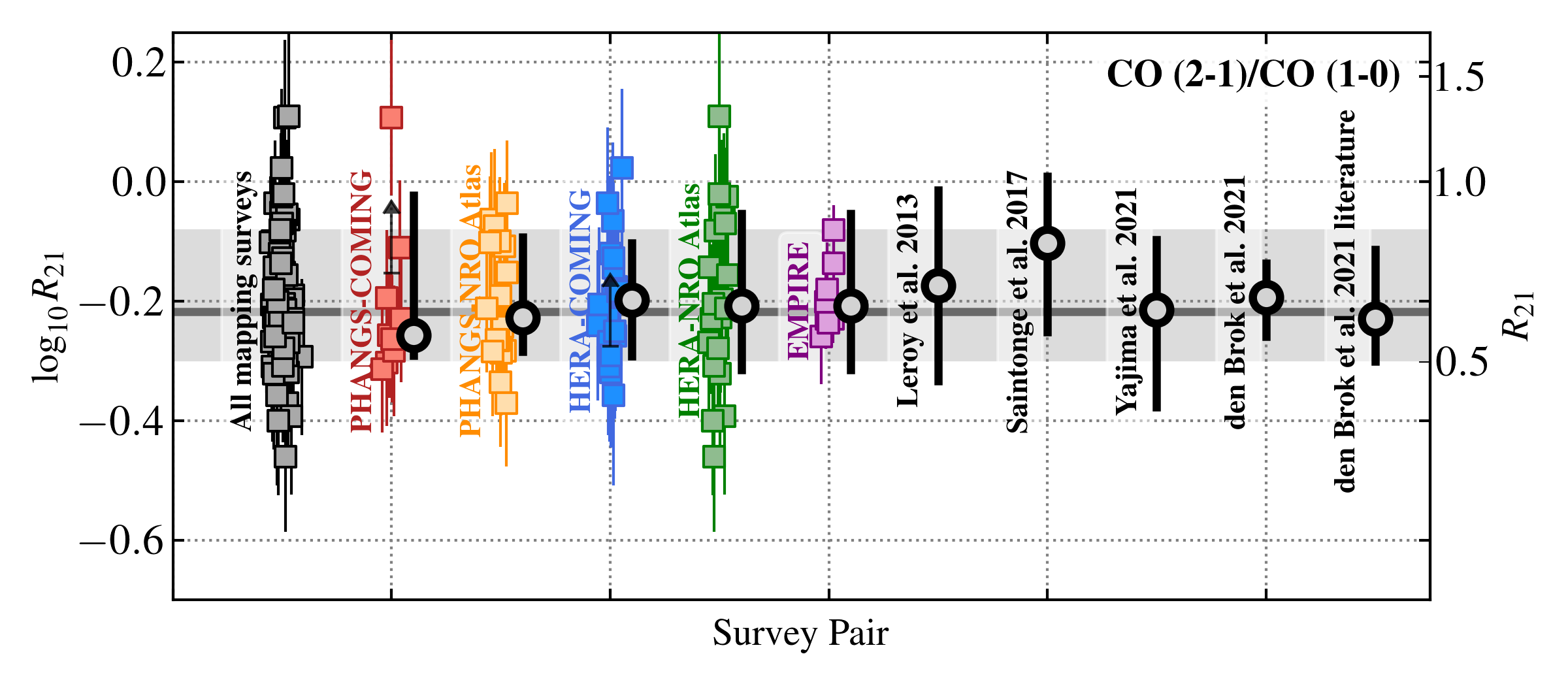}\\
\includegraphics[width=0.875\textwidth]{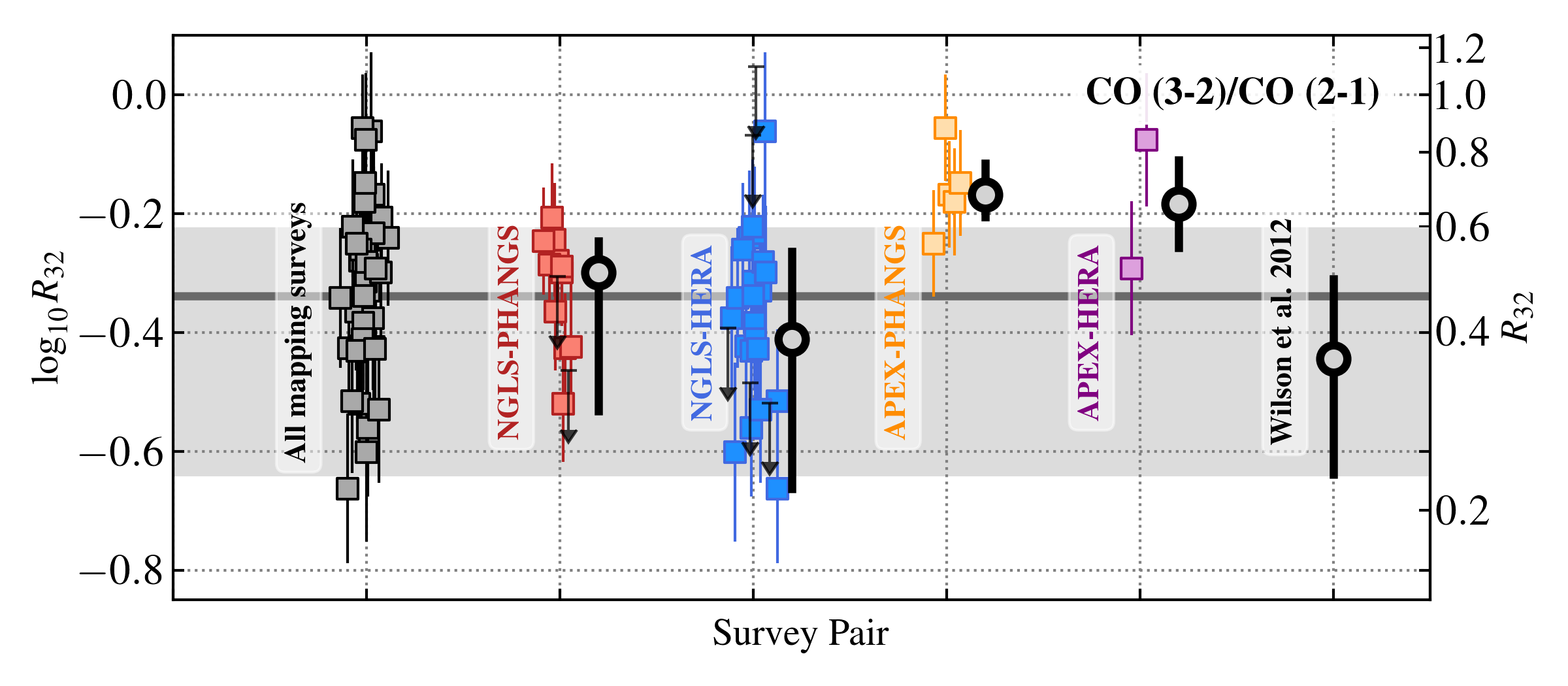}
\includegraphics[width=0.875\textwidth]{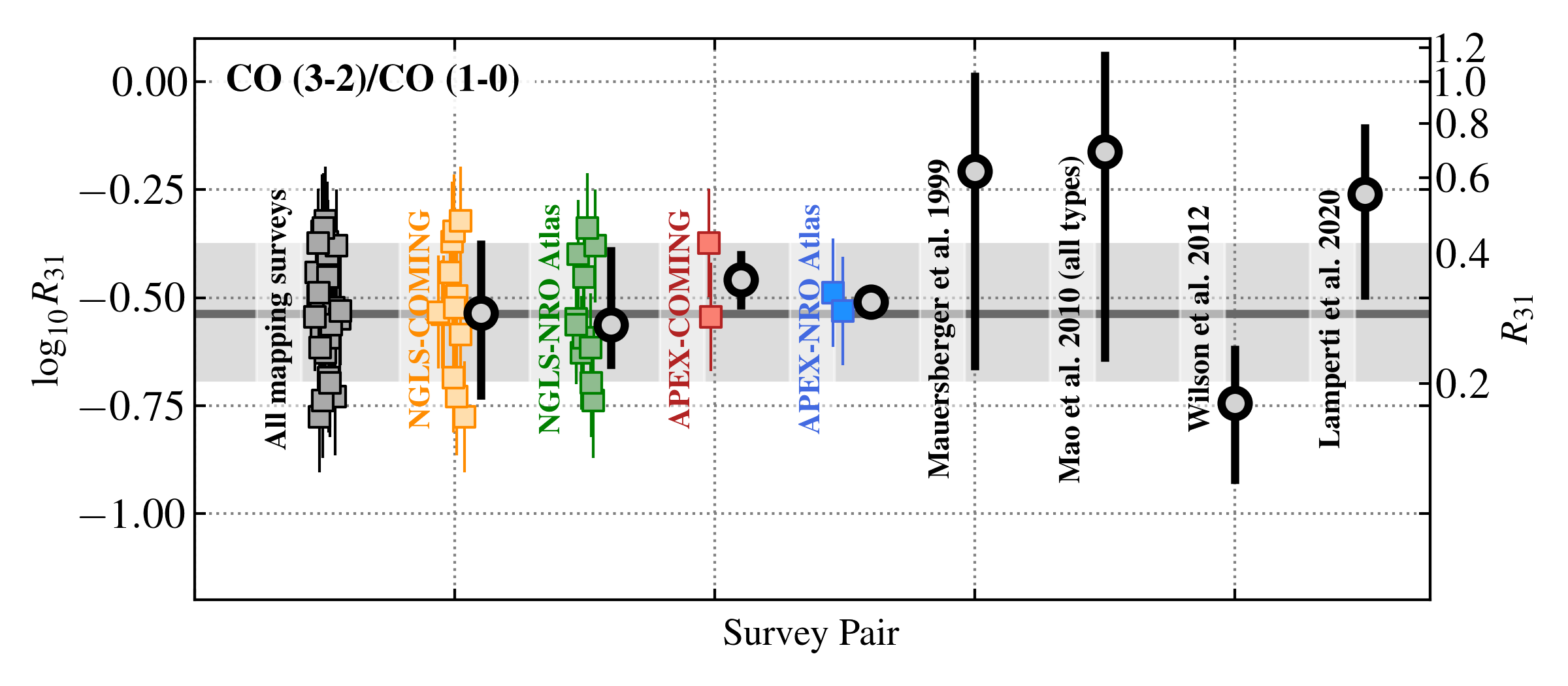}
\end{center}
\caption{\textit{Galaxy-integrated CO line ratios.} Colored data points show galaxy-integrated line ratios measured from maps, including limits. The error bars on the individual points reflect both statistical and calibration error as described in \S\ref{sec:meas}. For each survey pair or literature result, the black circle shows the median and the error bar shows the $\pm 1\sigma$ reported range for that survey pair or literature study. The \textit{top row} shows results for $R_{21}$. The \textit{middle row} shows results for $R_{32}$ and the \textit{bottom row} shows $R_{31}$. In each panel the thick gray horizontal line and shaded gray region show the median and $16{-}84^{\rm th}$ percentile range from all mapping data (see Table~\ref{tab:intrats}) in this study. 
\label{fig:hist}}
\end{figure*}

In total, as reported in Table~\ref{tab:sample}, we study $\pairs$ pairs of overlapping surveys. For each measured $R_{21}$, $R_{32}$, and $R_{31}$, Table~\ref{tab:meas} gives the name, survey pair, adopted distance to the galaxy, estimated \SFR, $M_\star$, and the CO luminosity, $L_{\rm CO}$, in each line. Note that as discussed above, we only quote one CO luminosity per line for each galaxy. This is the $L_{\rm CO}$ that we suggest to use as characteristic of the galaxy, not the value used in the calculation of the line ratio, because our quoted $L_{\rm CO}$ includes an aperture correction. When a galaxy was covered in the same line by multiple surveys, we chose which to use for $L_{\rm CO}$ following the same prioritization of surveys as noted in \S\ref{sec:methods}.

Figure~\ref{fig:hist} shows the distribution of the three measured line ratios, along with ranges reported for nearby galaxies in the literature. The first part of each plot shows results for all mapping surveys, then we separate the results according to the survey pair used for the measurement. Table~\ref{tab:intrats} reports basic results for the distributions combining all mapping surveys.

\medskip

\noindent $\bm{R_{21} \equiv \textbf{CO}\,(2{-}1) / \textbf{CO}\,(1{-}0)}$: Treating all data equally, we find a median $R_{21} = 0.61$ and a mean $R_{21} = 0.65$ with a $16{-}84^\mathrm{th}$ percentile range from $0.50{-}0.83$. This reflects $76$ mapping measurements from $43$ galaxies in this work and $9$ measurements from \citet{DENBROK21}. Note that this distribution allows repeat measurements when the same galaxy was targeted by multiple surveys, and will therefore weigh ``popular'' targets more heavily. Since our goal here is to synthesize the current literature we consider this approach reasonable, and we use only a single best value for each galaxy when fitting scaling relations below. As Figure~\ref{fig:sfms} shows, the targets of the surveys we consider emphasize high mass, high SFR galaxies (see also \S\ref{sec:biases}).

As Figure~\ref{fig:hist} shows, our $R_{21}$ values agree well with previous results for normal star-forming galaxies. We find an almost identical median value to the $0.64$ measure for EMPIRE galaxies in \citet{DENBROK21}, though our data show higher scatter than theirs. This likely reflects both the high quality of the EMPIRE \coone\ maps presented in \citet{DENBROK21} and the narrower range of galaxy properties sampled by EMPIRE, which we illustrate in Figure~\ref{fig:sfms}. We also find almost perfect agreement with \citet{YAJIMA21}, who also derive a median of $R_{21} = 0.61$ with a scatter of $\pm 0.19$. The sample in \citet{YAJIMA21} represents a subset of our own so we expect this close match. Finally, our measurements also agree reasonably well with previous HERACLES results by \citet{LEROY13}, who found median $R_{21}$ of $0.67$\footnote{Note that the earlier $0.8$ recommended by \citet{LEROY09} based on HERACLES was revised down by \citet{LEROY13} based on updated estimates of the IRAM \mbox{30-m} main beam efficiency.} with a scatter of $\pm 0.16$~dex or $\pm 44\%$, corresponding to a range of ${\sim} 0.46{-}0.97$. Finally, we also agree well with the $0.58$ median and ${\sim} 0.49{-}0.77$ $1\sigma$ range for literature single-pointing measurements compiled by \citet{DENBROK21}.

Our measurements appear slightly lower than the xCOLD~GASS measurements by \citet{SAINTONGE17}. We attribute this mostly to selection effects, though given that the line ratios drop with radius (\S\ref{sec:local}) there could also be some mild impact from the limited xCOLD~GASS beam size. As Figure~\ref{fig:sfms} shows, the xCOLD~GASS $R_{21}$ measurements target a wider range of stellar mass than our current $R_{21}$ sample. The lower mass galaxies and high $\SFR/M_\star$ galaxies in the xCOLD~GASS sample likely shift the median to the higher average value of $R_{21}=0.8$ that they report. For reference, if we include the \citet{SAINTONGE17} measurements in our sample, the combined data set has median $R_{21} = 0.64$, mean $R_{21} = 0.74$, and $16{-}84^{\rm th}$ percentile range of $0.52{-}0.91$.

This paper does not focus on individual targets, but we briefly note that the three high $R_{21} > 1.0$ values seen in Figure~\ref{fig:hist} are all consistent with $R_{21} \lesssim 1$ within the uncertainties. These are NGC~1087 in PHANGS--ALMA+COMING, NGC~2976 in HERA+NRO Atlas, and NGC~4536 in HERA+NRO Atlas. Given the sample size and magnitude of the uncertainties, we expect a few such outliers.

Summarizing, Figure~\ref{fig:hist} shows overall good convergence among recent studies of $R_{21}$. Adopting a typical value of $R_{21} = 0.65$ with a $30\%$ uncertainty that reflects scatter across the galaxy population represents a good assumption for high mass, $\log_{10} M_\star \, [{\rm M}_\odot] \gtrsim 10.25$ (see Figure~\ref{fig:sfms}), $z=0$ galaxies on the main sequence of star-forming galaxies \citep{LEROY13,DENBROK21,YAJIMA21}.

\medskip

\noindent $\bm{R_{32} \equiv \textbf{CO}\,(3{-}2) / \textbf{CO}\,(2{-}1)}$: We find a median $R_{32}= 0.46$ with mean $R_{32} = 0.50$ and a $16{-}84^{\rm th}$ percentile range of $0.23{-}0.59$. Because \cothree\ is comparatively faint and the \cotwo\ data have higher signal-to-noise, upper limits affect our $R_{32}$ distribution more than the other two ratios. We have treated the upper limits as equal to our minimum measured ratio for quantifying the distribution. This choice mainly affects our $16^{\rm th}$ percentile estimate. With $7$ of $40$ $R_{32}$ measurements being upper limits, the $16^{\rm th}$ percentile quoted is set to the lowest measured ratio.

Compared to $R_{21}$ and $R_{31}$, $R_{32}$ has the least extensive sample of previous beam-matched or mapping based studies of whole nearby galaxies; we are only aware of the work by \citet{WILSON12}, who found mean $R_{32}$ of $0.36$ with $\pm 0.13$ scatter using earlier versions of the same data that we use here.  This is moderately lower than our calculated mean value. We attribute part of the difference to revisions to the adopted IRAM main beam efficiency \citep[see above and][]{LEROY13B} after \citet{WILSON12} made their measurements, and to our ability to match the areas used for the calculations in this paper, which was not possible in \citet{WILSON12}. Taking these factors into account, the measurements appear roughly consistent. We also note that in Figure~\ref{fig:hist} the APEX data and JCMT NGLS data show hints of an offset. As far as we can tell, this reflects a mixture of small number statistics and perhaps the choice to focus the initial APEX LASMA mapping on bright, actively star-forming targets, which may have more excited molecular gas. Only one target overlaps between the two surveys, NGC~3627, and for that case we do find a ${\sim}30\%$ higher \cothree\ luminosity from APEX than the JCMT, but this is only a single target.

$R_{32}$ does appear to have a larger dynamic range than $R_{21}$. Because lower limits confuse the $16^{\rm th}$ percentile estimate for $R_{32}$, we compare the interquartile ($25{-}75^{\rm th}$ percentile) ranges for the two ratios, and find ${\sim}0.14$ dex range for $R_{21}$ and ${\sim}0.27$~dex for $R_{32}$. Though caveats related to a small sample size and the effect of lower limits still apply, this agrees with the expectation (see \S\ref{sec:expect}) that $R_{32}$ shows significant excitation variations across the range of real conditions found in molecular gas and indicates that the ratio has potential to act as a strong diagnostic of local excitation of molecular gas.

\medskip

\noindent $\bm{R_{31} \equiv \textbf{CO}\,(3{-}2) / \textbf{CO}\,(1{-}0)}$: We find a median $R_{31} = 0.29$ and a mean $R_{31} = 0.30$ with a $16{-}84^{\rm th}$ percentile range of $0.20{-}0.41$. Though the samples used to calculate the ratios vary, our $R_{31}$, $R_{32}$, and $R_{21}$ values approximately ``close'' as expected with $R_{31} \sim R_{32} \times R_{21}$, implying reasonable self-consistency within our measurements.

The literature reports a wide range of values. Our measurement is high compared to the $0.18\pm0.06$ reported by \citet{WILSON12} comparing NRO Atlas and JCMT NGLS data. Note, however, that \citet{WILSON12} caution that they do not match areas for the comparison, so their lower value could simply reflect a mismatch in apertures. Our measurements do agree well with the results for a smaller set of galaxies from \citet{WILSON09}. They used the NGLS and NRO Atlas to study regions in individual galaxies and found ratios in the ${\sim} 0.3{-}0.4$ range.

Our measurements lie on the low end of the range $R_{31} = 0.2{-}1.0$ found by \citet{MAUERSBERGER99}, but note that their sample also includes many starburst and active galaxies and only ${\sim} 7$ normal spiral galaxies. A similar case holds for \citet{MAO10}, who obtained $R_{31} = 0.7 \pm 0.5$. Partially based on those studies, we expect much higher, approaching thermal, $R_{31}$ in active galaxies and dense galaxies, so the contrast with our more quiescent targets seems reasonable. This appears to be mostly born out by our the results in \S\ref{sec:intcorr}. Our values also appear low compared to the mean $R_{31} = 0.55$ and $\pm 0.23$ scatter found for $28$ star-forming galaxies by \citet{LAMPERTI20} using single-pointing measurements. $R_{31}$ clearly shows a wide dynamic range and so good promise as a diagnostic. However the state of $R_{31}$ observations remains fairly limited, though not quite so much as for $R_{32}$.

\medskip

With $R_{31} < R_{32} < R_{21} < 1$ on average, these values appear consistent with the standard picture that most low-$J$ CO emission from nearby star-forming galaxies comes from optically thick clouds with the \cotwo\ and \cothree\ transitions moderately sub-thermally excited \citep[e.g., see \S\ref{sec:expect},][]{WEISS05,BOLATTO13B}. We discuss this more in Section~\ref{sec:discussion}.

\subsection{Comparison to integrated galaxy properties}
\label{sec:intcorr}

\begin{figure}[ht!]
\begin{center}
\includegraphics[width=0.49\textwidth]{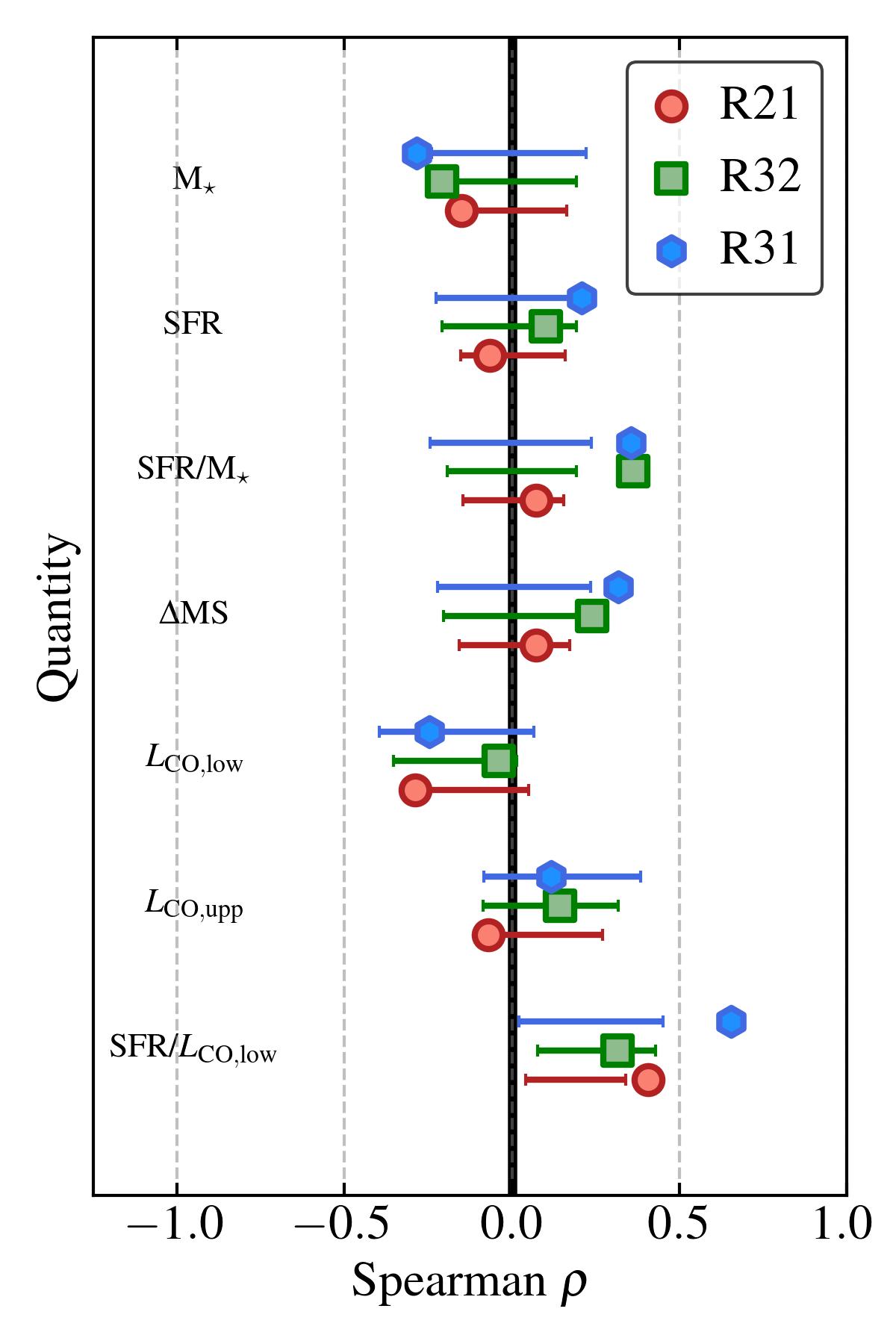}
\end{center}
\caption{\textit{Rank correlation of line ratios with galaxy properties.} Spearman rank correlation coefficients relating global galaxy properties to measured line ratios, neglecting limits. The bars show the $16{-}84$\% range of coefficients expected under the null hypothesis of no real correlation (N.B., in the cases with correlated axes, the rank correlation coefficient for no true correlation may not be centered on zero, see the text). $L_{\rm CO,low}$ refers to the luminosity of the lower CO transition in the ratio, e.g., \coone\ in $R_{21}$, while $L_{\rm CO, upp}$ refers to the upper transition, e.g., \cothree\ in $R_{31}$. As the contrast between results for $L_{\rm CO,upp}$ and $L_{\rm CO,low}$ demonstrates, the correlated axes play an important but hard-to-avoid role in determining correlations with $L_{\rm CO}$ and $\SFR/L_{\rm CO}$. Most correlations show marginal significance, but are consistent with low mass, high $\SFR/M_\star$, and high $\SFR/L_{\rm CO}$ galaxies showing higher line ratios, especially among the measurements involving \cothree. See also Figure~\ref{fig:intcorr}.
\label{fig:rankcorr}}
\end{figure}

\begin{figure*}[ht!]
\begin{center}
\includegraphics[width=0.4\textwidth]{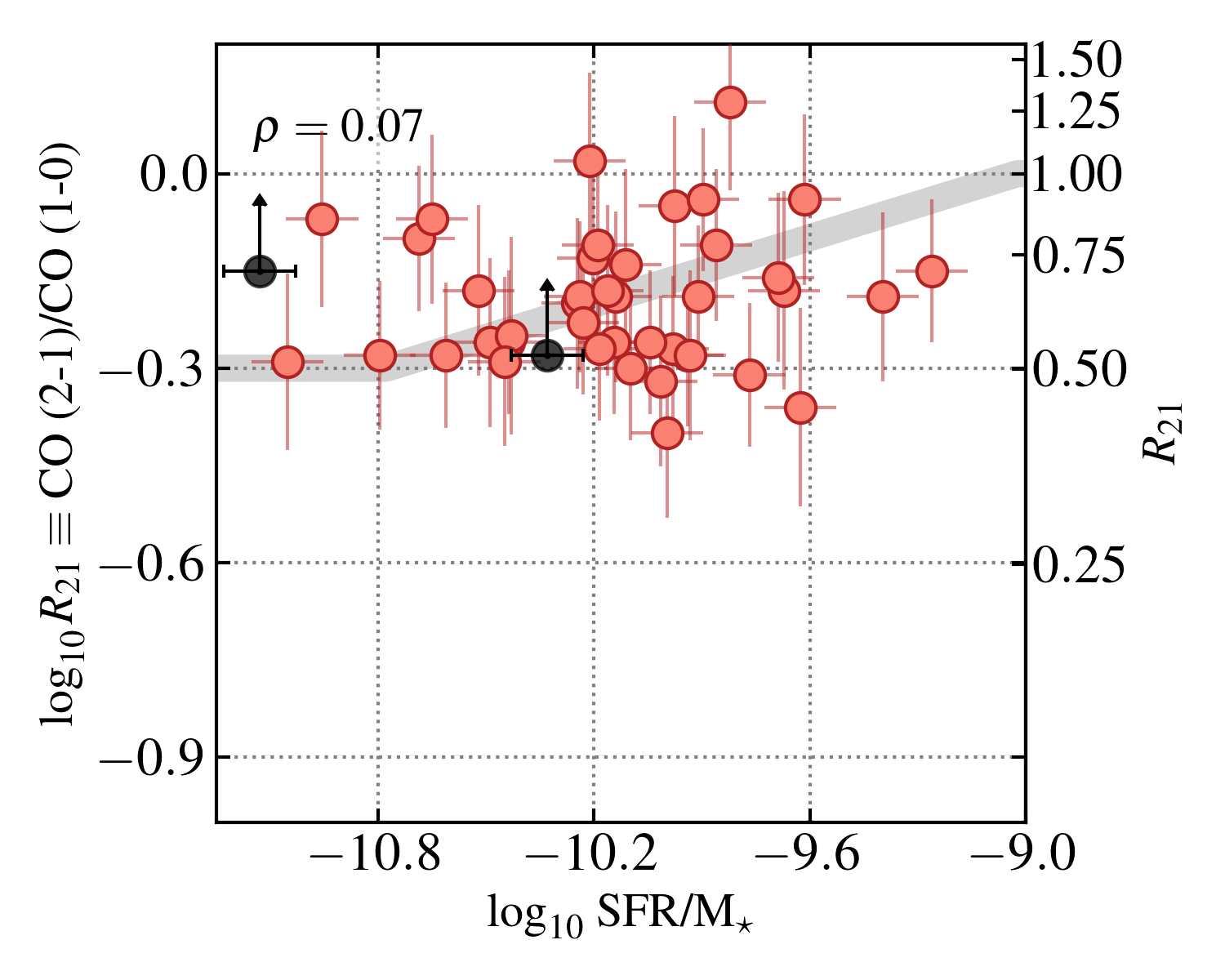}
\includegraphics[width=0.4\textwidth]{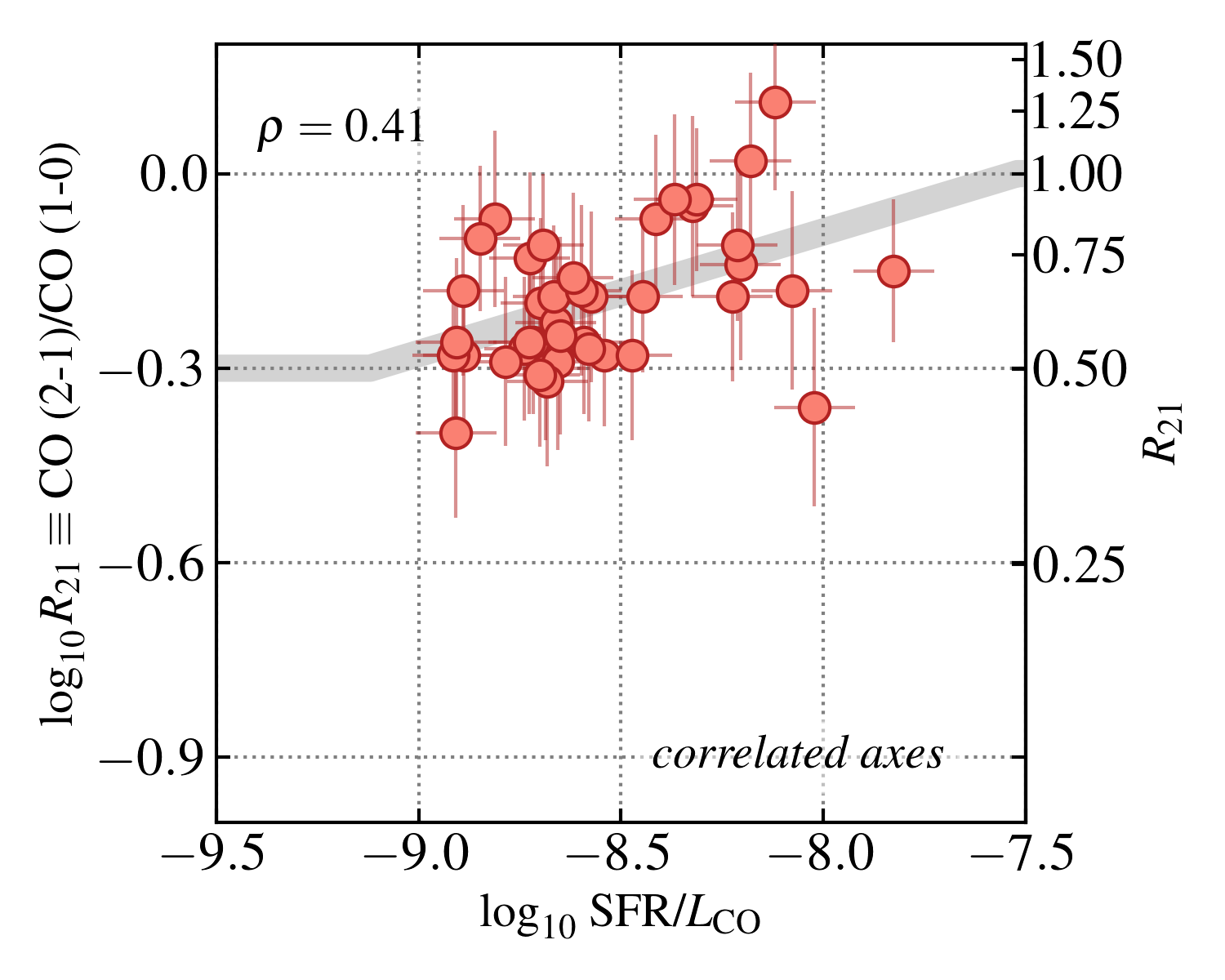} \\
\includegraphics[width=0.4\textwidth]{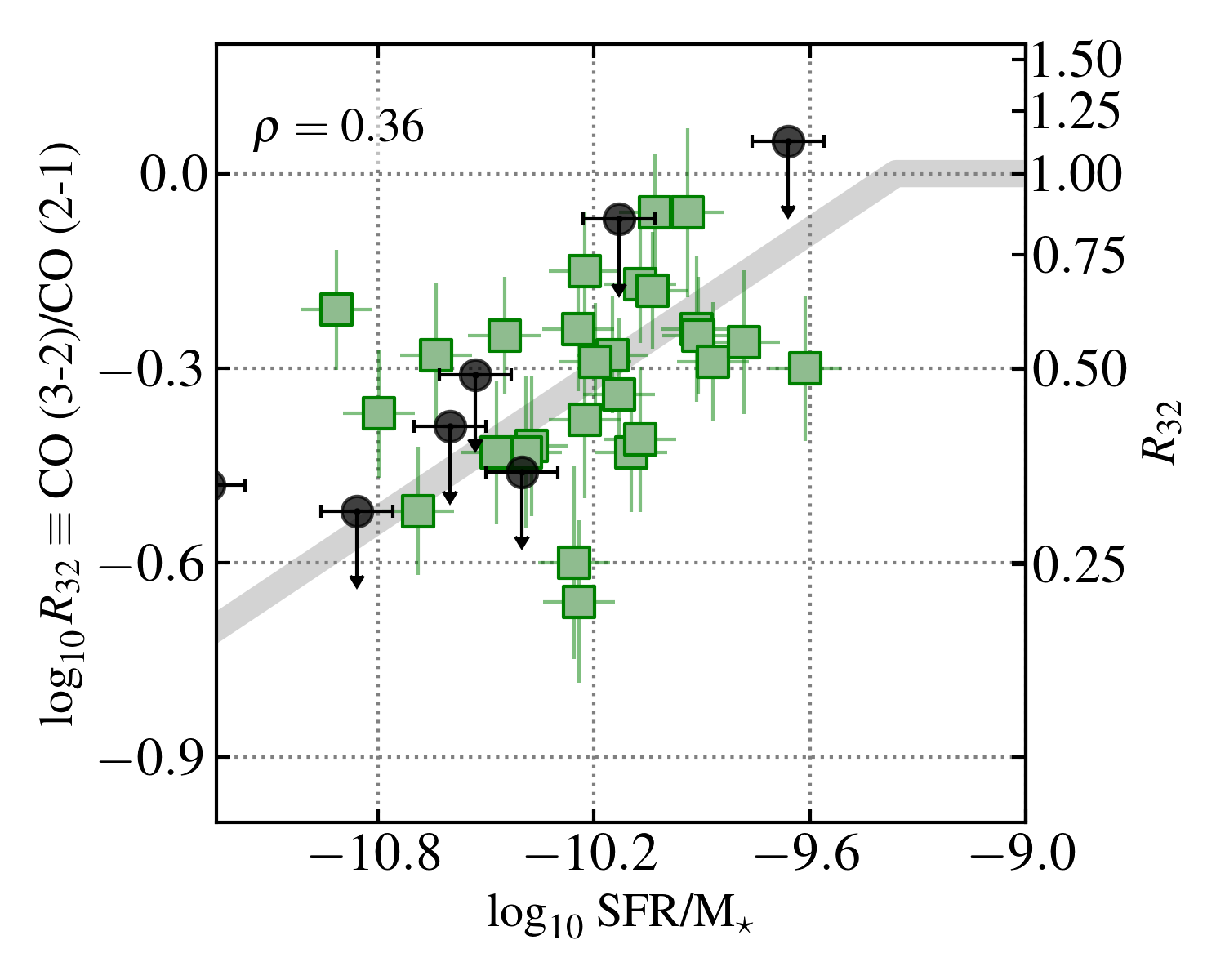}
\includegraphics[width=0.4\textwidth]{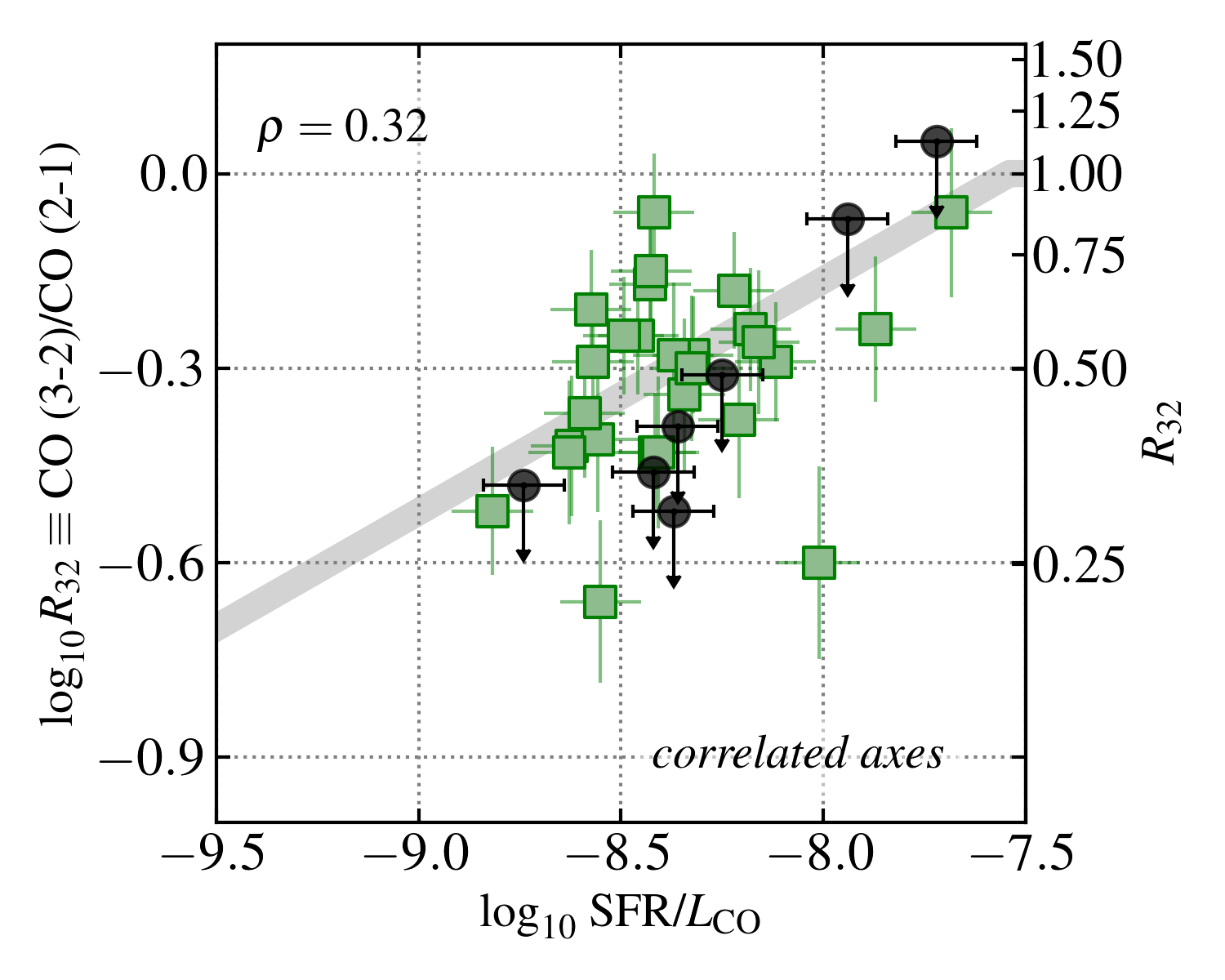} \\
\includegraphics[width=0.4\textwidth]{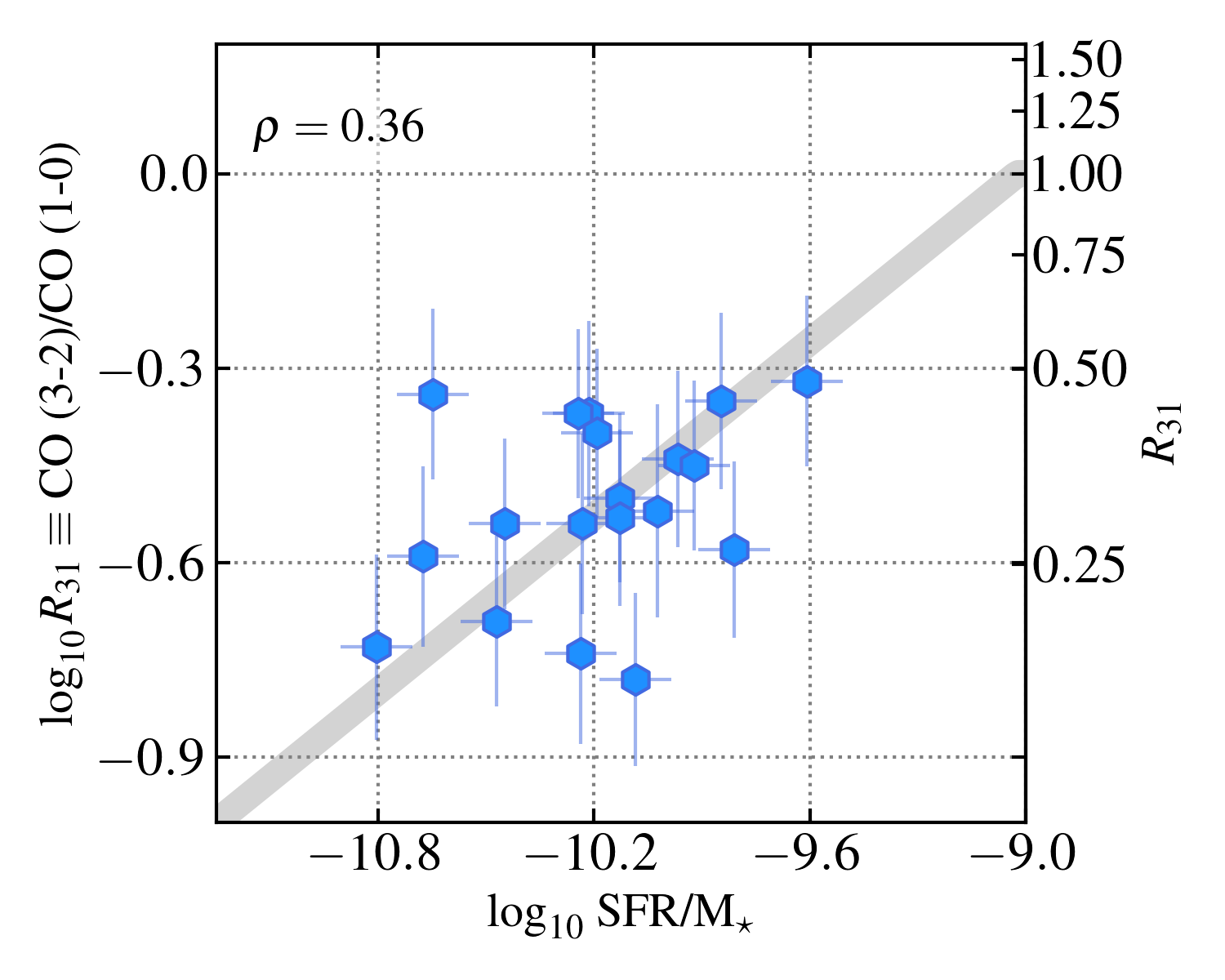}
\includegraphics[width=0.4\textwidth]{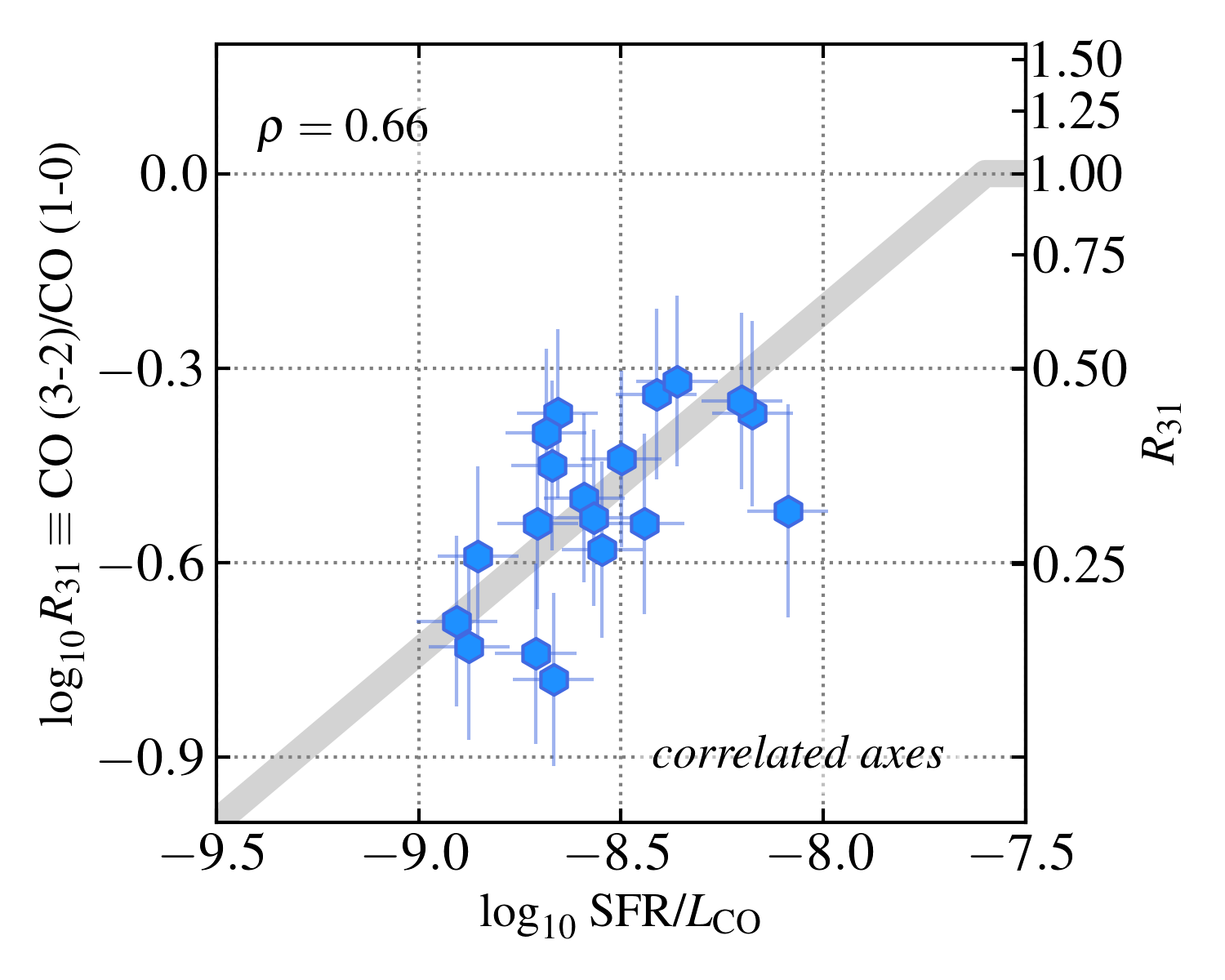}
\end{center}
\caption{
\textit{Correlations between line ratios and galaxy-integrated properties.} Galaxy-integrated $R_{21}$, $R_{32}$, and $R_{31}$ as function of quantities expected to relate to excitation, specifically $\SFR/M_\star$ and $\SFR/L_{\rm CO,low}$. Each point represents one map pair for one galaxy, with the black points showing limits. Error bars reflect both statistical and calibration uncertainties. The numbers in the top left corner report the Spearman $\rho$ relating the quantities on the $x$ and $y$ axes. The gray lines show highly approximate scaling relations that go through the data and then saturate at the thermal $\log_{10} R = 0$, see \S\ref{sec:fits},  Equation~\eqref{eq:pred_linerat}, and Table~\ref{tab:fits}. As discussed in \S\ref{sec:intcorr}, the axes in the right hand column are correlated and this may exert a significant effect on the observed relation.
\label{fig:intcorr}}
\end{figure*}

\begin{figure*}[ht!]
\begin{center}
\includegraphics[width=0.4\textwidth]{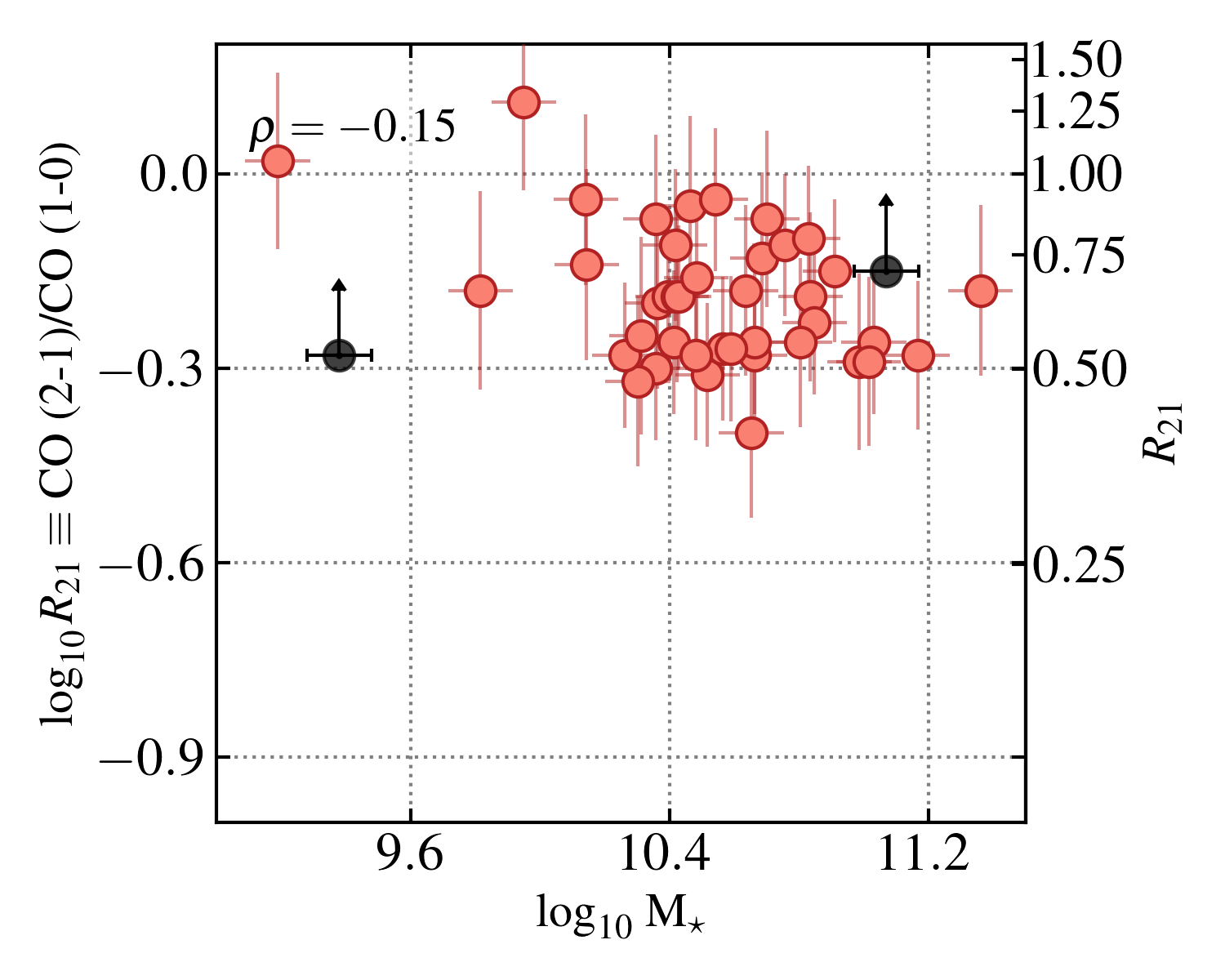}
\includegraphics[width=0.4\textwidth]{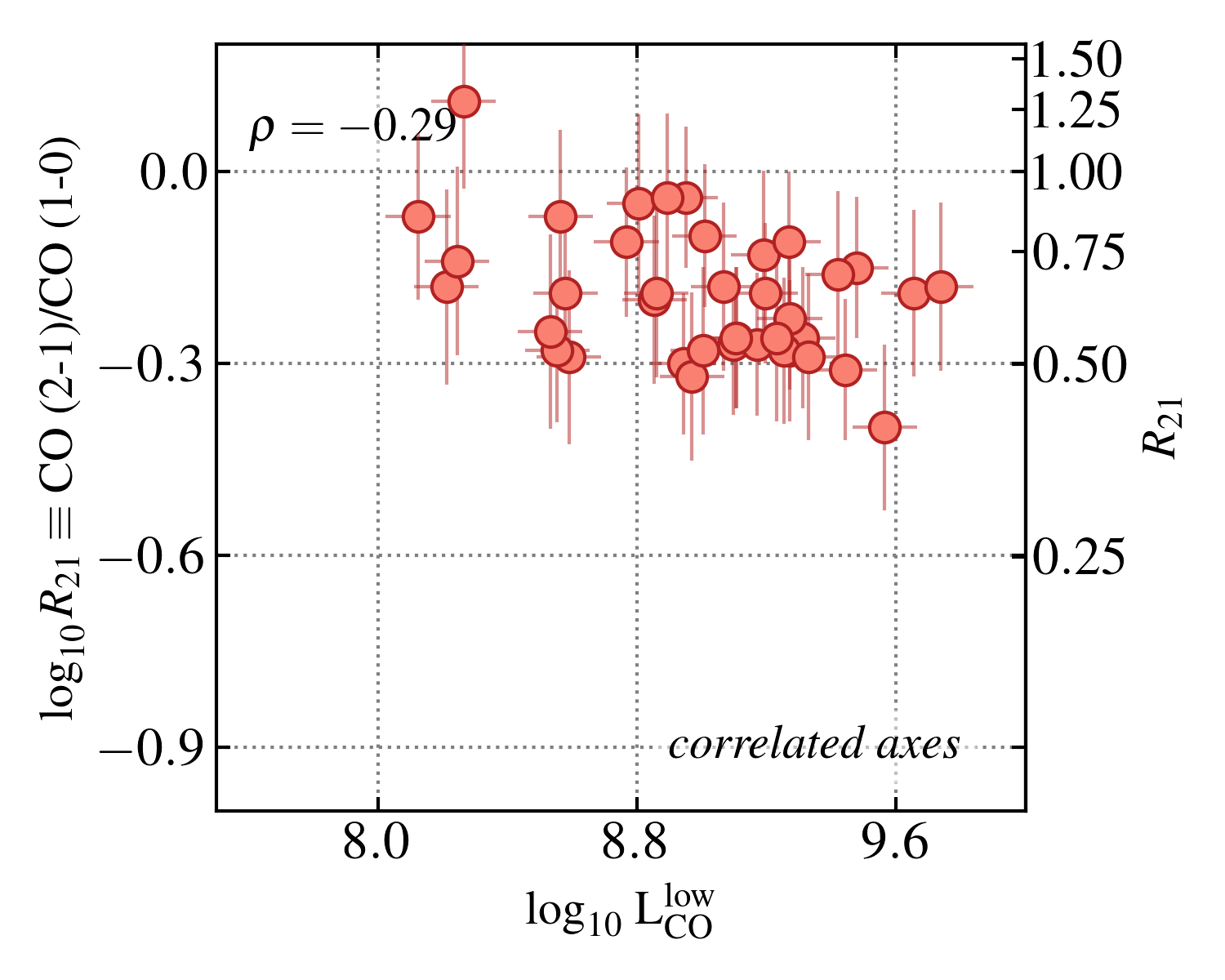} \\
\includegraphics[width=0.4\textwidth]{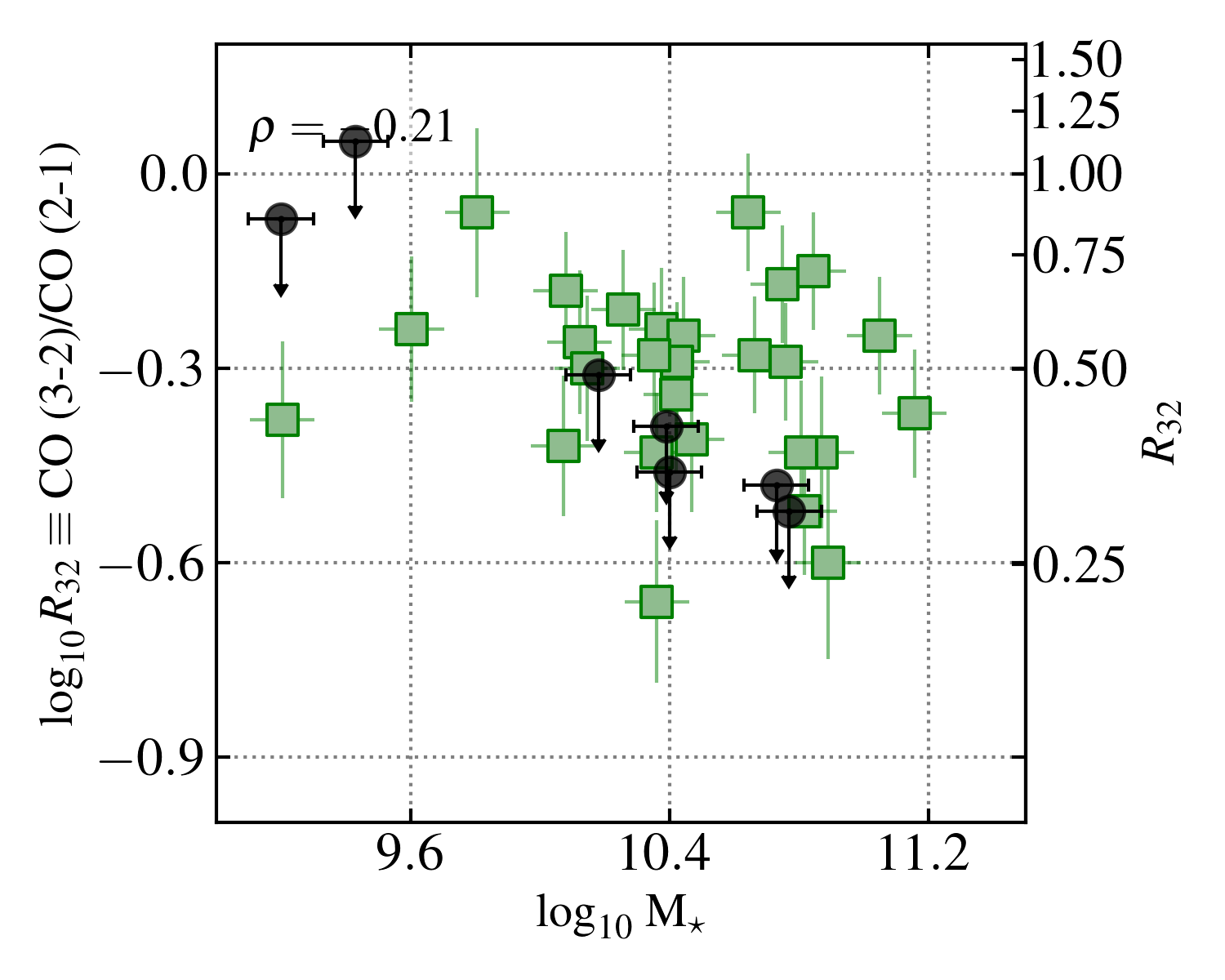}
\includegraphics[width=0.4\textwidth]{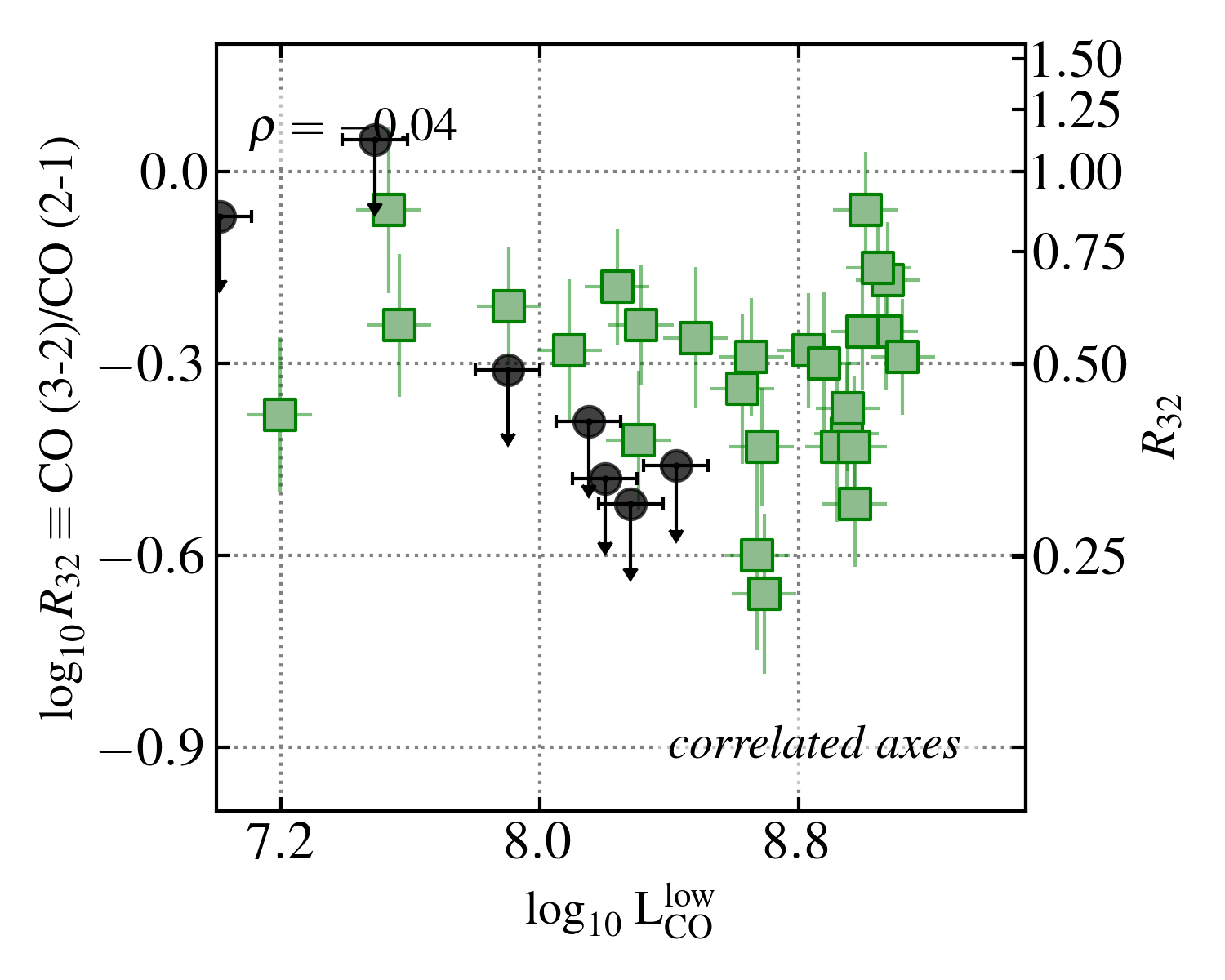} \\
\includegraphics[width=0.4\textwidth]{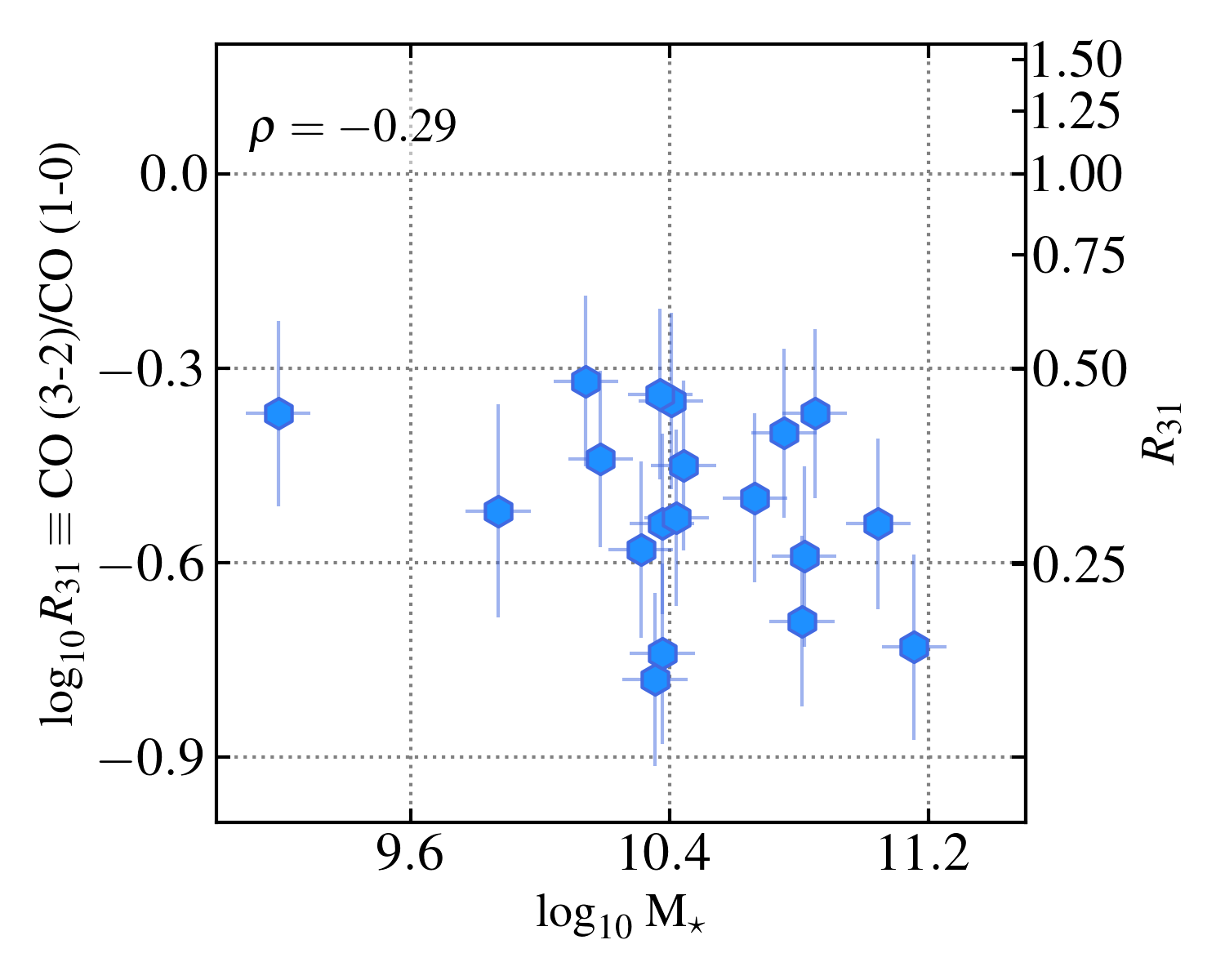}
\includegraphics[width=0.4\textwidth]{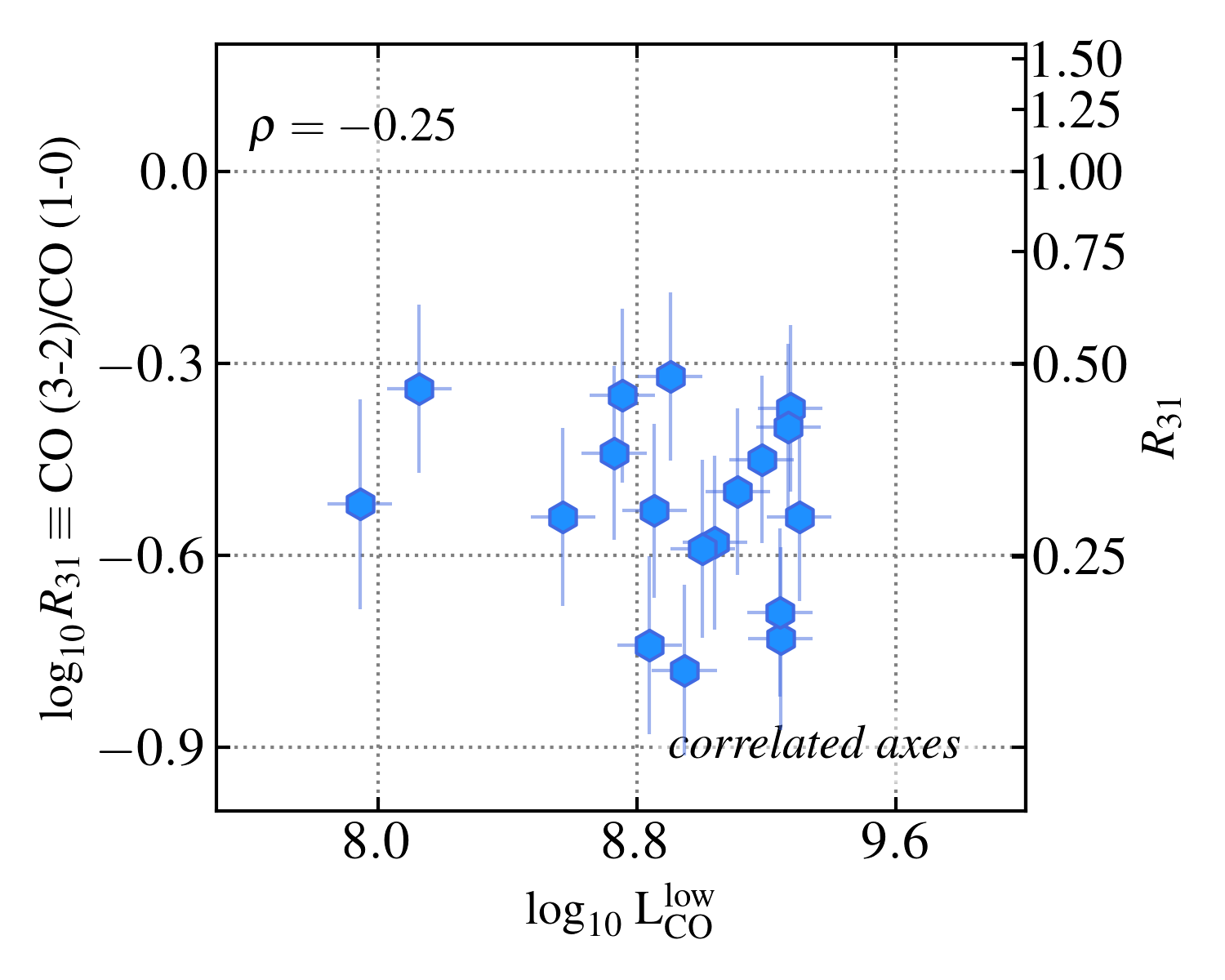}
\end{center}
\caption{
\textit{Correlations between line ratios and galaxy-integrated properties.} Galaxy-integrated $R_{21}$, $R_{32}$, and $R_{31}$ as function of $M_\star$ and $L_{\rm CO,low}$. Each point represents one map pair for one galaxy, with the black points showing limits. Error bars reflect both statistical and calibration uncertainties. The numbers in the top left corner report the Spearman $\rho$ relating the quantities on the $x$ and $y$ axes. As discussed in \S\ref{sec:intcorr}, the axes in the right hand column are correlated and this may exert a significant effect on the observed relation. The correlations here appear even weaker than those in Figure~\ref{fig:intcorr}, and we do not show any possible functional relations.
\label{fig:intcorr2}}
\end{figure*}

As discussed in \S\ref{sec:expect}, these ratios are both expected and observed to vary between galaxies. We test for correlations between all three line ratios and global galaxy properties in Figures~\ref{fig:rankcorr} and~\ref{fig:intcorr}. Before doing so, we again highlight the relatively narrow range of galaxy properties covered by current mapping surveys, visualized in Figure~\ref{fig:sfms} and discussed in \S\ref{sec:biases}. Especially for $R_{21}$, the current measurements focus on high-mass galaxies, mostly in the range $\log_{10} M_\star \, [{\rm M}_\odot] \sim 10.25{-}10.75$ with star formation rates just above the star-forming main sequence. Though the data are sparser for the ratios involving \cothree, these measurements do span a larger range of stellar mass and star formation rate at fixed stellar mass.

Figure~\ref{fig:rankcorr} shows the rank correlation coefficients relating each line ratio to $M_\star$, \SFR, specific star formation rate ($\SFR/M_\star$), offset from the main sequence of star-forming galaxies ($\Delta {\rm MS}$), CO luminosity, and \SFR-per-unit $L_{\rm CO}$ ($\SFR/L_{\rm CO}$). The colored bars show the expected correlation and $1\sigma$ scatter for the null hypothesis. In the cases of $M_\star$, \SFR , $\SFR/M_\star$, and $\Delta {\rm MS}$, we adopt the simple null hypothesis that the line ratio and the other variable are not correlated. Then, the expected correlation will be $0$ and the $1\sigma$ scatter reflects the range from randomly re-pairing the variables. Recall that here, unlike in Figure~\ref{fig:hist}, we use only a single best estimate of each line ratio for each galaxy.

In the case of $L_{\rm CO}$ and $\SFR / L_{\rm CO}$, the two variables used in the correlation will be correlated by construction. In this case, we construct the null hypothesis as follows. First, we measure the logarithmic scatter in the real line ratio across our data set. Then, as the null hypothesis, we assume a fixed underlying line ratio and that this scatter is evenly distributed across the variables involved. We generate the expectations for the null hypothesis by randomly applying this scatter repeatedly to each variable and calculating the rank correlation coefficient. Here, we consider the individual CO luminosities, not the line ratios, as the underlying variables for this exercise. This ensures that the null hypothesis captures the built-in correlation between the axes. For example, when we correlate the line ratio with $L_{\rm CO, upp}$, i.e., the luminosity in the numerator, we measure the scatter in the $\log_{10}$ of the line ratio, $\sigma$, and then we apply $\sqrt{2}$ times $\sigma$ in the model noise to each of $L_{\rm CO, upp}$ and $L_{\rm CO,low}$. Then, we construct the model line ratios $R = L_{\rm CO,upp}/L_{\rm CO,low}$ before calculating the expected correlation. That is, we define the null hypothesis to be the case where the line ratio is fixed and the variance matches the observed variance and is randomly distributed among the relevant variables.

Overall the figure shows a consistent sense of variations. Lower mass galaxies, which also have lower $L_{\rm CO}$, higher $\SFR/M_\star$, and higher $\SFR/L_{\rm CO}$, tend to show higher line ratios. The absent or weak correlation with \SFR\ can be understood as competing effects: low $M_\star$ galaxies have higher $\SFR/M_\star$ but also lower overall \SFR . The radiation field in low mass galaxies may be more intense due to a higher local $\Sigma_{\rm SFR}$, but the integrated \SFR\ will still be lower. In general, correlations with integrated galaxy properties appear stronger for $R_{32}$ and $R_{31}$ compared to $R_{21}$. This partially reflects the broader range of galaxy properties covered by those measurements (Figure~\ref{fig:sfms}), and may also reflect that $R_{32}$ and $R_{31}$ have more sensitivity than $R_{21}$ to the range of conditions found in normal galaxies (see \S\ref{sec:expect}).

These trends make physical sense and agree with the limited previous measurements. Physically, elevated $\SFR/M_\star$ may trace more intense radiation fields and stronger heating of the gas, suggesting higher temperatures. The anticorrelation with $M_\star$ may reflect the impact of dust shielding. Based on the existence of the mass--metallicity relation \citep[e.g.,][]{TREMONTI04,KEWLEY08}, we expect the low mass members of our sample to also have lower dust-to-gas ratios \citep[e.g.,][]{LEROY11,REMYRUYER14,CASASOLA19} and more intense radiation fields.  In literature studies, CO line ratios do appear enhanced in low metallicity regions or galaxies \citep[e.g.,][among many others]{LEQUEUX94,BOLATTO03,DRUARD14,KEPLEY16,CICONE17}. Higher $\SFR/L_{\rm CO}$ may indicate poorly-shielded, low metallicity gas in which the CO persists only in the core of a molecular cloud \citep[e.g., see discussion in][]{GLOVER12,SCHRUBA12,BOLATTO13A,RUBIO15}. Alternatively, higher $\SFR/L_{\rm CO}$ can indicate more efficiently star-forming gas, which will often be denser gas with more nearby heating sources. These are both factors that can lead to higher line ratios, especially $R_{32}$ and $R_{31}$ (see \S\ref{sec:expect}). Given that our sample skews towards relatively massive, and thus nearly solar metallicity targets, we expect that these density and heating effects likely represent the main drivers of the observed correlations. The correlations that we see agree with the results of \citet{LAMPERTI20} who showed a correlation between $\SFR/L_{\rm CO,low}$ and $R_{31}$ and with \citet{YAJIMA21}, who used a subset of the data we consider here and showed a correlation between $R_{21}$ and $\SFR/L_{\rm CO,low}$. Qualitatively, Figure~\ref{fig:rankcorr} echos other results  at low and high redshift that show a correlation between normalized star formation activity and excitation \citep[e.g.,][]{WEISS05,BOLATTO13A,LIU21}.

Although the pattern in Figure~\ref{fig:rankcorr} makes physical sense, the trends are not particularly significant. The $p$-values relating $M_\star$ to $R_{31}$ and $R_{32}$ are only $0.24{-}0.29$. For $\SFR/M_\star$ the $p$-values for $R_{31}$ and $R_{32}$ are ${\sim}0.06{-}0.13$, more significant but still indicating only weak correlations. The other significant correlations involve $L_{\rm CO, low}$, the luminosity of the lower CO transition in the line ratio. We report these because the results make physical sense and are interesting, but the line ratio (${\sim} L_{\rm CO,upp}/L_{\rm CO,low})$, and a quantity involving $L_{\rm CO,low}$ are correlated by construction. To see this, contrast the significant correlations seen in Figure~\ref{fig:rankcorr} for all line ratios and $L_{\rm CO,low}$ and the lack of similar significant correlations for $L_{\rm CO,upp}$. For the moment, we only caution that these results include the effects of correlated measurements and should be taken as indicative but likely overstate the significance of the correlation in the data.

With these caveats in mind, Figures~\ref{fig:intcorr} and~\ref{fig:intcorr2} visualizes the correlations between each line ratio and global quantities: $\SFR/M_\star$, $\SFR/L_{\rm CO,low}$, $M_\star$, and $L_{\rm CO,low}$. The data show large scatter, but do show evidence for overall correlation with the sense that higher line ratios emerge from galaxies with high $\SFR/M_\star$ and/or $\SFR/L_{\rm CO}$. The correlations of line ratios with $M_\star$ and $L_{\rm CO}^{\rm low}$ appear weaker.

\subsubsection{Approximate scaling relations}
\label{sec:fits}

\begin{deluxetable}{lccccc}
\tabletypesize{\small}
\tablecaption{Approximate Scaling Relations \label{tab:fits}}
\tablewidth{0pt}
\tablehead{
\colhead{Ratio} & 
\colhead{Quantity} & 
\colhead{$y_{\rm low}$} & 
\colhead{$x_{\rm low}$} & 
\colhead{$x_{\rm high}$} &
\colhead{Scatter\tablenotemark{a}}
}
\startdata
$R_{21}$ & $\SFR/M_\star$ & $-0.3$ & $-10.78$ & $-9.02$ & $0.07$ \\
$R_{21}$ & $\SFR/L_{\rm CO,low}$\tablenotemark{b} & $-0.3$ & $-9.12$ & $-7.52$ & $0.07$ \\
\hline
$R_{32}$ & $\SFR/M_\star$ & $-0.7$ & $-11.25$ & $-9.36$ & $0.10$ \\
$R_{32}$ & $\SFR/L_{\rm CO,low}$\tablenotemark{b} & $-0.7$ & $-9.50$ & $-7.54$ & $0.08$ \\
\hline
$R_{31}$ & $\SFR/M_\star$ & $-1.0$ & $-11.25$ & $-9.02$ & $0.08$ \\
$R_{31}$ & $\SFR/L_{\rm CO,low}$\tablenotemark{b} & $-1.0$ & $-9.50$ & $-7.60$ & $0.08$ \\
\enddata
\tablenotemark{a}{``Scatter'' reports the median absolute value of the residuals about the fit in dex.}
\tablenotetext{b}{$L_{\rm CO,low}$ refers to the CO luminosity of the lower transition in the line ratio, e.g., \coone\ in $R_{21}$.}
\tablecomments{Coefficients for the indicative scaling relations following Equation~\eqref{eq:pred_linerat} and illustrated in Figure~\ref{fig:intcorr}. Fits derived from assuming $y_{\rm low}$ and then fitting $x_{\rm low}$ and $x_{\rm high}$ by minimizing $\chi^2$. We report all quantities, including the scatter, in dex. As shown in the figure these should be taken as approximate and we expect future work to revise them considerably.}
\end{deluxetable}

Given the current state of the data, especially the large systematic uncertainties and uneven sampling of the $\SFR{-}M_\star$ space, rigorous scaling relations to predict the line ratios are probably not feasible. Nonetheless, we find it useful to illustrate current best fits and sketch a plausible approach to line ratio scaling relations.

We adopt the following principles:
\begin{itemize}
\item The line ratio should not go above $1$. Although there are a few measurements with $R > 1$ in the literature, such ratios likely reflect optically thin gas, which is not expected or observed to be the dominant emitting component over large parts of galaxies. One should cap any fit at $R \sim 1$ (see \S\ref{sec:expect} and Appendix~\ref{sec:corrections}).
\item The line ratio also should not drop to arbitrarily low values, even when a galaxy has low SFR, specific star formation rate, or stellar mass. The appropriate lower limit for the ratios is less immediately clear than the upper limit, but based on \S\ref{sec:expect} we suggest that $R_{21} \gtrsim 0.5$, $R_{32} \gtrsim 0.2$, and $R_{31} \gtrsim 0.1$ represent reasonable lower bounds.
\item Lacking any strong physical motivation for another functional form, we assume a power law intermediate between these two limits.
\end{itemize}

Thus we suggest that the following broken power law represents a reasonable way to predict $y = \log_{10} R$ from some variable $x$ that shows a positive correlation with $R$:
\begin{equation}
\label{eq:pred_linerat}
y =
\begin{cases}
    y_{\rm low}, & \text{if}~ x < x_{\rm low} \\
    y_{\rm low} + m \left( x - x_{\rm low}\right) , & \text{if}~x_{\rm low} < x < x_{\rm high} \\
    0.0 , & \text{if}~x > x_{\rm high}
\end{cases}
\end{equation}
\noindent where
\begin{equation}
\nonumber 
y \equiv \log_{10} R \quad {\rm and} \quad
m = \frac{0.0-y_{\rm low}}{x_{\rm high}-x_{\rm low}}~.
\end{equation}
\noindent Each relation has three parameters: $y_{\rm low}$, the $\log_{10}$ value of the line ratio at low values of the $x$-axis; $x_{\rm low}$, the $x$-axis value where the ratio begins to increase, and $x_{\rm high}$, the $x$-axis value where the line ratio reaches the thermal value $y \equiv \log_{10} R = 0.0$, i.e., $R=1$. We adopt values of $y_{\rm low}$ described above based on physical expectations, though we note that particularly for $R_{21}$ a lower value could be plausible. Then for both $x = \log_{10} \SFR/M_\star$ and $x = \log_{10} \SFR/L_{\rm CO,low}$, we fit $x_{\rm low}$ and $x_{\rm high}$ by minimizing $\chi^2$ while varying $x_{\rm low}$ and $x_{\rm high}$. We report the best fit values in Table~\ref{tab:fits} and illustrate them in Figure~\ref{fig:intcorr}, but we emphasize again that in our current data the measured correlations have only marginal significance, especially for $R_{21}$. Improvements to both the data quality and the sample studied are needed before we will have a strong predictive relation to estimate $R$ based on other galaxy properties.

\subsection{Correlations with local conditions}
\label{sec:local}

\begin{figure*}[ht!]
\begin{center}
\includegraphics[width=0.4\textwidth]{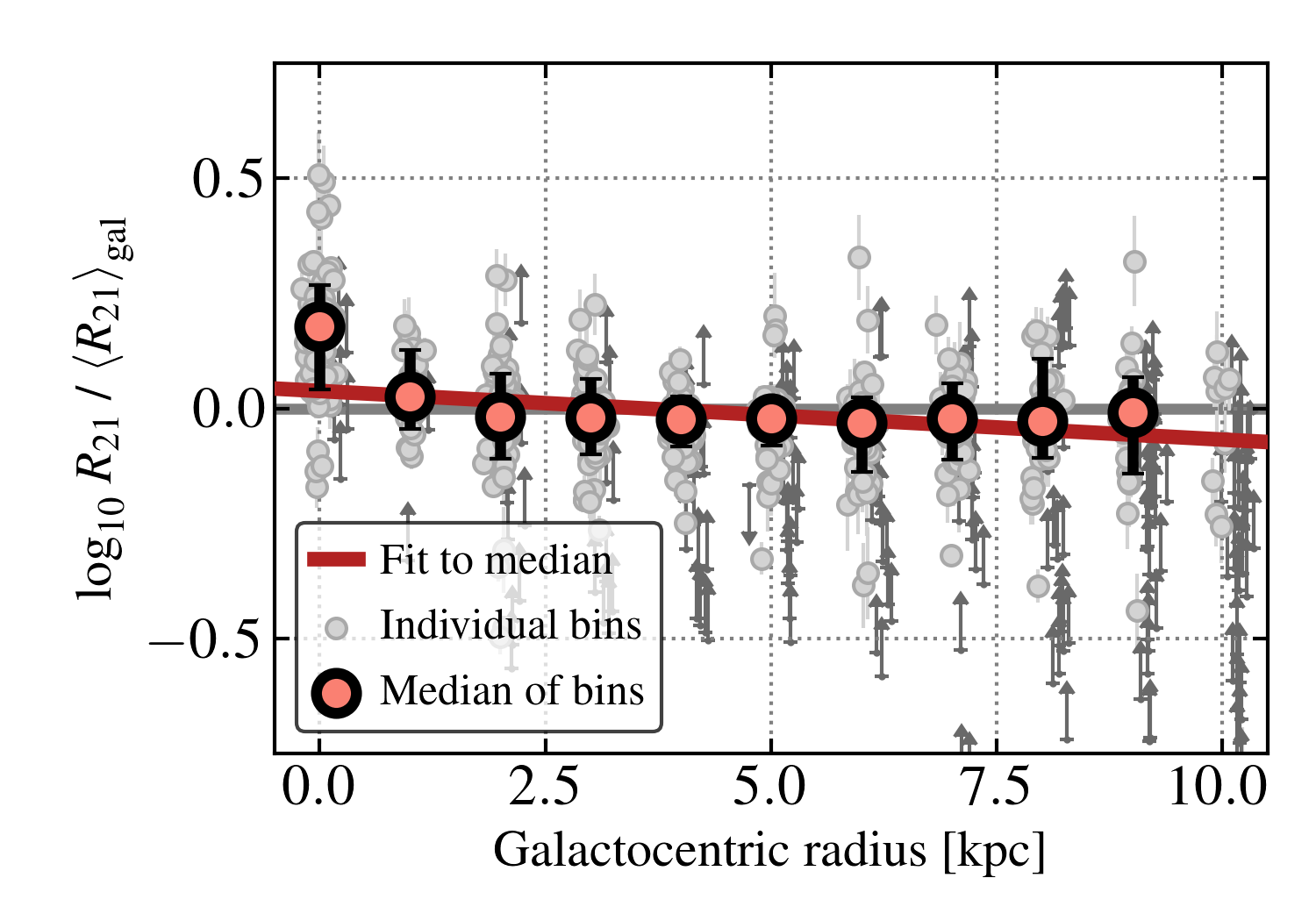}
\includegraphics[width=0.4\textwidth]{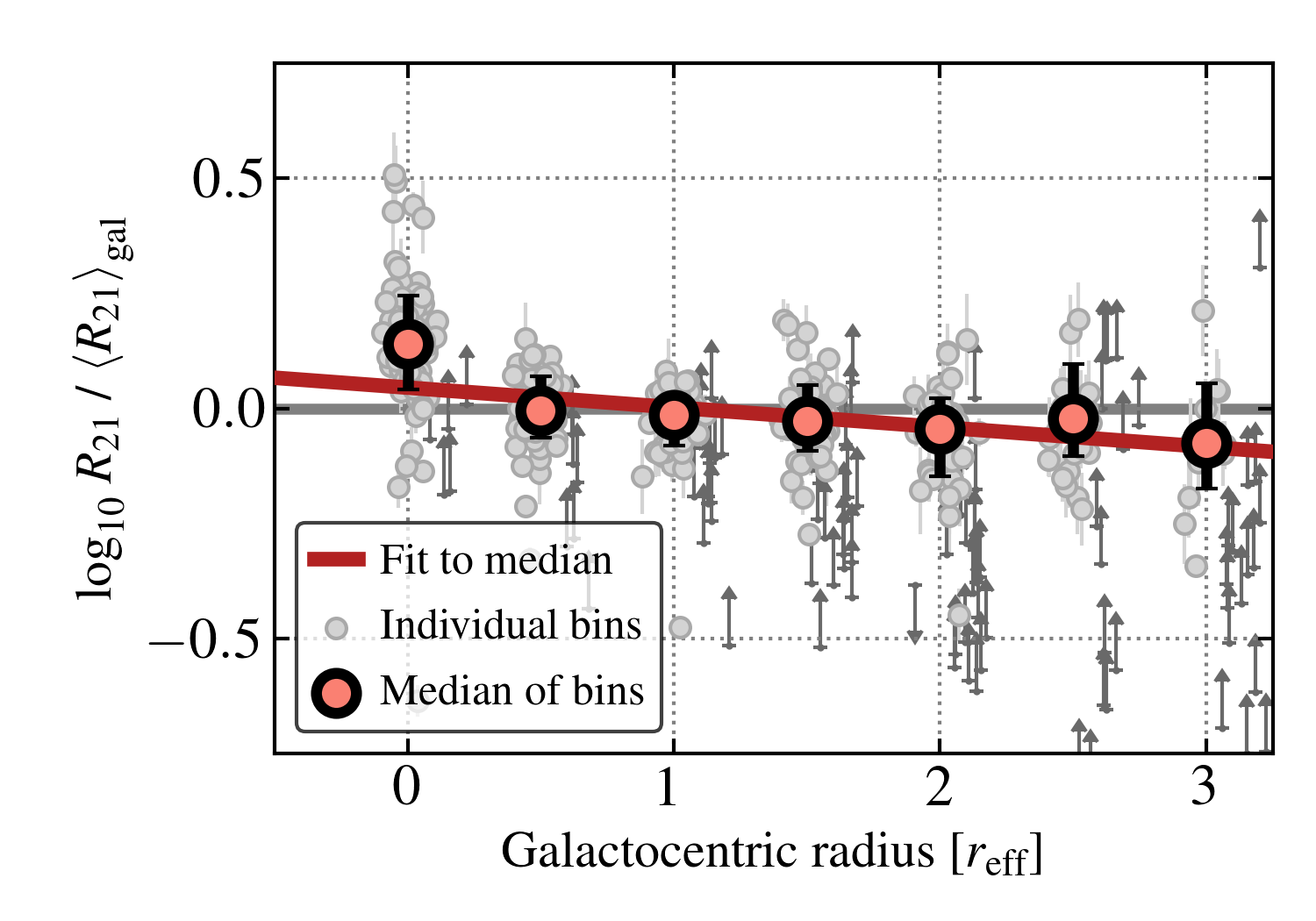} \\
\includegraphics[width=0.4\textwidth]{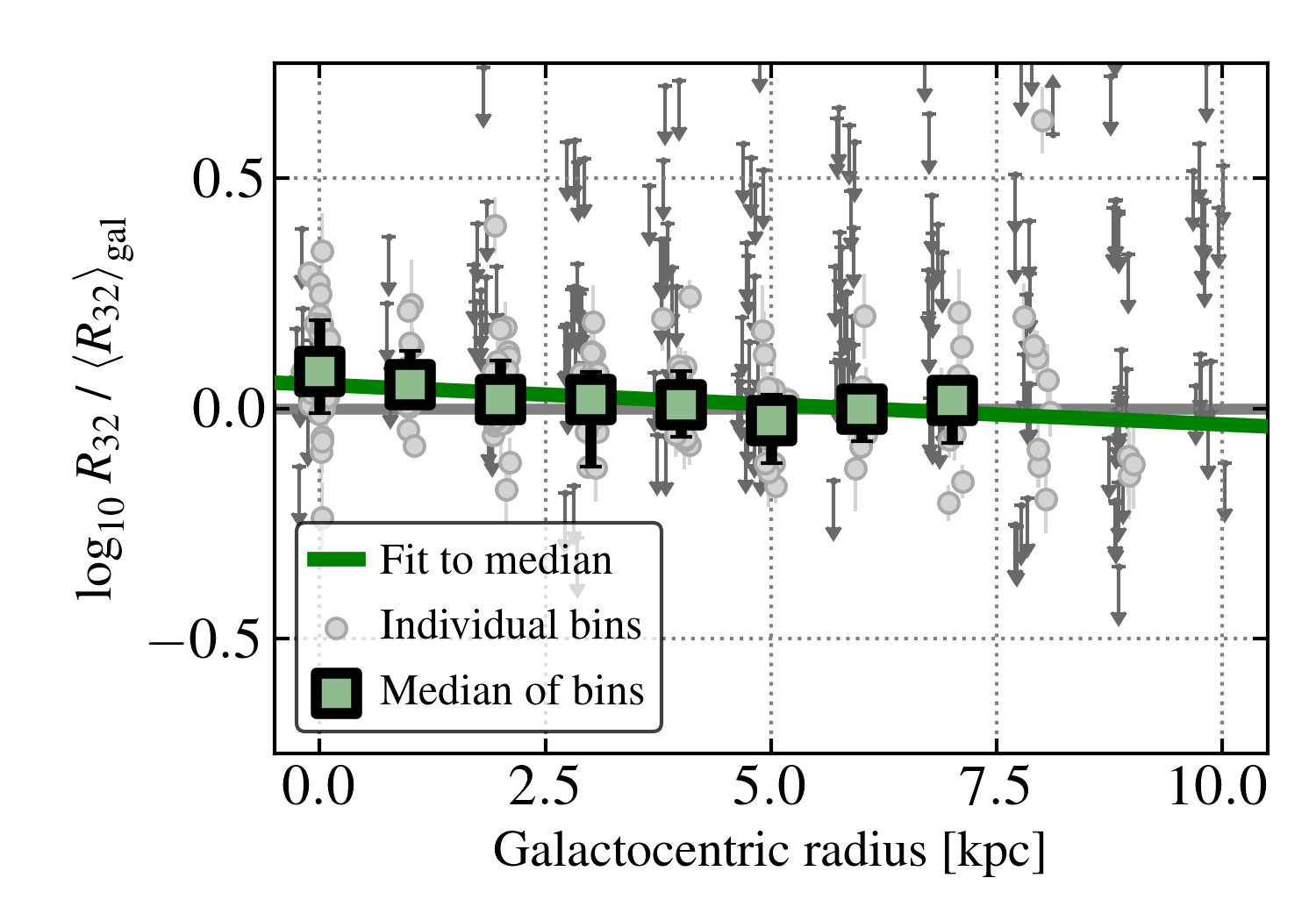}
\includegraphics[width=0.4\textwidth]{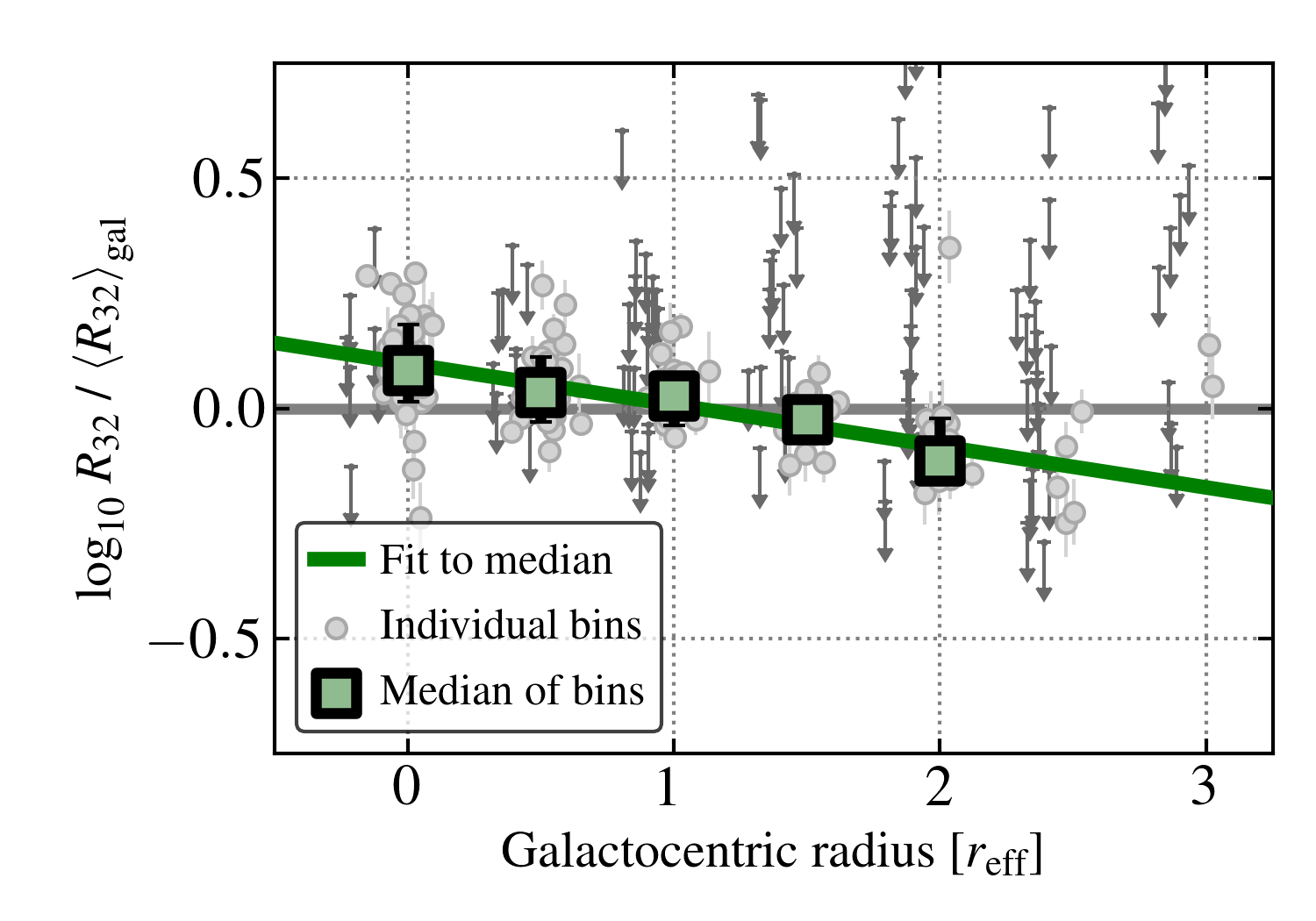} \\
\includegraphics[width=0.4\textwidth]{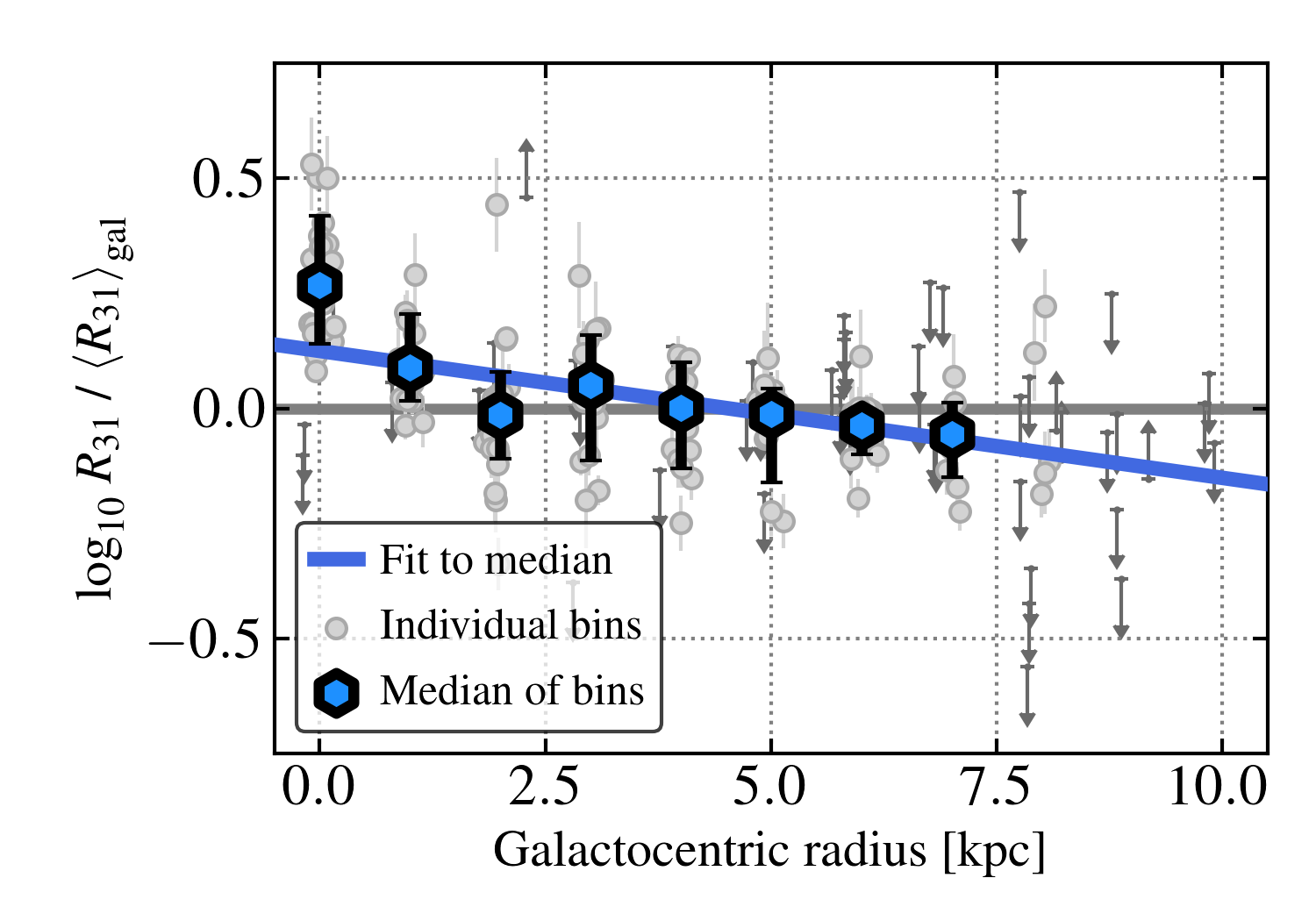}
\includegraphics[width=0.4\textwidth]{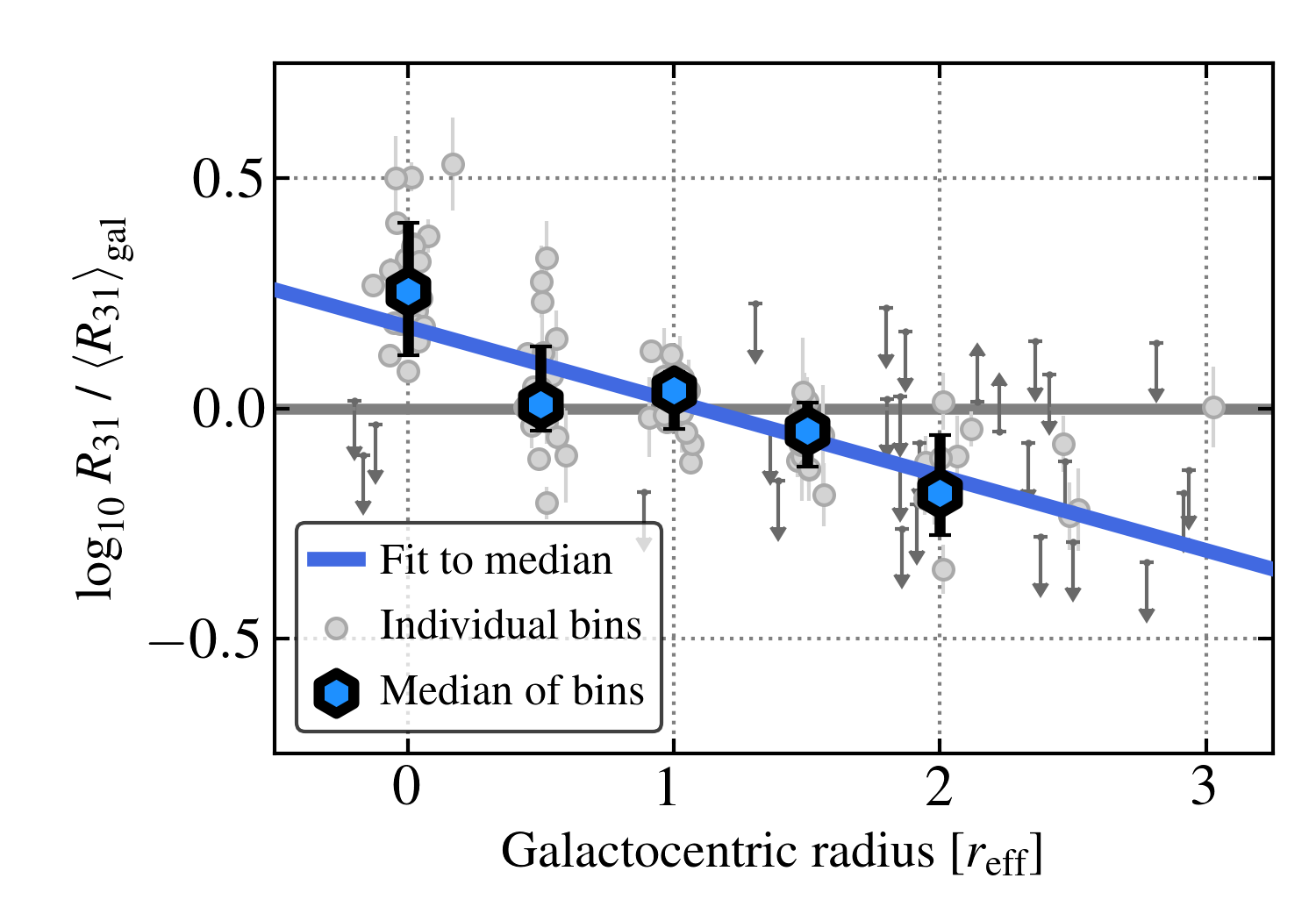}
\end{center}
\caption{
\textit{Correlations between line ratios and galactocentric radius.} Each plot shows line ratio enhancement or deficit relative to the galaxy average, $\log_{10} R / \!\langle R \rangle$, as a function of galactocentric radius. Individual points showing binned results for individual galaxy plus survey pair combinations calculated as described in \S\ref{sec:meas}. Arrows show $4\sigma$ upper or lower limits. Colored points indicate the median in each radial bin considering all galaxies and survey pairs with a measurement in that bin, including limits. The black vertical error bars show the $16{-}84^{\rm th}$ percentile of measurement in that bin. We only plot median values where a non-limit 16, 50, and 84\% value could be inferred from our measurements. The solid colored line shows a least squares fit to all bins with at least $5$ data points (see Table~\ref{tab:resolved}). The left panel shows bins constructed considering physical radius, while the right panel normalizes the radius by the half-mass radius, $r_{\rm eff}$.
\label{fig:resolved_rad}}
\end{figure*}

\begin{figure*}[ht!]
\begin{center}
\includegraphics[width=0.4\textwidth]{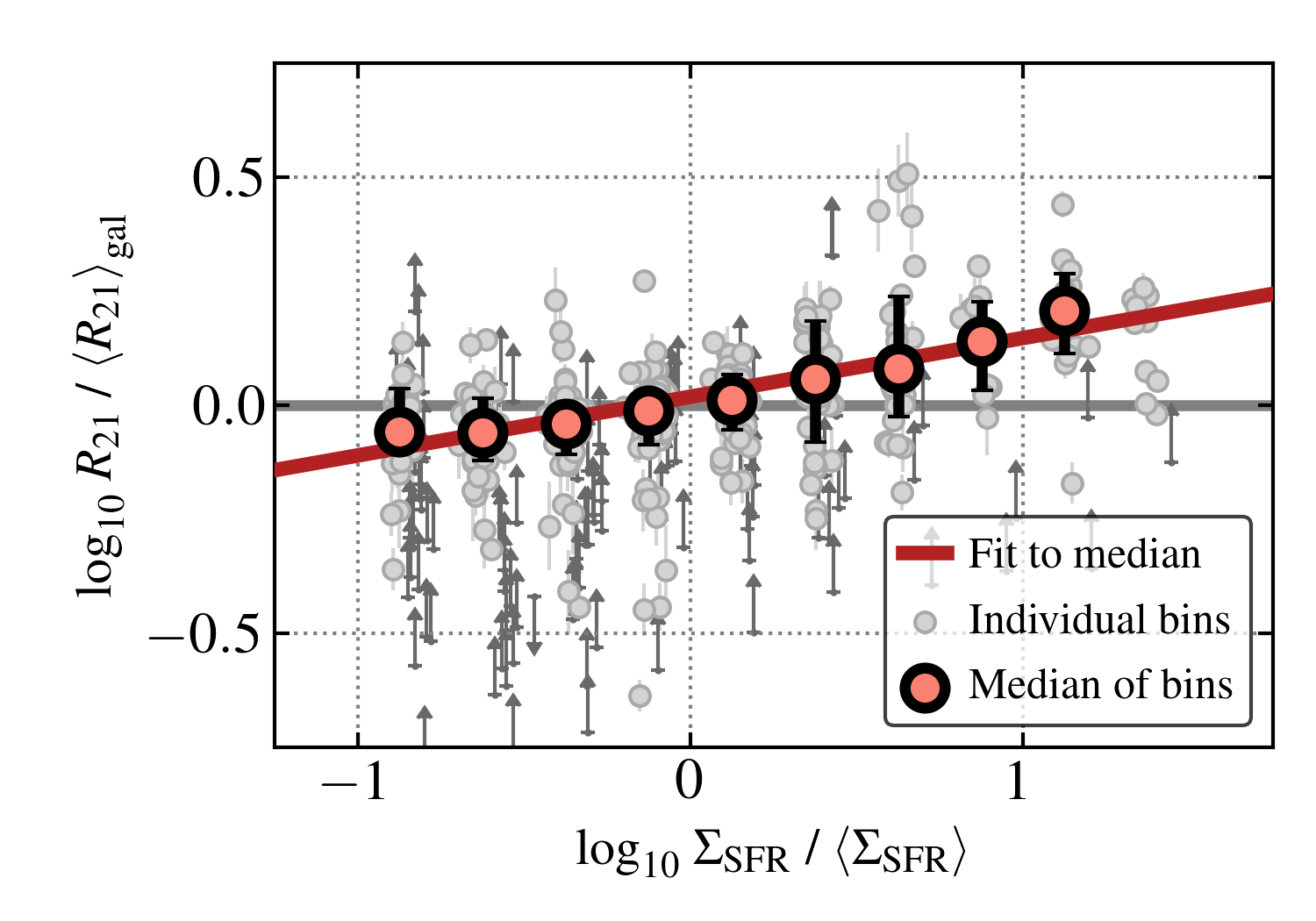}
\includegraphics[width=0.4\textwidth]{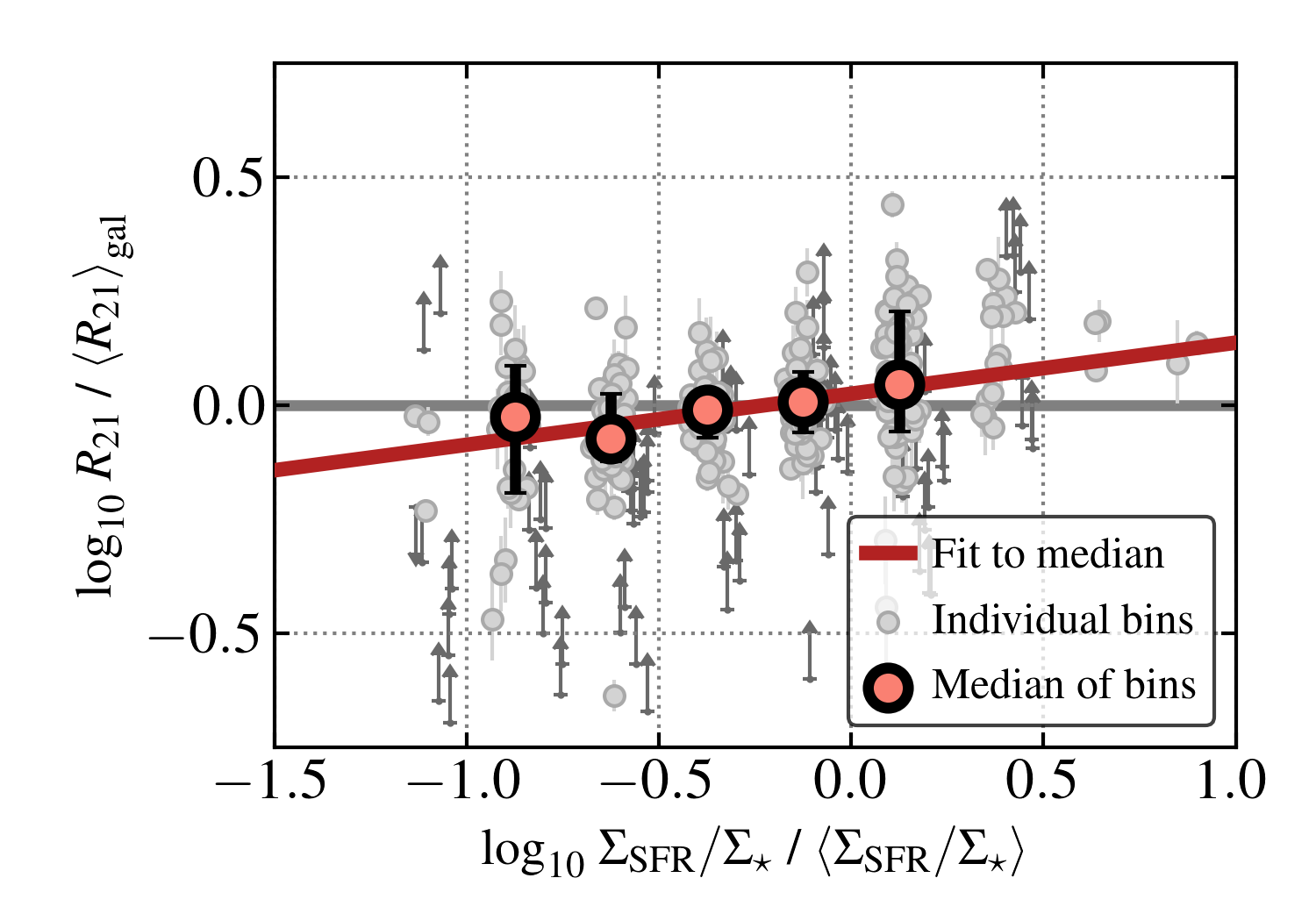} \\
\includegraphics[width=0.4\textwidth]{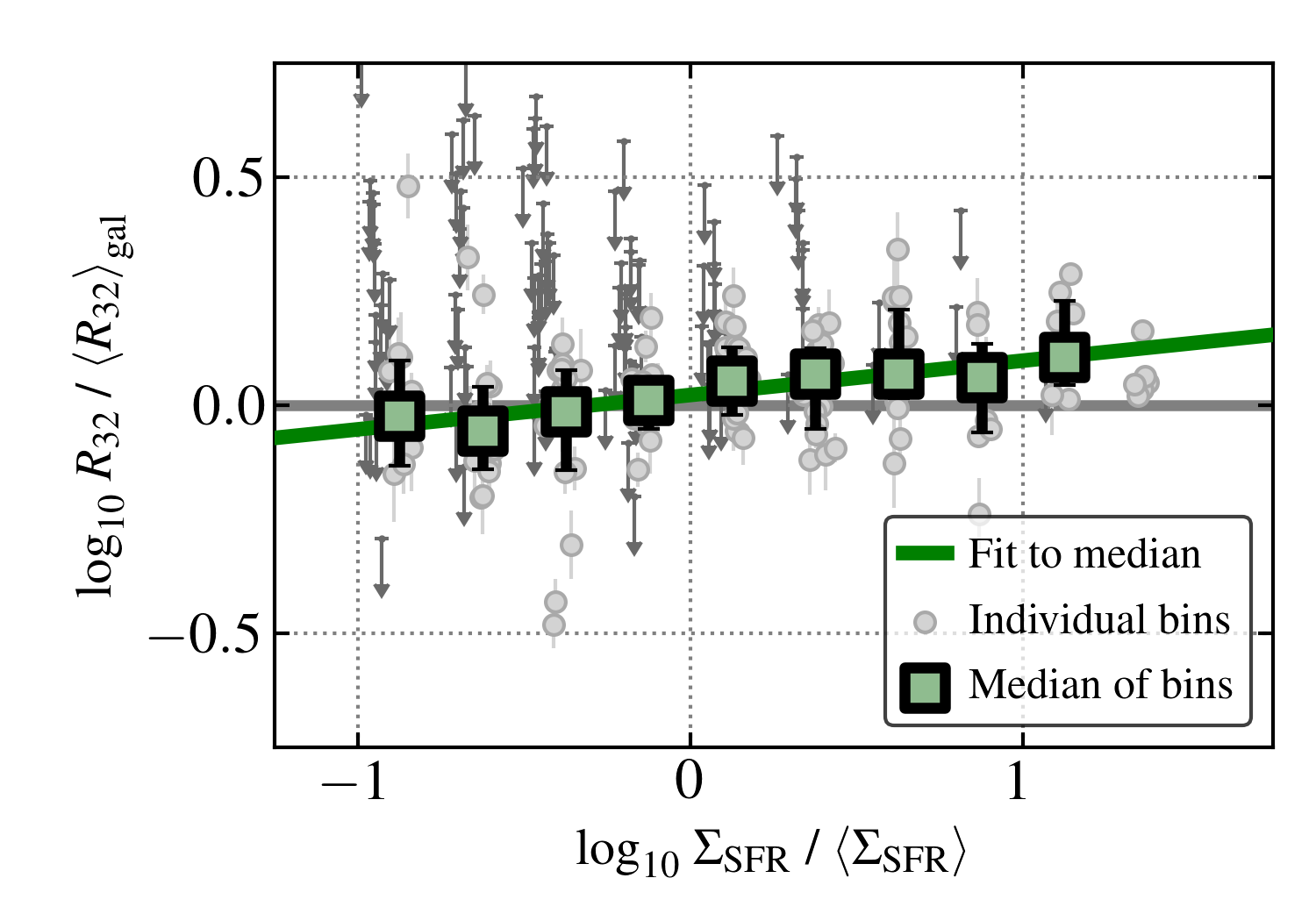}
\includegraphics[width=0.4\textwidth]{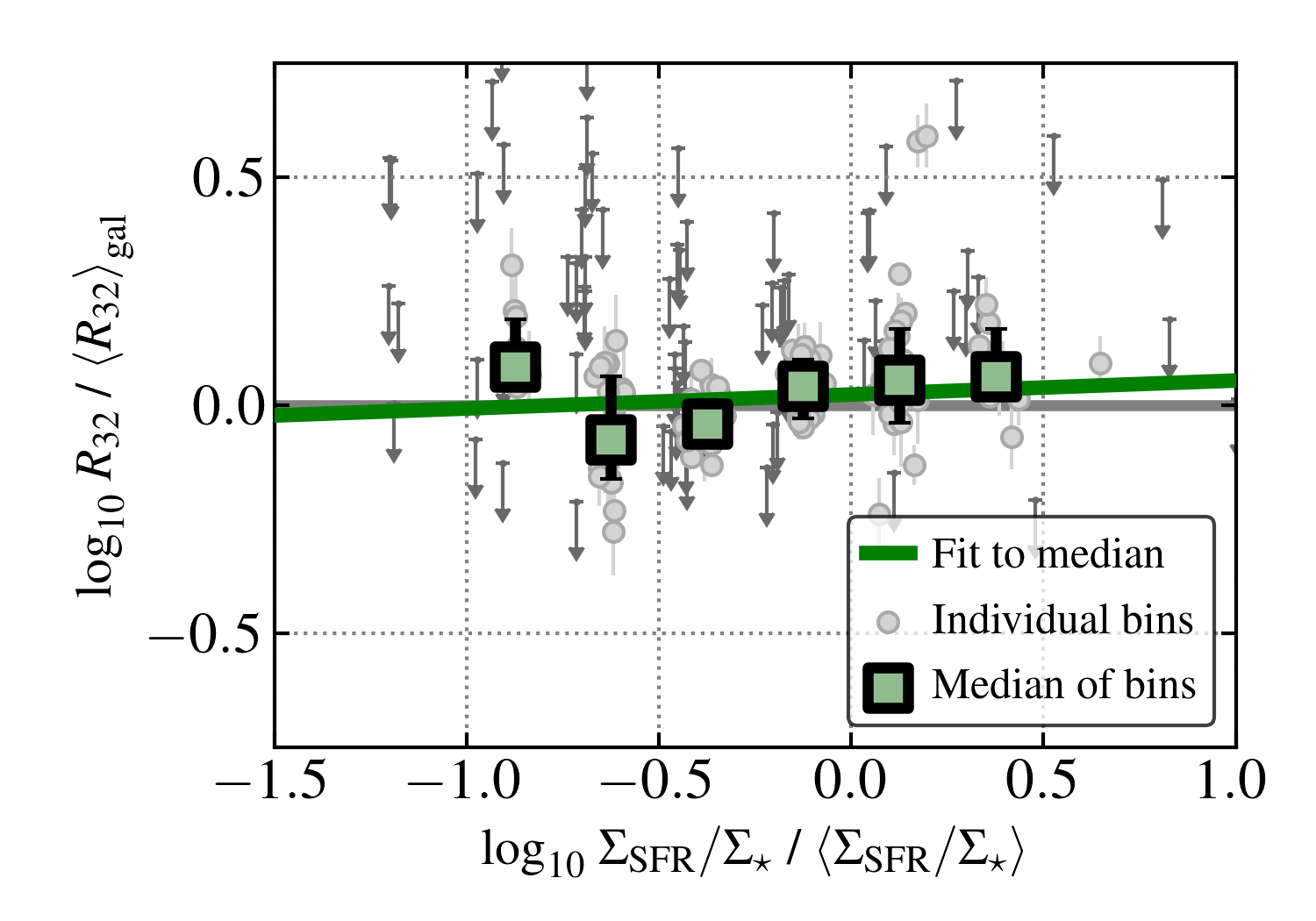} \\
\includegraphics[width=0.4\textwidth]{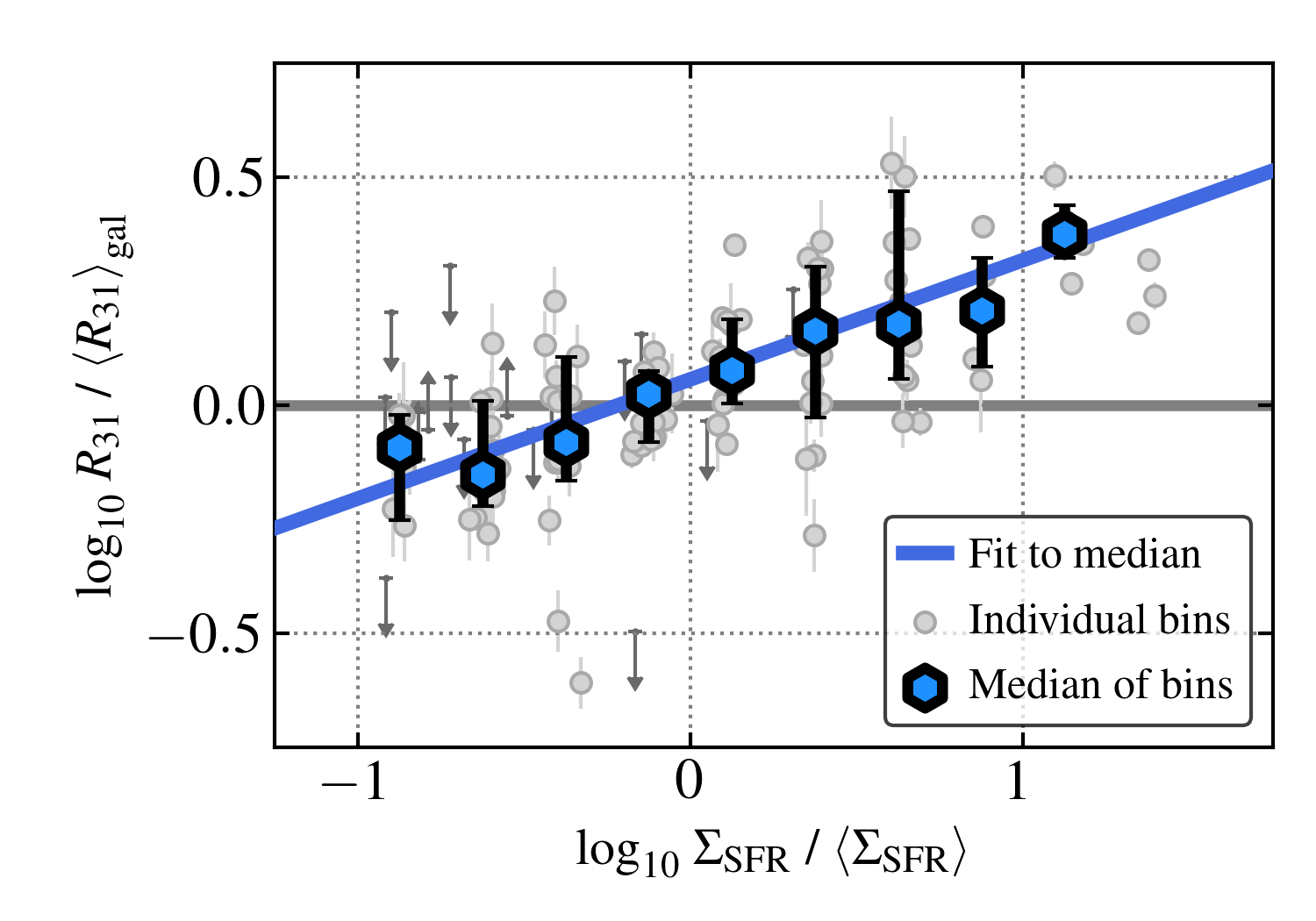}
\includegraphics[width=0.4\textwidth]{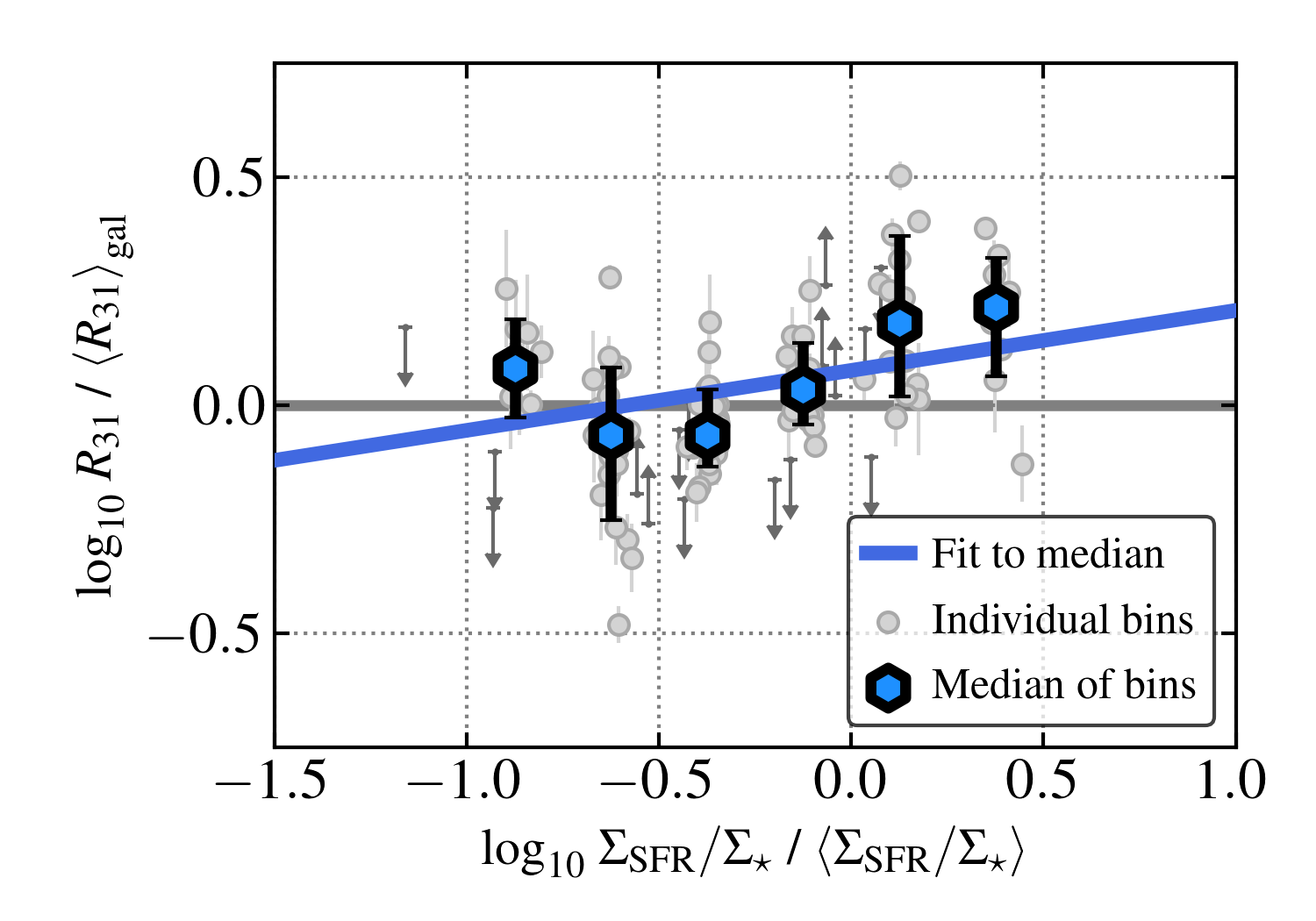}
\end{center}
\caption{
\textit{Correlations between line ratios and star formation activity.} Each plot shows line ratio enhancement or deficit relative to the galaxy average, $\log_{10} R / \!\langle R \rangle$, as a function of local star formation rate surface density and specific star formation rate. Before binning we normalize $\Sigma_{\rm SFR}$ and $\Sigma_{\rm SFR}/\Sigma_\star$ by their galaxy averages, so that the plots show how enhancements in the line ratio relative to the galaxy average correlate with enhancements in \SFR\ or $\SFR / M_\star$ relative to the galaxy average. As in Figure~\ref{fig:resolved_rad}, gray points show results for individual map pairs calculated as described in \S\ref{sec:meas} and arrows show $4\sigma$ upper or lower limits. Colored points indicate the median result for all map pairs with a measurement in that bin, including limits, and so distil the general trend for the whole data set. The black vertical error bars show the $16{-}84^{\rm th}$ percentile of individual measurements. The solid colored line shows a least squares fit to all median values with at least $5$ data points  (see Table~\ref{tab:resolved}). \label{fig:resolved_sfr}}
\end{figure*}

\begin{deluxetable}{lcccc}
\tabletypesize{\small}
\tablecaption{Correlations between line ratio variations and local conditions}
\label{tab:resolved}
\tablewidth{0pt}
\tablehead{
\colhead{Ratio} & 
\colhead{Quantity} & 
\colhead{$a$} & 
\colhead{$b$} & 
\colhead{$\rho$\tablenotemark{a}} 
}
\startdata
$R_{21}/\langle R_{21} \rangle$ & $r_{\rm gal}$ [kpc] & $-0.011$ & $0.038$ & $-0.37$ \\ 
$R_{21}/\langle R_{21} \rangle$ & $r_{\rm gal}$ --- no center\tablenotemark{b} & $-0.003$ & $-0.003$ & \nodata \\ 
$R_{32}/\langle R_{32} \rangle$ & $r_{\rm gal}$ [kpc] & $-0.009$ & $0.051$ & $-0.35$ \\ 
$R_{32}/\langle R_{32} \rangle$ & $r_{\rm gal}$ --- no center\tablenotemark{b} & $-0.007$ & $0.043$ & \nodata \\ 
$R_{31}/\langle R_{31} \rangle$ & $r_{\rm gal}$ [kpc] & $-0.028$ & $0.125$ & $-0.48$ \\ 
$R_{31}/\langle R_{31} \rangle$ & $r_{\rm gal}$ --- no center\tablenotemark{b} & $-0.019$ & $0.077$ & \nodata \\ 
\hline
$R_{21}/\langle R_{21} \rangle$ & $r_{\rm gal}/r_{\rm eff}$ & $-0.043$ & $0.045$ & $-0.48$ \\
$R_{21}/\langle R_{21} \rangle$ & $r_{\rm gal}/r_{\rm eff}$ --- no center\tablenotemark{b} & $-0.021$ & $0.006$ & \nodata \\
$R_{32}/\langle R_{32} \rangle$ & $r_{\rm gal}/r_{\rm eff}$ & $-0.090$ & $0.097$ & $-0.51$ \\
$R_{32}/\langle R_{32} \rangle$ & $r_{\rm gal}/r_{\rm eff}$ --- no center\tablenotemark{b} & $-0.099$ & $0.111$ & \nodata \\
$R_{31}/\langle R_{31} \rangle$ & $r_{\rm gal}/r_{\rm eff}$ & $-0.162$ & $0.177$ & $-0.60$ \\
$R_{32}/\langle R_{31} \rangle$ & $r_{\rm gal}/r_{\rm eff}$ --- no center\tablenotemark{b} & $-0.131$ & $0.135$ & \nodata \\
\hline
$R_{21}/\langle R_{21} \rangle$ & $\Sigma_{\rm SFR} / \langle \Sigma_{\rm SFR} \rangle $  & $0.129$ & $0.019$ & $0.55$ \\
$R_{32}/\langle R_{32} \rangle$ & $\Sigma_{\rm SFR} / \langle \Sigma_{\rm SFR} \rangle $ & $0.075$ & $0.023$ & $0.38$ \\
$R_{31}/\langle R_{31} \rangle$ & $\Sigma_{\rm SFR} / \langle \Sigma_{\rm SFR} \rangle $ & $0.261$ & $0.057$ & $0.63$ \\
\hline
$R_{21}/\langle R_{21} \rangle$ & $\Sigma_{\rm SFR} / \Sigma_\star / \langle \Sigma_{\rm SFR} / \Sigma_\star \rangle$ & $0.112$ & $0.025$ & $0.43$ \\
$R_{32}/\langle R_{32} \rangle$ & $\Sigma_{\rm SFR} / \Sigma_\star / \langle \Sigma_{\rm SFR} / \Sigma_\star \rangle$ & $0.031$ & $0.024$ & $0.27$ \\
$R_{31}/\langle R_{31} \rangle$ & $\Sigma_{\rm SFR} / \Sigma_\star / \langle \Sigma_{\rm SFR} / \Sigma_\star \rangle$ & $0.131$ & $0.077$ & $0.35$ \\
\enddata
\tablenotemark{a}{The Spearman rank coefficient only for bins with both lines detected at $\mathrm{S/N} > 4$.}
\tablenotetext{b}{These fits exclude the innermost radial bin.}
\tablecomments{The table reports correlations between each line ratio, normalized to the galaxy value, and various local conditions. These are calculated binning the data within individual galaxies to increase the $\mathrm{S/N}$ as described in \S\ref{sec:meas}. The columns $a$ and $b$ report the slope and intercept for a linear fit of the form in Equation~\eqref{eq:localfit} carried out on the median binned data. The column $\rho$ reports the Spearman rank correlation coefficient relating enhancements in the line ratio to the local quantity considering the individual bins (gray points in Figures~\ref{fig:resolved_rad} and~\ref{fig:resolved_sfr}).}
\end{deluxetable}

\begin{deluxetable}{lc}
\tabletypesize{\small}
\tablecaption{Central enhancements in line ratios \label{tab:central}}
\tablewidth{0pt}
\tablehead{
\colhead{Ratio} & 
\colhead{Central enhancement} \\
\colhead{} & 
\colhead{[dex]}
}
\startdata
$R_{21}/\langle R_{21} \rangle$  & $0.18$~($0.04~{-}~0.27$) \\
$R_{32}/\langle R_{32} \rangle$ & $0.08$~($-0.01~{-}~0.19$) \\
$R_{31}/\langle R_{31} \rangle$ & $0.27$~($0.14~{-}~0.42$) \\
\enddata
\tablecomments{Enhancement in line ratio at galaxy center relative to the galaxy-averaged value. The table reports the median value of $\log_{10} R / \langle R \rangle$ calculated in a $1$~kpc sized bin centered at $r_{\rm gal} = 0$~kpc (see Figure~\ref{fig:resolved_rad}). The quoted range gives the $16{-}84^{\rm th}$ range of measurements in that bin.}
\end{deluxetable}

As described in \S\ref{sec:measres}, we also separate galaxies into individual regions to measure how changes in local conditions inside a galaxy relate to variations in the line ratios. Figures~\ref{fig:resolved_rad} and~\ref{fig:resolved_sfr} and Tables~\ref{tab:intrats} and~\ref{tab:central} summarize the results of these measurements.

We conduct this analysis by binning the data within each galaxy according to several properties of interest: galactocentric radius, star formation rate surface density ($\Sigma_{\rm SFR}$), and local specific star formation rate ($\Sigma_{\rm SFR}/\Sigma_\star$). The binning allows us to improve the signal-to-noise of individual measurements. Despite the averaging, many bins still lack a detection, and Figures~\ref{fig:resolved_rad} and~\ref{fig:resolved_sfr} also show upper and lower limits. Here a detection has $\mathrm{S/N}>4$ in both lines, while a limit has $\mathrm{S/N} > 4$ in only one line. We do not analyze bins with neither line detected.

We focus this analysis on the \textit{relative} behavior of the line ratio and these local quantities. In order to do this, we normalize all line ratio measurements to their average value for the galaxy. We perform a similar normalization of $\Sigma_{\rm SFR}$ and $\Sigma_{\rm SFR}/\Sigma_\star$, normalizing the measurements for each individual galaxy by the galaxy-averaged values, $\langle \Sigma_{\rm SFR} \rangle$ and $\langle \SFR / M_\star \rangle$. We present the trends with galactocentric radius in both physical units of kpc and in radius normalized by the half mass radius, $r_{\rm eff}$. This normalization means that our analysis mostly controls for calibration uncertainties and galaxy-to-galaxy scatter. More details, including the exact bin definitions, can be found in \S\ref{sec:measres}.

\medskip

\textit{CO excitation and galactocentric radius:} All three line ratios show significant anticorrelation with radius and show enhanced values relative to the galaxy mean in the inner parts of galaxies. This has been seen before for samples of galaxies by \citet{LEROY09}, \citet{WILSON09}, \citet{LEROY13}, \citet{DENBROK21}, and \citet{YAJIMA21}, as well as for many individual galaxies, though note that only \citet{WILSON09} used \cothree\ among these studies. 

In Figure~\ref{fig:resolved_rad}, we see that radial gradients in all three line ratios appear to be a general feature. This is true whether we express the galactocentric radius in physical units (the left panel) or normalize to the effective radius (the right panel). To characterize the gradient, we fit a linear function  of the form
\begin{equation}
\label{eq:localfit}
\log_{10} \frac{R}{\langle R \rangle} = a \times Q + b~,
\end{equation}
\noindent where $R / \!\langle R \rangle$ is the line ratio relative to the galaxy mean and here $Q$ is $r_{\rm gal}$, either expressed in kpc or normalized to $r_{\rm eff}$, the half-light radius. We conduct a simple $\chi^2$ minimization to all of the median values (colored points in Figure~\ref{fig:resolved_rad}) in each bin. We calculate these medians accounting for lower limits in $R_{21}$ and upper limits in $R_{32}$ and $R_{31}$, and use bins that have at least five non-limit measurements and for which the 16\%, 50\%, and 84\% value is not a limit. We report $a$ and $b$ for the fits for each line in Table~\ref{tab:resolved}.

We find gradients, given by the $a$ values in Table~\ref{tab:resolved}, of $-0.011$, $-0.009$, and $-0.028$~dex per kpc or $-0.043$, $-0.090$, and $-0.162$~dex per $r_{\rm eff}$. This implies, on average, a mild but significant change in each of the line ratios across the disk. The gradients appear weaker for $R_{21}$ than for $R_{32}$ or $R_{31}$, but they are statistically significant in all cases based on the $p$-value associated with the Spearman rank correlation coefficient reported in the table.

\medskip

\textit{Central enhancements:} For $R_{21}$ and $R_{31}$ in Figure~\ref{fig:resolved_rad}, much of the apparent gradient is driven by high $R$ in the innermost parts of galaxies. In Table~\ref{tab:central}, we quantify the central enhancement in each line ratio relative to the disk-averaged value. We report the median and $16{-}84^{\rm th}$ percentile value in the innermost $1$~kpc-wide bin. We measure a median enhancement of $0.18$~dex for $R_{21}$, $0.27$~dex for $R_{31}$, and a weaker median enhancement of $0.08$~dex for $R_{32}$.

Recall that the samples are not matched among the three lines, but the qualitative point seems clear: both $R_{21}$ and $R_{31}$ appear enhanced in the inner parts of normal star-forming galaxies, and $R_{32}$ appears at least mildly enhanced. In addition to agreeing with previous measurements, this trend makes physical sense given that both gas density and star formation activity tend to rise towards the inner parts of galaxies \citep[e.g.,][for studies showing this via HCN observations to make this point]{USERO15,GALLAGHER18A,JIMENEZDONAIRE19}. Recall that any galaxy without central gas detected at $\mathrm{S/N}>4$ in both lines will not appear in this analysis.

Are the centers special or simply the endpoints of a steady radial gradient? To test this, Table~\ref{tab:resolved} also includes fits that exclude the central, $r_{\rm gal} = 0$, bin. Other than excluding these $r_{\rm gal} = 0$ data, the fits are identical. These ``no center'' fits still all show negative gradients. In other words, all of the line ratios decline with galactocentric radius outside of the galaxy center, and the radial trends do not appear to be purely products of central enhancements. However, for $R_{21}$ and $R_{31}$ the radial gradients excluding the center appear much weaker than those including the center. In fact, the $R_{21}$ trends appear almost flat outside the inner two data points. This demonstrates that much of the overall decline in the ratio with radius is driven by the behavior of the inner galaxy, rather than a smooth gradient. This also agrees with the visual impression from Figure~\ref{fig:resolved_rad}. In short, bright galaxy centers do appear special, showing evidence of enhanced excitation, and there also appear to be weak radial gradients outside just the central region.

\medskip

\textit{CO excitation and star formation activity:} Figure~\ref{fig:resolved_sfr} shows two closely related trends. We plot each line ratio as a function of normalized $\Sigma_{\rm SFR}$ and $\Sigma_{\rm SFR}/\Sigma_\star$. Both variables trace the local star formation activity, which should relate directly to the heating of the gas and indirectly to the gas density. Again we report rank correlation coefficients and $\chi^2$ minimization fits to the median values in Table~\ref{fig:resolved_sfr} (now $Q = \log_{10} \Sigma_{\rm SFR} / \!\langle \Sigma_{\rm SFR} \rangle$ or $\log_{10} \Sigma_{\rm SFR}/\Sigma_\star \, / \langle \Sigma_{\rm SFR}/\Sigma_\star \rangle$).

The left column of Figure~\ref{fig:resolved_sfr} shows more or less continuous trends relating enhancement in each line ratio to enhancement in $\Sigma_{\rm SFR}$. The figure does give some hint of ``flattening'' in $R_{21}$ and $R_{31}$ at least at low $\Sigma_{\rm SFR} / \!\langle \Sigma_{\rm SFR} \rangle$. This might relate to the structure seen in the radial profiles, in which the central enhancements appear strong while the ratio, at least for $R_{21}$ appears flatter outside the center. More high signal-to-noise observations of the extended, low $\Sigma_{\rm SFR}$, parts of galaxy disks are needed to pursue this further.

Table~\ref{tab:resolved} reports the gradient in each line ratio per dex change in $\Sigma_{\rm SFR} / \!\langle \Sigma_{\rm SFR}\rangle$. For $R_{21}$ this is a ${\sim} 0.13$~dex change in $R_{21} / \!\langle R_{21} \rangle$ per dex change in $\Sigma_{\rm SFR} / \!\langle \Sigma_{\rm SFR} \rangle$. This agrees well with the slope relating $R_{21}$ to total infrared surface brightness, $\Sigma_{\rm TIR}$, measured by \citet{DENBROK21} for $9$ EMPIRE targets.

The line ratios all appear at least moderately enhanced in regions with higher star formation activity, and again these trends make physical sense. More intense activity correlates with higher density gas \citep[e.g.,][]{USERO15,GALLAGHER18A,JIMENEZDONAIRE19}, stronger radiation fields, and higher cosmic ray densities, all leading to warmer gas. The results here show that after accounting for galaxy-to-galaxy variations, internal correlations between star formation activity and galactocentric radius are evident and widespread inside nearby, normal, star-forming galaxies.

The right column of Figure~\ref{fig:resolved_sfr} shows how changes in $R$ relate to variations in the specific star formation rate, $\Sigma_{\rm SFR}/\Sigma_\star$. In principle, this quantity has appealing properties as a predictor of line ratio variations: its behavior should be more scale-independent than $\Sigma_{\rm SFR}$ and the normalization by $\Sigma_\star$ removes some overall scaling effects to isolate the intensity of star formation. 

Figure~\ref{fig:resolved_sfr} and Table~\ref{tab:resolved} show that we do observe significant correlations with the expected sense between $R_{21}$ and $\Sigma_{\rm SFR}/\Sigma_\star$ and $R_{31}$ and $\Sigma_{\rm SFR}/\Sigma_\star$, but $R_{32}$ does not show a formally significant correlation. The correlations between $R_{21}$ and $R_{32}$ and $\Sigma_{\rm SFR}/\Sigma_\star$ are also weaker than those with $\Sigma_{\rm SFR}$ (see Table~\ref{tab:resolved}), which is reflected in the noisier appearance of the correlations in Figure~\ref{fig:resolved_sfr}, especially at extreme values.

Contrasting the two columns suggests that $\Sigma_{\rm SFR}$ may be a somewhat better predictor of line ratio variations within normal galaxies than $\Sigma_{\rm SFR}/\Sigma_\star$. This could make physical sense if gas density and the overall gas reservoir play more important roles than the strength of the interstellar radiation field and the intensity of star formation. Alternatively, it may simply reflect that $\Sigma_{\rm SFR}$ shows a larger dynamic range and more regular structure within our target galaxies than $\Sigma_{\rm SFR}/\Sigma_\star$.

\medskip

\textit{Overall:} Our resolved analysis offers a consistent first order picture. The inner parts of galaxies often host the most intense star formation activity and these regions appear enhanced in all line ratios in all plots. These enhancements also correlate with increased $\Sigma_{\rm SFR}$ and, perhaps to a lesser extent, higher $\Sigma_{\rm SFR}/\Sigma_\star$. The trends become weaker outside the inner parts of galaxies. Tables~\ref{tab:resolved} and~\ref{tab:central} give quantitative estimates of gradients and central enhancements.

\medskip

\textit{A note on correlations with $\Sigma_{\rm mol}$ and $\SFR / CO$:} Note that we deliberately avoid stacking by \SFR/CO because at the modest signal-to-noise present in the individual pixels, the effect of the correlated axes becomes overwhelming \citep[see][for more discussion]{DENBROK21}. We did explore this direction. As expected based on previous results, the resolved trends relating $R/\!\langle R \rangle$ to $\Sigma_{\rm SFR}/\Sigma_{\rm mol}$ or to $\Sigma_{\rm mol}$ appear highly significant \citep[see also][]{YAJIMA21}. However, as we also saw in the previous section, interpreting these trends is difficult due to the correlated axes and in these cases we lack an independent, high S/N quantity to stack the data (i.e., we would have to use the CO itself as the $x$-axis) or stack by a third quantity (e.g., radius) and carefully handle upper limits in the stacks. For now, we note only that these correlations appear significant but driven by correlated axes and that this will be a crucial trend to carefully analyze in data with higher signal-to-noise, excellent inter-line calibration, and well-understood uncertainties.

\section{Discussion}
\label{sec:discussion}

\begin{figure*}[ht!]
\begin{center}
\includegraphics[width=0.4\textwidth]{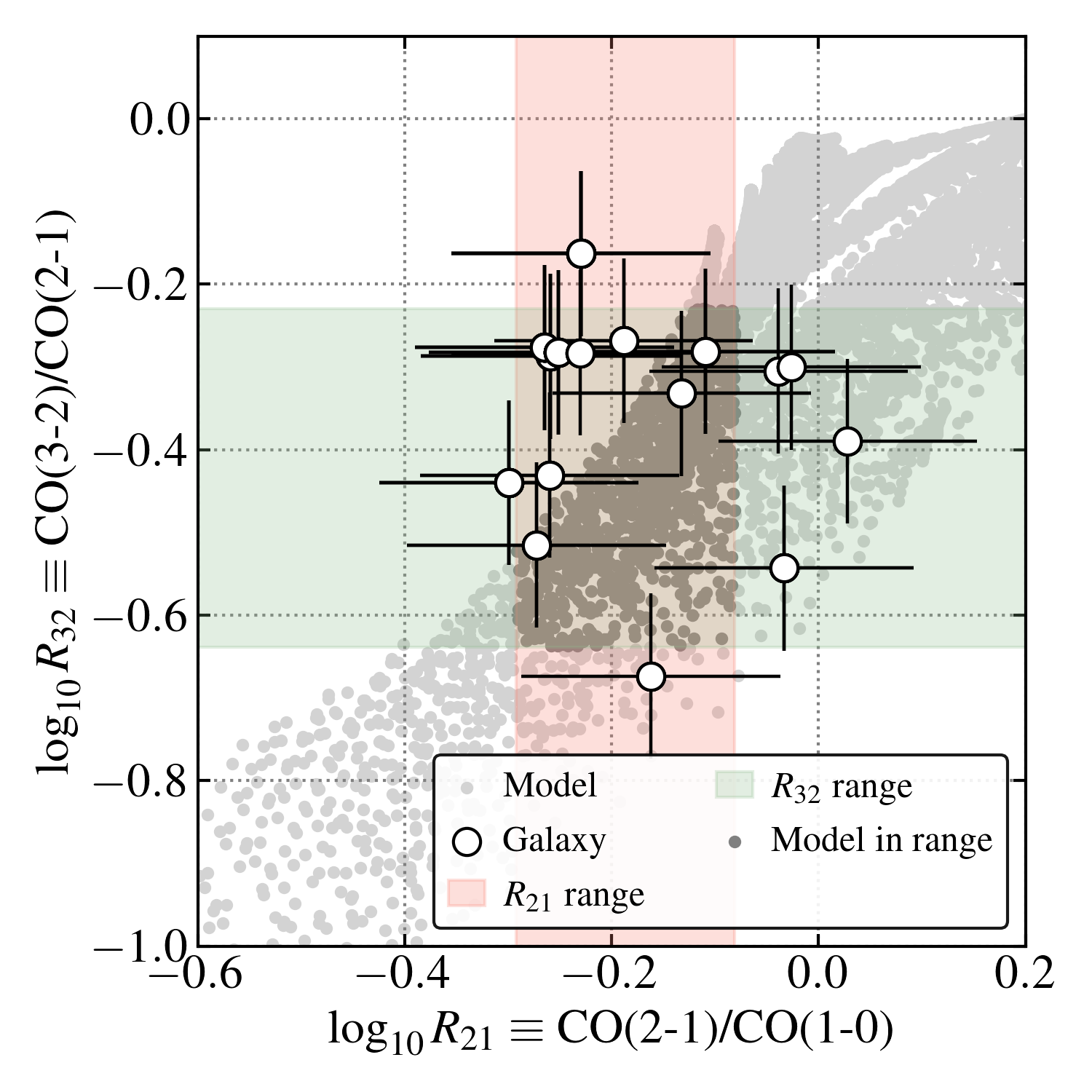}
\includegraphics[width=0.4\textwidth]{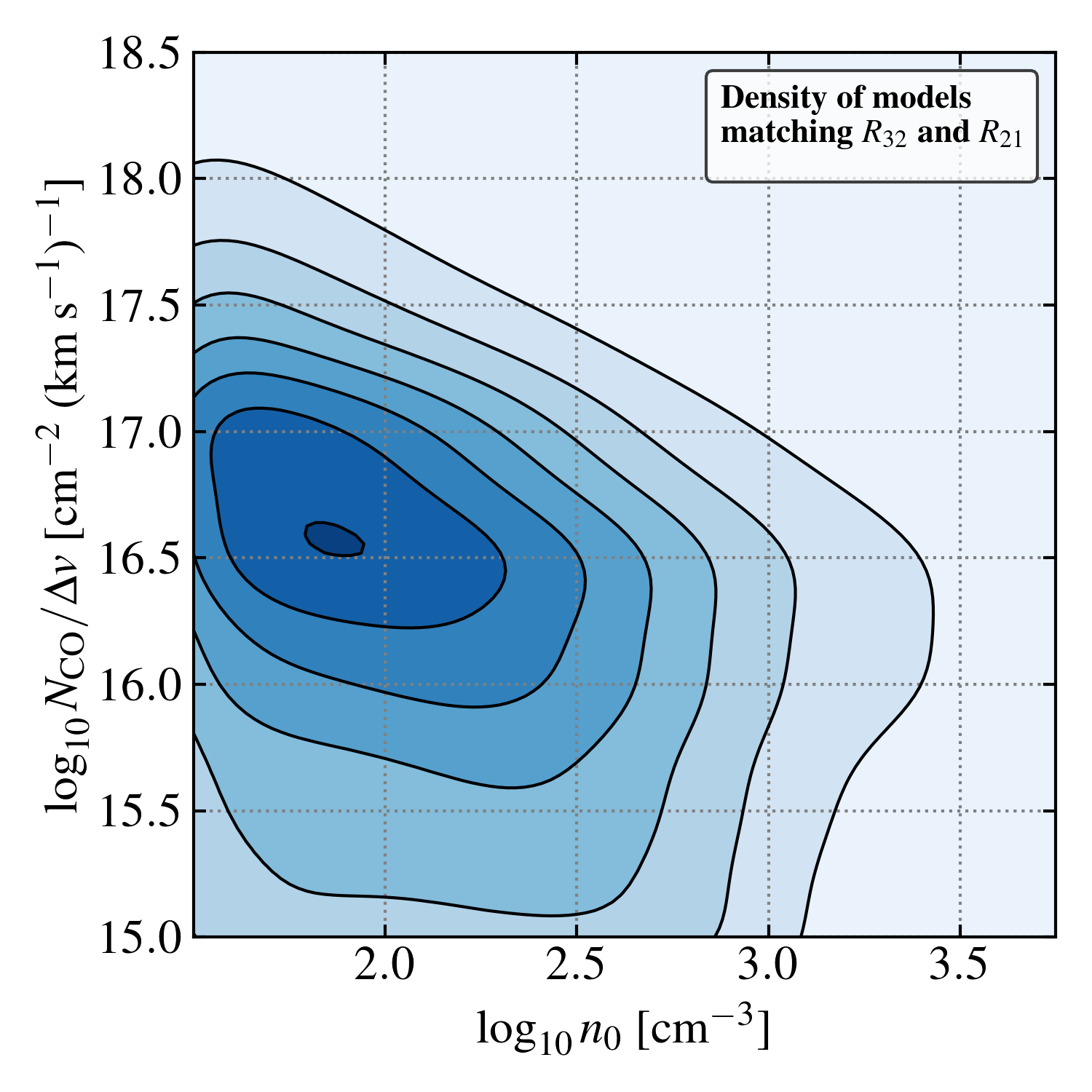} \\
\includegraphics[width=0.4\textwidth]{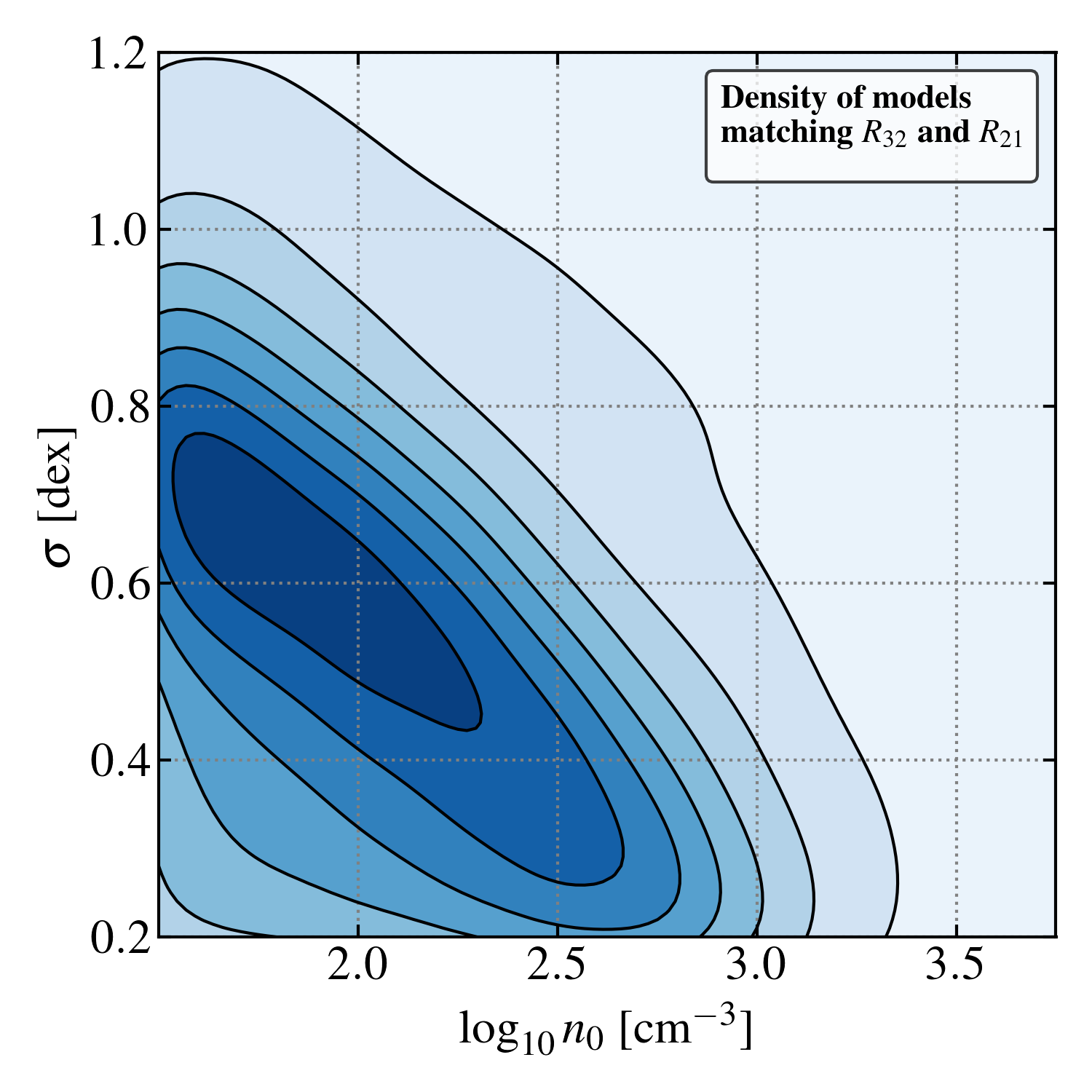}
\includegraphics[width=0.4\textwidth]{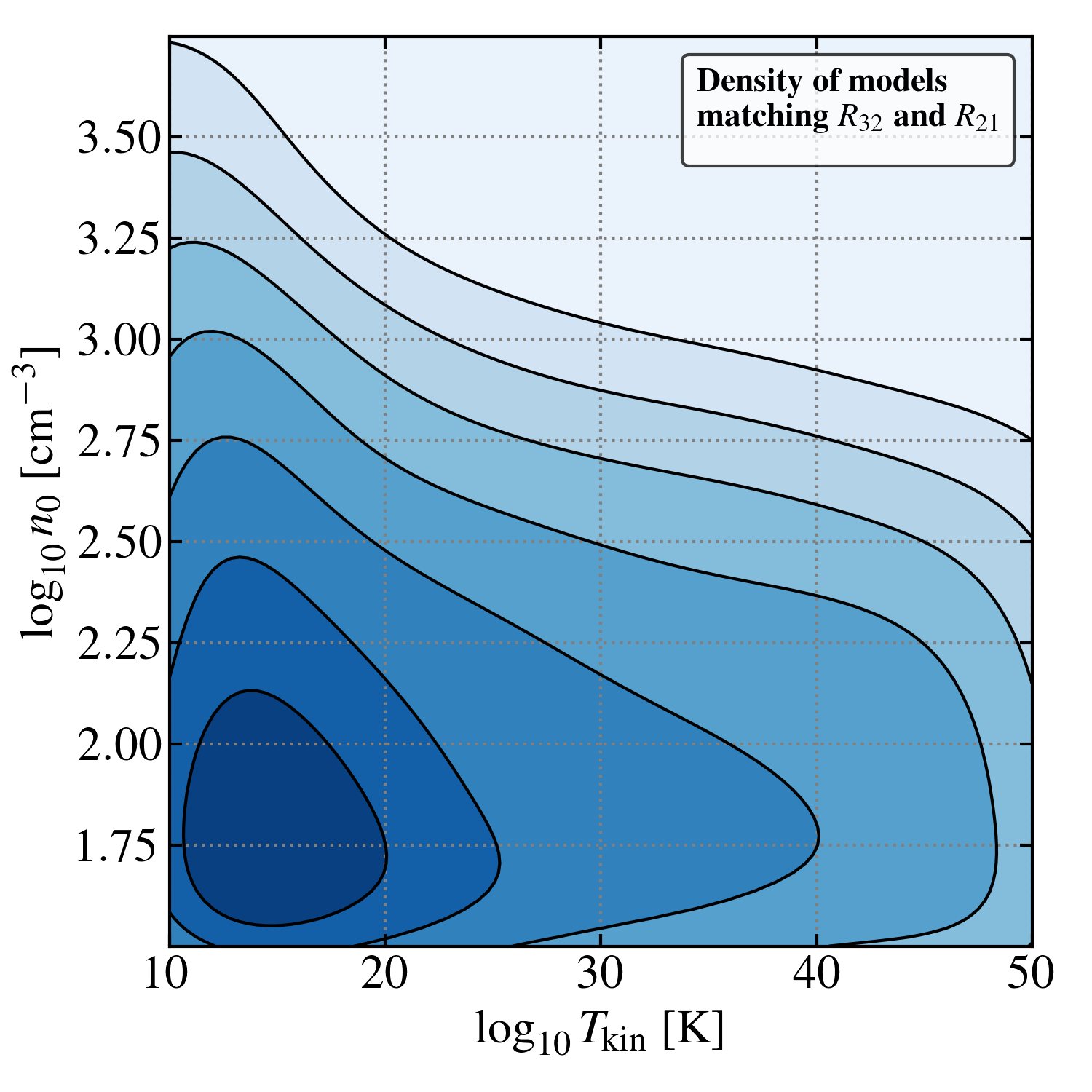} \\
\end{center}
\caption{
\textit{Comparison of measurements and model calculations.} The \textit{top left} panel shows the same model grid visualized in Figure~\ref{fig:model}, though now with logarithmic axes. Individual points show $L_{\rm CO}$-based estimates for $R_{21}$ and $R_{32}$ for galaxies with all three lines measured. Note that we use the $L_{\rm CO}$ values and now show only one point per galaxy to homogenize the area considered and use the best available data. Shaded regions show the $16^{\rm th}$ to $84^{\rm th}$ percentile range for each ratio from Table~\ref{tab:intrats}. Darker points highlight entries in the model grid that lie within the measured $16^{\rm th}$ to $84^{\rm th}$ percentile for both ratios. In the remainder of the panels, we visualize the properties of the model grid entries that lie in the overlap of average $R_{32}$ and $R_{21}$ constraints. The \textit{top right} panel shows the distribution of viable models in $N_{\rm CO}/\Delta v {-} n_0$ space, the \textit{lower left} shows $\sigma {-} n_0$ space, and the \textit{bottom right} shows $n_0 {-} T_{\rm kin}$ space. The line ratios can be most readily produced by cold, moderate opacity (inferred from the column per line width), moderate density gas (from $n_0$ and $\sigma$).
\label{fig:modelcomp}}
\end{figure*}

\subsection{Comparison to simple models assuming lognormal density distributions} \label{sec:modelcomp}

What do these measured ratios imply for physical conditions in the cold gas? Figure~\ref{fig:modelcomp} compares our measurements to the models described in \S\ref{sec:expect} and Figure~\ref{fig:model}. In the upper left panel, we replot the models from Figure~\ref{fig:model}, though now using a logarithmic grid. In that same panel, we also indicate the $16{-}84^{\rm th}$ percentile range of \textit{all} measured $R_{21}$ and $R_{32}$ (i.e., the values from Figure~\ref{fig:hist} and Table~\ref{tab:intrats}) as shaded, colored regions, and we plot points for all $16$ galaxies that have measured \coone , \cotwo , \textit{and} \cothree\ luminosities. Note that here we plot only the ratios implied by the luminosities in Table~\ref{tab:meas}, because this allows us to ensure matched area among all measurements.

The other three panels of Figure~\ref{fig:modelcomp} visualize the set of conditions in the model grids that produce both $R_{21}$ and $R_{32}$ in the measured $16{-}84^{\rm th}$ percentile range. That is, these are the model grid entries that lie inside the square overlap region in the top left panel and appear as darker shaded points in that panel. The most common conditions in the model grid appear comparable to conditions expected in the cold ISM. We find $\log_{10} (N_{\rm CO}/\Delta v) \, [{\rm cm}^{-2}~({\rm km~s}^{-1})^{-1}] \approx 16.5{-}17$. For a standard CO abundance $N({\rm CO})/N({\rm H_2}) \sim 10^{-4}$ \citep[e.g.,][]{VANDISHOECK88} and a typical molecular cloud line full width of ${\sim}10$~km~s$^{-1}$ \citep[e.g.,][]{SUN18,SUN20B,ROSOLOWSKY21}, this $N_{\rm CO}/\Delta v$ implies a total $\log_{10} N({\rm H_2}) \, [{\rm cm}^{-2}] \approx 21.5{-}22$, in reasonable agreement with column densities for molecular clouds or resolved surface densities in maps of cold gas in galaxies.

The models that produce the observed line ratios generally have low temperatures (see bottom right panel in Figure~\ref{fig:modelcomp}), preferring $T_{\rm kin} < 20$~K and frequently $T_{\rm kin} = 10{-}15$~K, the lowest temperatures that we modeled.

Meanwhile, the bottom left panel in Figure~\ref{fig:modelcomp} shows the mean, $n_0$, and width, $\sigma$, of the lognormal distribution of densities used in the models. Each individual model combines an ensemble of single density models in a way that mimics a lognormal distribution of densities. Because a wider distribution with higher $\sigma$ includes higher density gas, it can produce higher excitation and higher line ratios \citep[e.g., see][]{LEROY17B}. As a result, the mean density, $n_0$, and the width, $\sigma$, somewhat trade off.\footnote{The mean $\log n$ depends on both $n_0$ and $\sigma^2$. Following \citet{PADOAN02} for a lognormal distribution of densities $\ln n = \ln n_0 + \sigma^2/2$.} The overall sense of the panel is that the viable models tend to include gas with densities in the range of $n_{\rm H_2} \sim 300{-}1000$~cm$^{-3}$. We expect that the addition of a power law tail of densities \citep[e.g., as expected for self-gravitating gas, see][]{KRUMHOLZ07B,FEDERRATH13,BURKHART18} could affect the \cothree\ emission, but would likely yield mostly similar mean densities.

In short, our observed line ratios can broadly be produced by cold, intermediate density, intermediate column density gas. Though our models include density distributions and so exhibit a range of optical depth, $\tau$, we find it useful to note the implied optical depth for a single zone model with the same average properties that we consider. A one zone RADEX model with $T_{\rm kin}=15$~K, $n_{\rm H_2}=1000$~cm$^{-3}$, and $N_{\rm CO}/\Delta v = 3 \times 10^{16}$ cm$^{-2}$~(km~s$^{-1}$)$^{-1}$ yields $\tau \sim 4$ for \coone , $\tau \sim 9$ for \cotwo , and $\tau \sim 5$ for \cothree\ (suggesting heavy excitation to the $J=2$ state). For \coone , these are also in good agreement with the opacities implied by contrasting $^{13}\coone$ and $^{12}\coone$ measurements \citep[e.g., see summary in][]{ROMANDUVAL10,CAO17,CORMIER18}.

Though these conditions appear reasonable, we note that not all of our measurements can be readily explained by the models that we consider. This appears entirely reasonable, given that the simple models we use are most appropriate for individual molecular clouds or parts of galaxies. In the top left panel of Figure~\ref{fig:modelcomp}, we see observations with higher $R_{32}$ at low $R_{21}$ than the model readily produces, i.e., up and to the left of the model points, though note that most of the data still lie within ${\sim}1\sigma$ of the model grid.

More sophisticated models may be able to explain the observed line ratios by mixing different models, considering, e.g., multi-modal density distributions or distributions of $T_{\rm kin}$ or column density. This would naturally reflect the blending of clouds and regions in different physical states that we expect to be averaged together in our whole-galaxy measurements.

Alternatively, abundance variations, the impact of cosmic rays, geometry, and coupling between zones can all broaden the range of parameter space covered by the models \citep[e.g.,][]{BEMIS19,BISBAS19}. Comparisons to numerical models of galaxy disks that include chemistry and radiation transfer \citep[e.g.,][]{TRESS20,GONG20} will help illuminate line ratios that we might expect from a realistic blend of conditions.

\subsection{Line ratios, the CO-to-\texorpdfstring{H\textsubscript{2}}{H2} conversion factor, and scaling relations} 
\label{sec:alphaco}

\begin{figure}[ht!]
\begin{center}
    \includegraphics[width=0.45\textwidth]{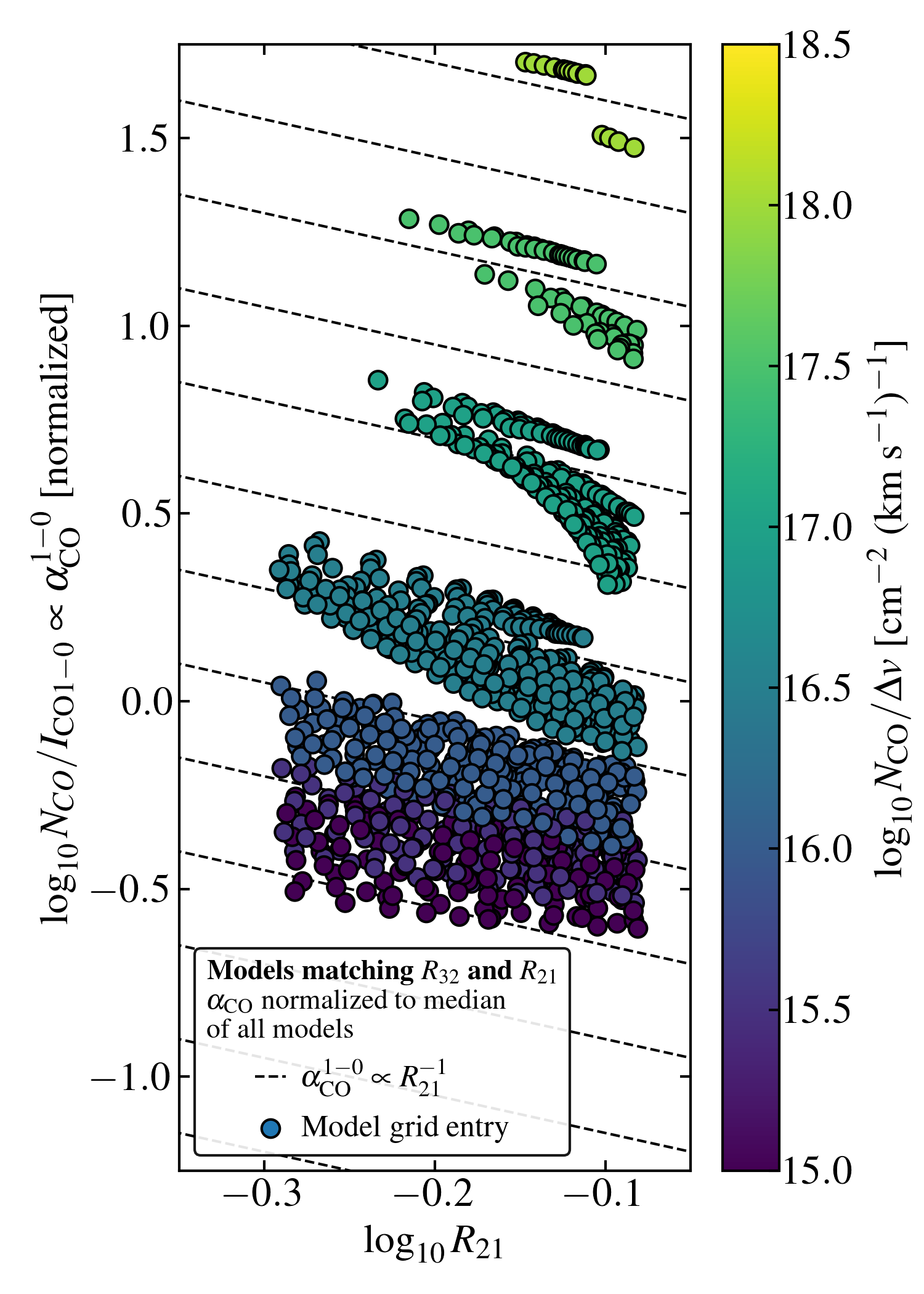}
\end{center}
\caption{
\textit{$\alpha_{\rm CO}^{1-0}$ tracks line ratio variations.} Inverse of \coone\ emissivity in the model grid, $N_{\rm CO}/I_{\coone}$, as a function of $\log_{10} R_{21}$. Points show model grid entries that satisfy the mean $R_{32}$ and $R_{21}$ constraints in Figure~\ref{fig:modelcomp}, colored by the column density per line width, $N_{\rm CO}/\Delta v$, of that model. For fixed CO abundance, $N_{\rm CO} / N_{\rm H_2}$, the inverse of the emissivity is proportional to $\alpha_{\rm CO}^{1-0}$, the CO-to-H$_2$ conversion factor for \coone. The offsets among model families with different $N_{\rm CO} / \Delta v$ reflect that the column per line width directly relates to the opacity of the line. The plot shows that for approximately fixed $N_{\rm CO} / \Delta v$, increasing $\log_{10} R_{21}$ corresponds to decreasing $\alpha_{\rm CO}^{1-0}$, with the proportionality roughly linear (the dashed lines) or slightly steeper. As emphasized by \citet{GONG20}, the low-$J$ CO line ratios offer one of our best prospects to empirically map out likely changes in $\alpha_{\rm CO}^{1-0}$.
\label{fig:alphaco10}}
\end{figure}

As discussed by both \citet{DENBROK21} and \citet{YAJIMA21}, an immediate implication of a systematic dependence of the line ratios on local conditions is that the slopes of scaling relations, e.g., measured between $\Sigma_{\rm SFR}$ and $\Sigma_{\rm mol}$, will differ depending on the observed transition. This is in addition to offsets in normalization that reflect $R_{21}$, $R_{32}$, and $R_{31}$.

At the simplest level, our mean line ratios represent factors that can be used to place relations measured using different transitions on the same scale. That is, our median $R_{21}$ and $R_{31}$ from Table~\ref{tab:intrats} can and should be used to renormalize relations derived using \cotwo\ or \cothree\ onto a consistent scale with those derived using \coone . Equivalently, our line ratios can be combined with the standard Milky Way $\alpha_{\rm CO}^{1-0} = 4.35$~\acounits\ to yield:
\begin{equation}\label{eq:acolines}
\begin{split}
\alpha_{\rm CO}^{2-1} &= 6.7~\acounits~, \\
\alpha_{\rm CO}^{3-2} &= 14.0~\acounits~.
\end{split}
\end{equation}

Equation~\eqref{eq:acolines} helps set the normalization of any relation involving $\Sigma_{\rm mol}$ or $M_{\rm mol}$. Beyond this, Table~\ref{tab:resolved} and Figure~\ref{fig:resolved_sfr} suggest that, when considering the slope of any power law-style scaling relation involving $\Sigma_{\rm SFR}$, one should expect differences of ${\sim} 0.1$ between \cothree\ and \cotwo, ${\sim} 0.15$ between \cotwo\ and \coone, and ${\sim} 0.25{-}0.3$ between \cothree\ and \coone. These values come from the slope of the gradient fit between each line ratio and $\Sigma_{\rm SFR}$. The most aggressive application of these trends would be to use them to adjust the power law index of scaling relations measured using different lines to place them on a common scale. In this case, e.g., the slope of the $\Sigma_{\rm SFR}$ vs.\ $\Sigma_{\rm mol}$ relation measured with \cothree\ would be adjusted \textit{up} by ${\sim} 0.3$ to compare to one estimated using \coone. Similarly, one would adjust the slope of $\Sigma_{\rm SFR}$ vs.\ $\Sigma_{\rm mol}$ measured from \cotwo\ \textit{up} by $\sim 0.15$ to compare to a \coone-based measurement.

Two issues complicate this approach. First recall that, as discussed in \S\ref{sec:expect}, \S\ref{sec:fits}, and Appendix~\ref{sec:corrections}, the line ratios likely saturate at or near~$1$. They also seem unlikely to drop to arbitrarily low values. As a result, the scaling of the line ratios with $\Sigma_{\rm SFR}$ likely occurs only over a bounded range. Because the scaling likely only occurs over a limited range of $\Sigma_{\rm SFR}$, simply adjusting the index of a power law fit by these values will likely over-correct when the scaling relation is measured over high dynamic range. For example, $R_{21}$ variations seem unlikely to strongly affect the slope of scaling relations measured for highly excited major mergers. A more conservative approach would be to only apply such a correction to more quiescent, ``normal'' galaxies and to add a one-sided component to the systematic uncertainty estimate to reflect the uncertainty in the magnitude of the term. E.g., say one measures a slope $\alpha$ for $\Sigma_{\rm SFR}$ vs.\ $\Sigma_{\rm mol}$ using \cotwo ; it would be reasonable and conservative to report a best estimate $\alpha^{\rm corr} = \alpha / (1 - 0.15 \alpha)$ with an additional component $\Delta \alpha \sim \alpha^{\rm corr} - \alpha$ added to the systematic uncertainty, e.g., as $^{+0.0}_{-\Delta \alpha}$.

Second, and perhaps even more important, the changes in physical conditions associated with variations in $R_{21}$, $R_{32}$, or $R_{31}$ also imply changes in the CO-to-H$_2$ conversion factor for \coone. As density, temperature, and opacity vary, so will the emissivity of the gas in \coone, generally with the sense that gas with higher $R_{21}$ and $R_{31}$ will show higher emissivity.

Modulo changes in the CO abundance, $N_{\rm CO}/N_{\rm H_2}$, the emissivity of the gas relates directly to $\alpha_{\rm CO}^{1-0}$. Of course, abundance variations do occur, and the amount of ``CO-faint'' gas represents a major consideration in the behavior of $\alpha_{\rm CO}$ across the galaxy population \citep[e.g.,][]{WOLFIRE10,GLOVER11,LEROY11,BOLATTO13B,SANDSTROM13}. In fact, one of the findings in \S\ref{sec:intcorr} is that CO excitation appears to increase in low mass galaxies where CO-faint gas will be more prevalent. The same diminished shielding, that leads to the CO abundance variations, may relate to the gas heating \citep[e.g.,][]{PENALOZA18}, so emissivity can change non-linearly with abundance variations. But for many purposes it is useful to think about $\alpha_{\rm CO}^{1-0}$ as being a separable problem, with the emissivity of CO and the abundance of CO-faint gas representing distinct factors that combined set $\alpha_{\rm CO}^{1-0}$ \citep[e.g., see][]{BOLATTO13B}. We proceed discussing only the emissivity portion of the problem.

Figure~\ref{fig:alphaco10} illustrates the behavior of \coone\ emissivity $\propto \alpha_{\rm CO}^{-1}$ in the model grid entries that match our measured ratios in Figure~\ref{fig:modelcomp}. Simply knowing the $^{12}$CO line ratios does not determine $\alpha_{\rm CO}$, but the figure shows that variations in $R_{21}$ (or the other ratios) will correlate with variations in $\alpha_{\rm CO}^{1-0}$ as long as other factors like $N_{\rm CO}/N_{\rm H_2}$ and $N_{\rm CO}/\Delta v$ remain approximately fixed. As discussed in \S\ref{sec:expect}, $N_{\rm CO}/\Delta v$ relates closely to optical depth, and is expected to vary across galaxies and especially to vary some between galaxies and galaxy centers, e.g., in response to a changing dynamical state of the molecular gas or the emergence of a diffuse CO component \citep[e.g.,][]{DOWNES98,BOLATTO13B}. Still observations of the $^{13}$CO/$^{12}$CO ratio in normal galaxies show a relatively narrow range \citep[e.g.,][and references therein]{CORMIER18}, implying a relatively stable optical depth, and it may be reasonable to imagine that within any given galaxy one can focus on a single ``color'' in Figure~\ref{fig:alphaco10}.

This simple illustration agrees qualitatively with the more detailed discussion and numerical results in \citet{GONG20}. \citet{GONG20} simulated portions of galaxies with realistic chemistry and radiative transfer to predict scaling relations relating $\alpha_{\rm CO}^{2-1}$ and $\alpha_{\rm CO}^{1-0}$, finding $\alpha_{\rm CO}^{1-0} \propto R_{21}^{-0.9}$, similar to the scaling at fixed $N_{\rm CO}/\Delta v$ in Figure~\ref{fig:alphaco10}.

This relationship between emissivity and the line ratio implies that the trends in Figures~\ref{fig:resolved_rad} and~\ref{fig:resolved_sfr} and Tables~\ref{tab:resolved} and~\ref{tab:central} can also be interpreted as closely relating to variations in $\alpha_{\rm CO}^{1-0}$, albeit with plenty of caveats. In this case, $\alpha_{\rm CO}^{1-0}$ drops by ${\sim} 0.2$~dex, or a factor of ${\sim} 1.6$, on average towards the centers of galaxies, and $\alpha_{\rm CO}^{1-0}$ shows gradients of order $0.1{-}0.2$~dex per decade in $\Sigma_{\rm SFR}$. Note that these variations all refer to $\alpha_{\rm CO}^{1-0}$. Translating, e.g., \cotwo\ intensity into $\Sigma_{\rm mol}$ further requires multiplying by $R_{21}^{-1}$, so that $\Sigma_{\rm mol} \propto \alpha_{\rm CO}^{1-0} R_{21}^{-1} I_{\rm CO}^{2-1} \sim R_{21}^{-2} I_{\rm CO}^{2-1}$. In practice, e.g., as discussed by \citet{GONG20}, there will be distinct relationships between $\alpha_{\rm CO}^{2-1}$ and $R_{21}$ or $R_{31}$. The key point here is not the exact prescription, but that we would expect the relationship between $\alpha_{\rm CO}^{2-1}$ or $\alpha_{\rm CO}^{3-2}$ and the line ratios to be steeper than that for $\alpha_{\rm CO}^{1-0}$. The conservative recommendations above would account for this additional steepness by normalizing scaling relations to effectively operate in terms of \coone , but they would \textit{not} account for the implied variations of $\alpha_{\rm CO}^{1-0}$.

Qualitatively, this highlights that mapping the low-$J$ CO line ratios offers one of our best options to trace out $\alpha_{\rm CO}^{1-0}$ variations in detail. Despite ambiguities, these line ratios represent a handle on excitation that can be surveyed across many galaxies. Because of the brightness of CO emission, ALMA or NOEMA can map these lines with high angular resolution, e.g., the PHANGS--ALMA survey prompted this analysis. Though ambiguities in the interpretation of these ratios exist, progress on numerical simulations \citep[e.g.,][]{SZUCS16,PENALOZA18,TRESS20,GONG20} are cause for optimism and many of the alternative approaches to map out $\alpha_{\rm CO}$ variations, e.g., isotopologue mapping or dust observations, involve \emph{a~priori} unknown additional free parameters, like the isotopologue abundance or coupling between dust opacity, dust-to-gas ratio, and environment \citep[see][for extensive discussion]{BOLATTO13A}.

\subsection{Biases, selection effects, and next directions}
\label{sec:biases}

To the best of our knowledge, our sample reflects the current state of the literature. However, our analysis also highlights that the field of extragalactic line ratios remains in development. We highlight three important issues, already visible from the analysis above:

\medskip

\textit{Most map pairs target massive galaxies on the star-forming main sequence.} As shown in Figures~\ref{fig:sfms}, \ref{fig:intcorr}, and~\ref{fig:intcorr2}, our measurements still span only a limited range of galaxy properties. To be concrete, galaxies with $R_{21}$ measurements (not limits) span a 16\%{-}84\% range of $\log_{10} M_\star \, [{\rm ~M}_\odot] = 10.3$ to $10.8$  (median $10.5$) and $\log_{10} {\rm SFR}/M_\star \, [{\rm yr}^{-1}] = -10.5$ to $-9.8$. Those for $R_{32}$ span a 16\%{-}84\% range of $\log_{10} M_\star \, [{\rm ~M}_\odot] = 10.1$ to $10.8$ (median $10.4$)  and $\log_{10} {\rm SFR}/M_\star \, [{\rm yr}^{-1}] = -10.5$ to $-9.9$. And galaxies with measured $R_{31}$ have a 16\%{-}84\% range of $\log_{10} M_\star \, [{\rm ~M}_\odot] = 10.2$ to $10.8$ (median $10.4$) and $\log_{10} {\rm SFR}/M_\star \, [{\rm yr}^{-1}] = -10.5$ to $-9.9$. 

These values represent only a narrow range concentrated near the high mass end of the star-forming main sequence. Combined with the already low number of measurements, this limits the ability to fit robust scaling relations and contributes to the weakness of the correlations measured between integrated galaxy properties and the line ratios (\S\ref{sec:intcorr}). An obvious path to make progress will be to expand the set of well-measured, beam-matched line ratios (either from mapping or carefully constructed galaxy-integrated experiments) to include lower mass, more actively star-forming, and more quiescent galaxies. In the near future, populating the regime of $\log_{10} M_\star \, [{\rm M}_\odot] \sim 9.5{-}10.2$ with high quality measurements would already dramatically improve our understanding of how line ratios vary in the $z=0$ galaxy population (see the sparse coverage in Figure~\ref{fig:intcorr2}).

\medskip

\textit{Many maps that do exist have limited sensitivity, especially \cothree .} Figures~\ref{fig:resolved_rad} and~\ref{fig:resolved_sfr} show many limits at large radii in the stacked profiles of the maps that do exist. For $R_{32}$, there are a large number of unconstraining upper limits at large radii, reflecting the poor sensitivity of the \cothree\ compared to the \cotwo\ maps. For $R_{21}$ the situation is a bit better, but more extended, sensitive \coone\ maps are needed. Meanwhile the radial extent of $R_{31}$ coverage remains very poor and clearly more sensitive mapping is needed in both transitions.

\medskip

\textit{Calibration issues induce scatter of the same order as the dynamic range of the physical variations in the ratios.} In Table~\ref{tab:intrats}, the 16-84\% range of $R_{21}$ is $0.22$ dex, with $R_{31}$ and $R_{32}$ showing slightly wider ranges of $0.32$ dex and $0.41$ dex. As discussed in \S\ref{sec:codata} and shown in Figure~\ref{fig:consistency}, current calibration uncertainties are imperfectly known but likely 20{-}25\% for the non-ALMA data. This uncertainty in the calibration alone introduces $\approx 0.1{-}0.15$~dex rms scatter in the line ratios. Expressed as a 16{-}84\% range this is $0.2{-}0.3$~dex, of the same order as the range in line ratios themselves. The best path forward appears to be larger internally consistent or carefully cross-calibrated data sets.

\section{Summary}
\label{sec:summary}

We combine a large set of publicly available maps of \coone , \cotwo , and \cothree\ emission from nearby galaxies with the new PHANGS--ALMA \cotwo\ survey to measure low-$J$ CO line ratios for \pairs\ CO map pairs (see Table~\ref{tab:sample}). The full sample spans $M_\star \sim 10^9{-}10^{11}~\mathrm{M}_\odot$ but consists mostly of relatively massive ($\log_{10} M_\star \, [\mathrm{M}_\odot] \approx 10.25{-}11.0$), star-forming ($\SFR \sim 1{-}5~\mathrm{M}_\odot$~yr$^{-1}$) galaxies that lie near the star-forming main sequence (see Figure~\ref{fig:sfms}). 

These maps of low-$J$ CO line ratios across normal galaxies complement earlier detailed multi-transition studies \citep[e.g.,][]{PAPADOPOULOS02,ISRAEL01,ISRAEL03,PAPADOPOULOS12,ISRAEL15,BAYET04,BAYET06,KAMENETZKY14,KAMENETZKY17}, which tended to focus on galaxy-integrated measurements, galaxy centers, and starburst galaxies, but often employed a wider set of transitions than we have available here, including CO isotopologues and higher $J$ lines. We also extend earlier mapping work on samples by \citet{WILSON12,LEROY13,YAJIMA21,DENBROK21}, which used smaller samples, earlier versions of some of the same data, and focused mostly on a single line ratio. For this study a key addition, and the motivating data set for this study, is the new PHANGS--ALMA \cotwo\ mapping survey \citep{LEROY21b}.

Tables~\ref{tab:meas}, \ref{tab:intrats}, and~\ref{tab:fits} report our results for whole galaxies. Briefly:

\begin{enumerate} 

\item Integrating over whole map pairs, we find  $R_{21} = 0.50{-}0.83$ with mean $0.65$, $R_{32} = 0.22{-}0.58$ with mean $0.47$, and $R_{31} = 0.16{-}0.44$ with mean $0.32$. We compare these to literature measurements of the same ratios studying nearby galaxies (\S\ref{sec:global}, Figure~\ref{fig:hist}) and find overall consistency, though our $R_{31}$ values appear somewhat higher than previously reported for nearby galaxies. Area-matched $R_{32}$ and $R_{31}$ measurements for nearby galaxies remain relatively scarce, dominated by the JCMT NGLS \citep{WILSON12} and we note this as a productive area for new observations.

\end{enumerate}

We search for correlations between the low-$J$ CO line ratios and integrated galaxy properties (\S\ref{sec:intcorr}, Figures~\ref{fig:rankcorr} and~\ref{fig:intcorr}, Table~\ref{tab:fits}). Such correlations remain hard to discern, partially due to the limited diversity of galaxies with measured ratios (Figure~\ref{fig:sfms}) and partially because calibration uncertainties associated with the data are of the same order as the dynamic range of the ratio in the nearby galaxy population (\S\ref{sec:meas}, Figure~\ref{fig:consistency}). A secondary issue is that some of the most physically meaningful comparisons (e.g., between the line ratios and \SFR-per-CO) involve correlated axes that can strongly influence the inferred trends (see \S\ref{sec:global}). Despite this:

\begin{enumerate}
\setcounter{enumi}{1}
\item We identify a consistent set of marginally significant correlations between all three line ratios, $R_{21}$, $R_{32}$, and $R_{31}$, and quantities that trace normalized star formation activity. The line ratios anti-correlate with stellar mass ($M_\star$) and CO luminosity ($L_{\rm CO}$), and positively correlate with specific star formation rate ($\SFR/M_\star$) and \SFR-per-CO ($\SFR/L_{\rm CO}$). These correlations have the sense that both low mass dwarf galaxies and starburst galaxies should show high line ratios. This agrees with physical expectations and previous observations that dwarf galaxies have high excitation, poorly shielded molecular gas and that starburst galaxies have high CO excitation.
\end{enumerate}

We measure local variations of each line ratio within galaxies (\S\ref{sec:local}, Figures~\ref{fig:resolved_rad} and~\ref{fig:resolved_sfr}, Tables~\ref{tab:resolved} and~\ref{tab:central}). In this analysis, we control for galaxy-to-galaxy scatter and global calibration uncertainties by focusing on the line ratio normalized to the galaxy-average. We examine how line ratio variations within galaxies correlate with galactocentric radius, local star formation surface density, and local specific star formation rate, also normalized to the galaxy average. We find:

\begin{enumerate}
\setcounter{enumi}{2}
\item Most galaxies with measurements show enhanced values of all ratios in the central $1$~kpc-wide or $0.5~r_{\rm eff}$-wide bin in our analysis (Table~\ref{tab:central}). $R_{31}$ shows the strongest central enhancements, $0.27$~dex on average, followed by $R_{21}$ with median $0.18$~dex, and then $R_{32}$ with $0.08$~dex. These central enhancements have been noted before, especially in $R_{21}$, but this study represents the largest systematic measurement for all three lines to date. The sense of these variations agrees with the expectation that compared to the gas in galaxy disks, the gas in galaxy centers is denser, more actively star-forming, can be heated by active galactic nuclei, and perhaps includes an optically thinner diffuse component due to high turbulence.

\item Within galaxies, all three line ratios also show significant internal gradients as a function of radius, and $\Sigma_{\rm SFR}$. $R_{21}$ and $R_{31}$ also show significant gradients as a function of $\Sigma_{\rm SFR}/\Sigma_\star$. We report fits to the magnitude of these gradients in Table~\ref{tab:resolved}. This behavior agrees with the expectation that the more active parts of galaxies host hotter, denser gas, and reflects the same underlying trend as the central enhancements.

\end{enumerate}

Finally, we note some implications of our measurements:

\begin{enumerate}
\setcounter{enumi}{5}

\item We consider a set of simple non-LTE models that treat density distributions (\S\ref{sec:expect} and Appendix~\ref{sec:comodels}) and note the physical conditions in the models that satisfy our measurements of all three lines (\S\ref{sec:modelcomp}). Our observed ratios can broadly be reproduced by cold gas with moderate density and intermediate column density per line width, $N_{\rm CO}/\Delta v$.

\item Following \citet{GONG20} and illustrated using our own model grid, we highlight that these line ratio variations also imply corresponding variations in $\alpha_{\rm CO}^{1-0}$ because the changes in physical conditions tracked by the CO line ratio imply changes in the CO emissivity (\S\ref{sec:alphaco}). These $\alpha_{\rm CO}$ variations will compound with those caused by ``CO-faint'' gas to produce the overall variations of $\alpha_{\rm CO}$ in galaxies.

\end{enumerate}

Together these results paint a basic picture of how the low-$J$ CO line ratios vary across the local galaxy population, with dwarf galaxies and starburst galaxies showing enhanced excitation, and galaxy centers and high $\Sigma_{\rm SFR}$ regions also being more excited. 

We close by emphasizing that much more work is needed. The major limitations of the present study, and the field in general, are signal-to-noise in the \coone\ and \cothree\ data, uncertainties in the calibration, and the limited sampling of the full galaxy population. The PHANGS--ALMA \cotwo\ maps offer a high quality, high signal-to-noise, well-calibrated starting point that spans the local galaxy population. One next major step will be to measure a large, diverse sample of galaxies in all three lines with well-controlled calibration, excellent signal-to-noise, and good resolution. This will sharpen our knowledge of integrated galaxy scaling relations and allow us to investigate resolved line excitation variations across galaxies without the need for stacking or the aggressive normalizations performed here.

\acknowledgments

We thank the anonymous referee for a constructive and thorough report that significantly improved the quality of this work.

This work was carried out as part of the PHANGS collaboration.

The work of AKL and JS was partially supported by the National Science Foundation (NSF) under Grants No.1615105, and 1653300, as well as by the National Aeronautics and Space Administration (NASA) under ADAP grants NNX16AF48G and NNX17AF39G. 

ER acknowledges the support of the Natural Sciences and Engineering Research Council of Canada (NSERC), funding reference number RGPIN-2017-03987, and computational support from Compute Canada.

AU acknowledges support from the Spanish funding grants PGC2018-094671-B-I00 (MCIU/AEI/FEDER) and PID2019-108765GB-I00 (MICINN). 

KS and IDC acknowledge funding support from National Science Foundation grant No. 1615728 and NASA ADAP grants NNX16AF48G and NNX17AF39G.

ES, HAP, TS, and TGW acknowledge funding from the European Research Council (ERC) under the European Union’s Horizon 2020 research and innovation programme (grant agreement No. 694343).

JPe acknowledges support by the Programme National ``Physique et Chimie du Milieu Interstellaire'' (PCMI) of CNRS/INSU with INC/INP, co-funded by CEA and CNES.

MC and JMDK gratefully acknowledge funding from the Deutsche Forschungsgemeinschaft (DFG) in the form of an Emmy Noether Research Group (grant number KR4801/1-1) and the DFG Sachbeihilfe (grant number LR4801/2-1), as well as from the European Research Council (ERC) under the European Union’s Horizon 2020 research and innovation programme via the ERC Starting Grant MUSTANG (grant agreement number 714907).

RSK and SCOG acknowledge funding from the European Research Council via the ERC Synergy Grant ``ECOGAL -- Understanding our Galactic ecosystem: From the disk of the Milky Way to the formation sites of stars and planets'' (project ID 855130). They also acknowledges support from the DFG via the Collaborative Research Center (SFB 881, Project-ID 138713538) ``The Milky Way System'' (sub-projects A1, B1, B2 and B8) and from the Heidelberg cluster of excellence (EXC 2181 - 390900948) ``STRUCTURES: A unifying approach to emergent phenomena in the physical world, mathematics, and complex data'', funded by the German Excellence Strategy.  

ATB, FB and JdB would like to acknowledge funding from the European Research Council (ERC) under the European Union’s Horizon 2020 research and innovation programme (grant agreement No.726384/Empire).

CE acknowledges funding from the Deutsche Forschungsgemeinschaft (DFG) Sachbeihilfe, grant number BI1546/3-1.

MQ acknowledges support from the research project PID2019-106027GA-C44 from the Spanish Ministerio de Ciencia e Innovaci\'on.

AS is supported by an NSF Astronomy and Astrophysics Postdoctoral Fellowship under award AST-1903834.

This work is based on observations carried out under project numbers 169-06, 053-07, 122-07, 160-06, 218-05, 058-08, 212-08, 196-13, 078-14, and 190-14 with the IRAM 30 m telescope. IRAM is supported by INSU/CNRS (France), MPG (Germany) and IGN (Spain).

This work is based on COMING anbd the Nobeyama Atlas of Nearby Spiral Galaxies, which are both legacy programs of the Nobeyama 45 m radio telescope, which is operated by Nobeyama Radio Observatory, a branch of National Astronomical Observatory of Japan. We gratefully acknowledge both teams for making their data public and acknowledge the hard work of the teams and staff to obtain the data.

We also thank the JCMT NGLS team and the JCMT staff for their hard work and making their data public, and we acknowledge helpful correspondence with the JCMT staff regarding calibration uncertainties. The James Clerk Maxwell Telescope is operated by The Joint Astronomy Centre on behalf of the Science and Technology Facilities Council of the United Kingdom, the Netherlands Organisation for Scientific Research and the National Research Council of Canada.

This paper makes use of the following ALMA data, which have been processed as part of the PHANGS--ALMA \cotwo\ survey: \\
\noindent ADS/JAO.ALMA\#2012.1.00650.S, \linebreak 
ADS/JAO.ALMA\#2013.1.00803.S, \linebreak 
ADS/JAO.ALMA\#2013.1.01161.S, \linebreak 
ADS/JAO.ALMA\#2015.1.00121.S, \linebreak 
ADS/JAO.ALMA\#2015.1.00782.S, \linebreak 
ADS/JAO.ALMA\#2015.1.00925.S, \linebreak 
ADS/JAO.ALMA\#2015.1.00956.S, \linebreak 
ADS/JAO.ALMA\#2016.1.00386.S, \linebreak 
ADS/JAO.ALMA\#2017.1.00392.S, \linebreak 
ADS/JAO.ALMA\#2017.1.00766.S, \linebreak 
ADS/JAO.ALMA\#2017.1.00886.L, \linebreak 
ADS/JAO.ALMA\#2018.1.01321.S, \linebreak 
ADS/JAO.ALMA\#2018.1.01651.S, \linebreak 
ADS/JAO.ALMA\#2018.A.00062.S, \linebreak 
ADS/JAO.ALMA\#2019.1.01235.S, \linebreak 
ADS/JAO.ALMA\#2019.2.00129.S, \linebreak 
ALMA is a partnership of ESO (representing its member states), NSF (USA), and NINS (Japan), together with NRC (Canada), NSC and ASIAA (Taiwan), and KASI (Republic of Korea), in cooperation with the Republic of Chile. The Joint ALMA Observatory is operated by ESO, AUI/NRAO, and NAOJ. The National Radio Astronomy Observatory is a facility of the National Science Foundation operated under cooperative agreement by Associated Universities, Inc.


\begin{appendix}

\section{CO line emission from density distributions}
\label{sec:comodels}

 \begin{deluxetable*}{cccccccc} 
 \tablecaption{CO Emission from Multi-Density Models \label{tab:comodel}} 
 \tablewidth{0pt} 
 \tabletypesize{\footnotesize} 
 \tablehead{ 
 \colhead{$T_{\rm kin}$} & 
 \colhead{$N_{\rm CO}/\Delta v$} & 
 \colhead{$n_{\rm 0,H2}$} & 
 \colhead{$\sigma$} & 
 \colhead{$\epsilon_{1-0}$} & 
 \colhead{$R_{21}$} & 
 \colhead{$R_{32}$} & 
 \colhead{$R_{31}$}  
 \\ 
 \colhead{(K)} & 
 \colhead{(cm$^{-2}$ (km s$^{-1}$)$^{-1}$)} & 
 \colhead{(cm$^{-1}$)} & 
 \colhead{(dex)} & 
 \colhead{(K~km~s$^{-1}$ (cm$^{-2}$)$^{-1}$)} & 
 \colhead{} & 
 \colhead{} &  
 \colhead{} 
 }  
\startdata 
10 & 1.000E+15 & 3.160E+01 & 0.2 & 7.779E-21 & 0.17 & 0.05 & 0.01\\ 
10 & 1.000E+15 & 3.160E+01 & 0.3 & 9.791E-21 & 0.18 & 0.05 & 0.01\\ 
10 & 1.000E+15 & 3.160E+01 & 0.4 & 1.317E-20 & 0.19 & 0.05 & 0.01\\ 
10 & 1.000E+15 & 3.160E+01 & 0.5 & 1.818E-20 & 0.22 & 0.06 & 0.01\\ 
10 & 1.000E+15 & 3.160E+01 & 0.6 & 2.490E-20 & 0.26 & 0.07 & 0.02\\ 
10 & 1.000E+15 & 3.160E+01 & 0.7 & 3.306E-20 & 0.32 & 0.10 & 0.03\\ 
10 & 1.000E+15 & 3.160E+01 & 0.8 & 4.208E-20 & 0.40 & 0.12 & 0.05\\ 
10 & 1.000E+15 & 3.160E+01 & 0.9 & 5.130E-20 & 0.48 & 0.16 & 0.07\\ 
10 & 1.000E+15 & 3.160E+01 & 1.0 & 6.012E-20 & 0.56 & 0.19 & 0.11\\ 
10 & 1.000E+15 & 3.160E+01 & 1.1 & 6.811E-20 & 0.66 & 0.23 & 0.15\\ 
\enddata 
\tablecomments{This table is a stub. The full version of the table appears as a machine readable table 
in the online version of the paper. Columns give: 
$T_{\rm kin}$ --- kinetic temperature of all zones in the model; 
$N_{\rm CO}/\Delta v$ --- column of CO per line width in all zones in the model; 
$n_{\rm 0,H2}$ --- mean collider density for the lognormal distribution of densities in the model; 
$\sigma$ --- r.m.s. width $\sigma$ of the lognormal distribution of densities in the model; 
$\epsilon_{1-0}$ --- emissivity of H$_2$ in the \coone\ transition assuming a fixed CO/H$_2$ abundance of 1E-4; 
$R_{21}$ --- $\cotwo / \coone$ line ratio for the model; 
$R_{32}$ --- $\cothree / \cotwo$ line ratio for the model; 
$R_{31}$ --- $\cothree / \coone$ line ratio for the model; 
see \citet{LEROY17B} for more details regarding the calculation.} 
\end{deluxetable*}

Table~\ref{tab:comodel} presents predicted line ratios, $R_{21}$, $R_{32}$, and $R_{31}$, and \coone\ emissivity, $\epsilon_{1-0}$, for models with distributions of densities and fixed temperature, $T_{\rm kin}$, and column of CO per unit line width, $N_{\rm CO}/\Delta v$. Following \citet{LEROY17B}, we consider lognormal density distributions with mean density $n_0$ and width in log density space, $\sigma$. The predicted ratios represent the sum over an ensemble of zones that share $T_{\rm kin}$ and $N_{\rm CO}/\Delta v$ but each have a distinct density. 

The methods mostly follow \citet{LEROY17B} with atomic data from LAMDA \citep{SCHOIER05} and calculations via RADEX \citep{VANDERTAK07}. Distinct from \citet{LEROY17B}, we assume a single fixed $N_{\rm CO}/\Delta v$ across all zones. That paper assumes a fixed optical depth, $\tau$, and selects $N/\Delta v$ to match that opacity. The current approach is better suited to treat multi-transition measurements like CO line ratios, while adopting a fixed $\tau$ as in \citet{LEROY17B} may be more appropriate to single transition studies, e.g., \hcnone\ only, where $\tau$ may be at least roughly constrained from isotopologue studies \citep[e.g.,][]{JIMENEZDONAIRE17B}. We note that for fixed abundance, $N_{\rm CO}/N_{\rm H2}$, $N_{\rm CO}/\Delta v$ will relate to the collider density $n_{\rm H2}$ within a layer via a combination of size and line width that can vary from zone to zone. Since we model only a single species here, there is no implied inconsistency other than the zones having variable size or structure. This makes no difference to our modelling, which does not consider zone size as an important variable. However, as discussed by \citet{LEROY17B} when modeling multiple species fixing optical depth can lead to implied zone-to-zone abundance variations.

The models have the advantage of incorporating a realistic mixture of densities, which is certainly present in any low resolution observations of galaxies. They have a suite of caveats, discussed at length in \citet{LEROY17B}. Here we only note that we have assumed that the zones all share fixed $T_{\rm kin}$, fixed $N_{\rm CO}/\Delta v$, and that they do not ``shadow'' one another, i.e., that we observe the linear combination of emission from all zones. We also only consider a lognormal density distribution in this paper, because the modeling is not a central focus of this work. Beyond this, all of the usual caveats related to RADEX modeling apply, and we refer the reader to \citet{VANDERTAK07} for more details.

\section{Illustration of the effect of the cosmic microwave background and the Rayleigh--Jeans approximation}
\label{sec:corrections}

\begin{figure*}[ht!]
\begin{center}
\includegraphics[width=0.4\textwidth]{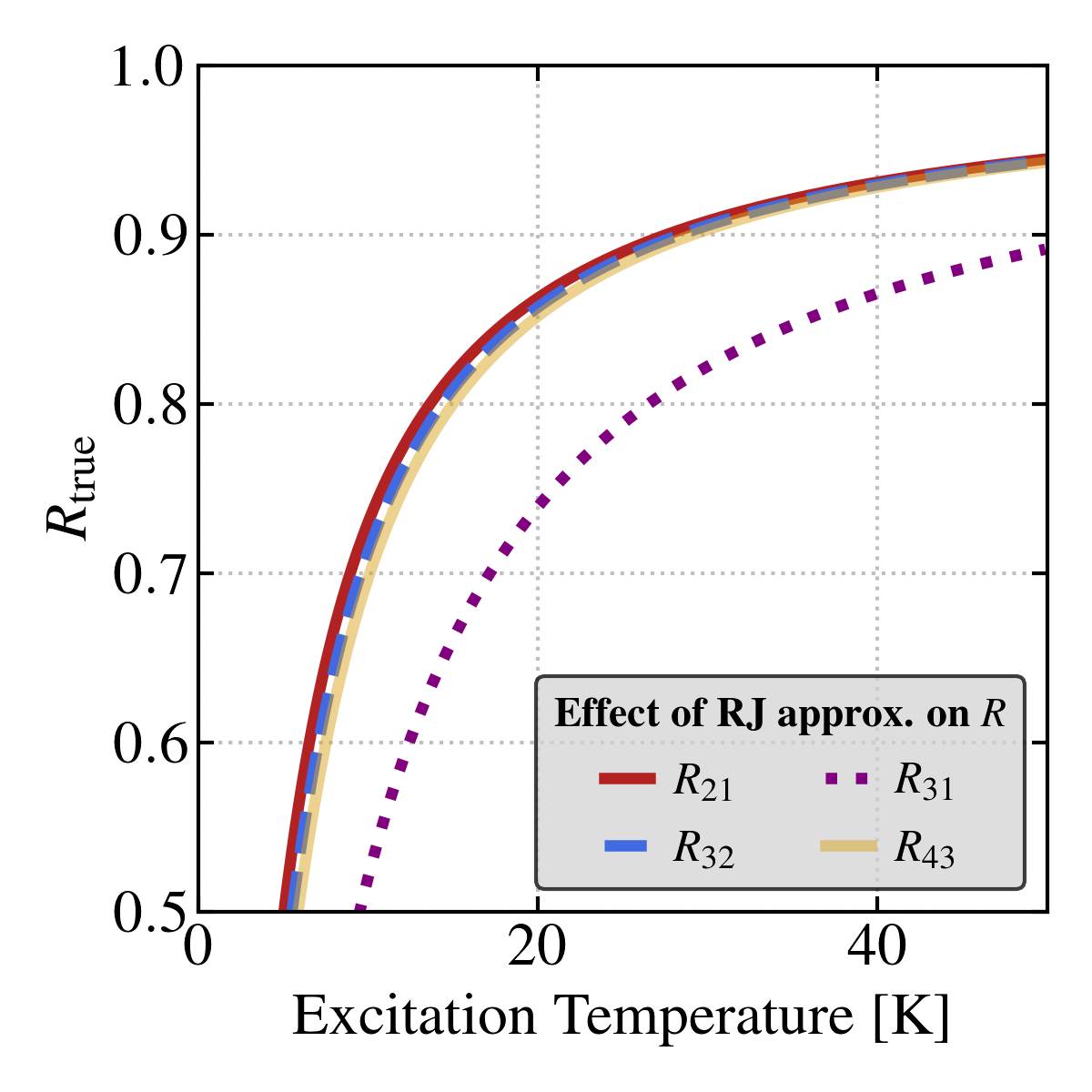}
\includegraphics[width=0.4\textwidth]{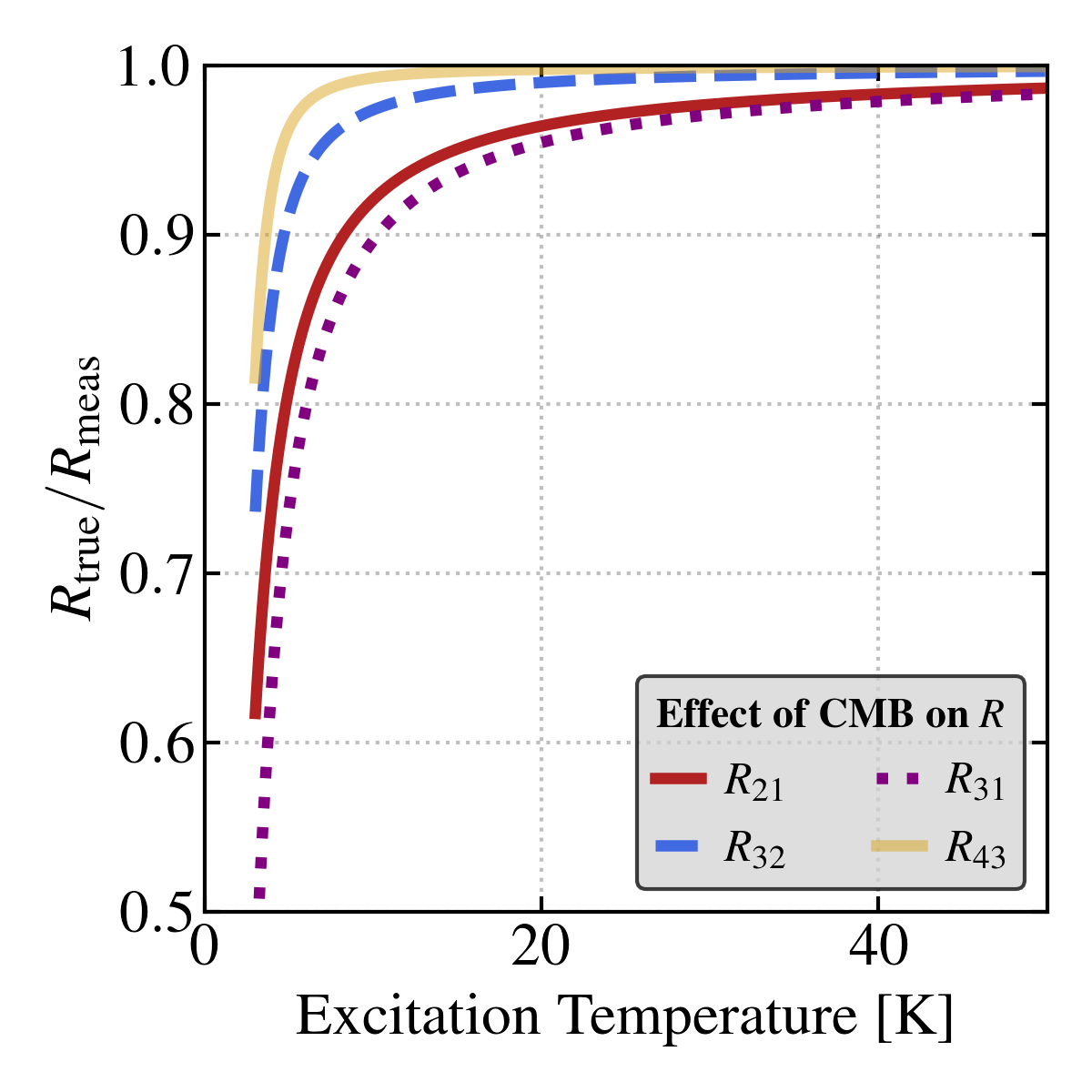}
\end{center}
\caption{\textit{Effect of the cosmic microwave background and Rayleigh--Jeans approximation on measured ratios.} For reference, we illustrate the effect of the Rayleigh--Jeans approximation (\textit{left}) and cosmic microwave background (CMB) (\textit{right}) for line ratios measured from perfectly opaque sources ($\tau \gg 1$). The left panel shows the expected $R_{\rm true}$ for a thermalized, opaque source when $R$ is expressed in Rayleigh--Jeans (RJ) brightness temperature. The increased inaccuracy of the RJ approximation at low temperature leads to ``thermal'' line ratios with values $R_{\rm true} < 1$. The right panel shows how the measured $R_{\rm meas}$ deviates from $R_{\rm true}$ due to the effects of the CMB, plotting $R_{\rm true}/R_{\rm meas}$ as a function of the excitation temperature for a CMB temperature of $2.73$~K. The CMB tends to increase the observed line ratio because it affects low-$J$ lines more. The two effects act in opposite directions, with $R_{\rm true} < 1$ but then $R_{\rm meas}$ increased by the effects of the CMB. Both are well-handled by modeling but important to bear in mind when interpreting measured line ratios. Correcting observations with low physical resolution for the effects of the CMB can be particularly challenging.
\label{fig:corrections}}
\end{figure*}

Figure~\ref{fig:corrections} illustrates the impact of the CMB and use of the Rayleigh--Jeans approximation on measured CO line ratios. These effects are well known, but we are not aware of a clean illustration of the impact of both on this set of ratios and we found these plots useful to interpret our measurements, so we include them here \citep[as above, we note discussions in][]{DACUNHA13,ZSCHAECHNER18,BOLATTO13B,ECKART90}.

The right panel shows the effect of measuring line ratios in contrast against the 2.73~K CMB \citep{FIXSEN96}. The effect of the CMB is to selectively suppress low-$J$ emission, leading the measured ratio to be higher than the true value expected for the source without any CMB. By contrast, the effect of using the Rayleigh--Jeans approximation, shown in the left panel, is that for sources with temperatures in range of real molecular clouds, we expect ``thermal'' line ratios (i.e., the value for an opaque source in LTE) $<1$, with lower values for colder objects. The two effects somewhat cancel out. 

As noted above, the magnitude of the CMB effect on real observations of galaxies can be difficult to gauge without information on the small-scale structure of the emission. The radiative transfer involving the CMB plays out on the scale of individual clouds and the intensity is then subject to a large beam dilution effect before entering the sort of measurements presented in this paper. Both effects are accounted for by models \citep[e.g., RADEX;][]{VANDERTAK07}, but coupling those models to observations can require estimates (or addition of a free parameter) of beam filling.

\end{appendix}

\end{document}